%% file: main.tex
\documentclass[a4,12pt]{article}

\usepackage[utf8]{inputenc} % allow utf-8 input
\usepackage[T1]{fontenc}    % use 8-bit T1 fonts
\usepackage[hidelinks]{hyperref}       % hyperlinks
\usepackage{url}            % simple URL typesetting
\usepackage{booktabs}       % professional-quality tables
\usepackage{amsfonts}       % blackboard math symbols
\usepackage{nicefrac}       % compact symbols for 1/2, etc.
\usepackage{microtype}      % microtypography
\usepackage{lipsum}
\usepackage{fancyhdr}       % header
\usepackage{graphicx}       % graphics
\graphicspath{{media/}}     % organize your images and other figures under media/ folder
\usepackage{natbib}
\usepackage{amsmath}
\usepackage{amsmath}
\usepackage{amssymb}
\usepackage{blkarray}
\usepackage{multirow}
\usepackage{graphicx}
\usepackage{subfigure}
\usepackage{subcaption}
\usepackage[linesnumbered,ruled,english]{algorithm2e}
\title{Generation and analysis of synthetic data via Bayesian networks: a robust approach for uncertainty quantification via Bayesian paradigm
}

\author{
  Larissa N. A. Martins, Fl\'{a}vio B. Gonçalves, Thais P. Galletti \\
  Universidade Federal de Minas Gerais, Brazil\\
}

\begin{document}
\maketitle

\renewcommand{\abstractname}{Abstract}
\begin{abstract}
Safe and reliable disclosure of information from confidential data is a challenging statistical problem. A common approach considers the generation of synthetic data, to be disclosed instead of the original data. Efficient approaches ought to deal with the trade-off between reliability and confidentiality of the released data. Ultimately, the aim is to be able to reproduce as accurately as possible statistical analysis of the original data using the synthetic one. 
Bayesian networks is a model-based approach that can be used to parsimoniously estimate the underlying distribution of the original data and generate synthetic datasets. These ought to not only approximate the results of analyses with the original data but also robustly quantify the uncertainty involved in the approximation. This paper proposes a fully Bayesian approach to generate and analyze synthetic data based on the posterior predictive distribution of statistics of the synthetic data, allowing for efficient uncertainty quantification. The methodology makes use of probability properties of the model to devise a computationally efficient algorithm to obtain the target predictive distributions via Monte Carlo. Model parsimony is handled by proposing a general class of penalizing priors for Bayesian network models. Finally, the efficiency and applicability of the proposed methodology is empirically investigated through simulated and real examples.


\end{abstract}
%%%%%%%%%%%%%%%%%%%%%%%%%%%%%%%%%%%%%%%%%%%%%%%%%%%%%%%%%%%%%%%%%%%%%%%%%%%%%%
\input{Intro}

\input{3.0main}
\input{3.3Priori}

\input{3.1Modelo}
\input{3.2Simula_d}

\input{4.0main}
\input{4.1.1Prioris}
\input{4.1.2Estima_rede}

\input{4.1.3Dados_sint}
\input{4.2Dados_reais}

\input{Consideracoes}

\newpage
%Bibliography
\bibliographystyle{apalike}  
\bibliography{references}  

 \section*{Appendix A - Further results from the simulations}

\input{Z_apendice}

\end{document}

%% file: Intro.tex
\section{Introduction}

One of the goals of agencies and institutions dealing with information dissemination is the secure publication of data \citep{data2001confidentiality, surendra2017review, raghunathan2021synthetic}. These institutions have been investing in research to improve the way data is released, increasing the level of disseminated information while preserving the confidentiality and privacy of the data \citep{karr_Reiter2014, reiter2023synthetic}.
Some of the simplest alternatives for altering data to be disclosed include: data aggregation, where multiple observations are condensed according to categorical variables; data swapping, where sensitive information is exchanged between pairs of records with similar characteristics; and suppression, where high-risk values are excluded from the database \citep{reiter_2003}.

Although these methods of data alteration can solve the problem of secure disclosure, the statistical analyses with the altered data are often limited and can lead to results that are very different from the analyses conducted on the original data \citep{kennickell2006measuring}. This motivates the development of methodologies that can protect the data but also allow for the performance of reliable statistical analyses.

One technique that allows for secure disclosure and also addresses statistical analyses for shared data is the synthetic data methodology proposed by \cite{rubin_1993}. The generation of synthetic data can be seen as an imputation method, where the ``missing'' values are actually the sensitive values that cannot be disclosed \citep{reiter2005}. In the usual synthetic data generation approach, one or more datasets with imputed values are created from the estimated probability distributions in the original data. Thus, the imputed data will have probability distributions similar to those of the real data, whilst preserving the confidentiality of the latter. Additionally, it is possible to make inferences very close to those that would be made on the original data \citep{rubin_1993}.

A variety of techniques has been used to describe the relationship between sensitive variables and generate synthetic data. Some examples can be found in \cite{raghunathan2003multiple, drechsler_Reiter_2011, drechsler2011synthetic}. In these works, the authors use methodologies such as multiple imputation and non-parametric methods like classification trees and random forests. 

One general methodology to describe the relationship between variables are Bayesian networks (BN) \citep{koller2009probabilistic}. The goal of BNs is to describe the joint distribution of random variables in a robust and parsimonious manner by considering conditional independence structures among them \citep{friedman_eal2003}. A Bayesian network has two components: a graphical model, that defines the conditional independence structures, and the probability distributions for the respective conditional distributions used to decompose the joint one \citep{friedman_1997bayesian}.

\cite{young2009using} and \cite{sun_Erath2015} use Bayesian network models to generate synthetic discrete data.
The former uses an existing software to estimate the network under a Bayesian approach with strong prior assumptions on the network. Synthetic data is then generate either from the highest posterior probability network or by model averaging a couple of estimated networks. In the latter,
the authors estimate a network by approximately maximizing a reasonable score function via Tabu search. After this step, synthetic data is generated from the estimated network.

The literature on estimation of Bayesian networks consider both classical and Bayesian methods. The former usually consider a score function that involves both fitting and complexity structures and searches the network that maximizes such function \citep[see, for example][]{cooper1992bayesian,cano2004applications,di2004bayesian,hruschka2004feature,mihaljevic2021bayesian}. The estimation of an ``optimal'' network, however, compromises the quantification of the uncertainty involved in the estimation process and may lead to considerably poor analyses, in particular, in a synthetic data context.
Bayesian approaches, on the other hand, offer suitable uncertainty quantification through the posterior distribution of the network. \cite{heckerman1995learning} analyzes methods to search for the highest posterior probability networks under strong prior assumptions. \cite{koller2009probabilistic}, \cite{grzegorczyk2008improving}, and \cite{goudie2016gibbs}, on the other hand, use Markov Chain Monte Carlo (MCMC) algorithms to sample from the posterior distribution of the network. The algorithm from \cite{goudie2016gibbs} is a random scan Gibbs sampling algorithm with blocks defined by subsets of the network variables. Their algorithm is compared to other MCMC ones, that often rely on Metropolis-hasting moves, and is shown to have a considerably superior performance.

In the context of synthetic data, quantifying uncertainty is a fundamental task for studying the quality of the generated data. Although point estimate networks may be relevant, they do not fulfill this task. It is then reasonable to consider a full Bayes approach in which all the uncertainty involved in the inference process, from the network estimation to the synthetic data generation, is properly quantified. Furthermore, it is also desired that no strong prior assumptions on the network are required and robust model complexity penalization can be made through the prior distribution. A methodology with all of these features has not yet been considered in the literature.

The aim of this paper is to propose a fully Bayesian methodology for the generation and analysis of synthetic data using Bayesian networks in the context of binary data. The primary motivation for this approach is to enable efficient and robust quantification of all the uncertainty involved in the analysis of synthetic data, without the need for strong assumptions about the data-generating model.
The estimation of the Bayesian network is performed with the MCMC algorithm from \cite{goudie2016gibbs} and a general class of penalizing prior distributions for the network is proposed and discussed. The generation and analysis of synthetic data is performed through the posterior predictive distribution of statistics of those data. The methodology uses probabilistic properties of the model to achieve computational efficiency in all the steps of the analysis, including the ones performed by the final user, who does not have access to the original data.

This paper is organized as follows. Section 2 presents the proposed methodology for the generation and analysis of synthetic data via Bayesian networks. Simulated and real data analyses are presented in Section 3. Section 4 brings some final remarks.

%% file: 3.0main.tex
\section{Proposed Methodology}

A Bayesian network can be described as a probabilistic model that represents relationships between variables. These relationships can be presented in the form of a directed acyclic graph (DAG), in which the nodes represent the variables, and the arrows represent the conditional dependence between them \citep{heckerman_2008}.

Let $(\Omega, \mathcal{F}, P)$ be the probability space in which $d$ univariate random variables $X=(X_1,\dots,X_d)$ are defined and consider the inference problem of estimating $P$ based on an i.i.d. sample of $X$. Define the statistical model $\mathcal{P}$ as the collection of all the probability measures in $(\Omega, \mathcal{F})$ admitted for $X$, including the true one $P$. In particular, let $\mathcal{P}$ be a Bayesian network with a specific form for the respective conditional distributions. The Bayesian network model considers all possible factorizations of the joint distribution of $X$, which can be represented by the space of DAGs with each $X_j$ being a node. Finally, define $\theta$ as the set of parameters indexing at least one model in $\mathcal{P}$.

A particular DAG $G$ defines the conditional independence relations among the $X_j$ variables. The most complex $G$ is given by the complete network, i.e.,
\begin{equation}
p(\mathbf{X})=p(X_1)p(X_2|X_1)p(X_3|X_1,X_2)\dots p(X_d|X_1\dots X_{d-1}),
    \label{eq1}
\end{equation}
for this or any other ordering of the $X_j$ variables, where $p$ is a density or a probability function.
Simpler networks simplify any possible ordering of (\ref{eq1}) by assuming conditional independence structures to at least one of the conditional distributions in the factorization.

Each network $G$ can be expressed as a $d \times d$ matrix of zeros and ones such that the $ij$-th element is 1, if $ X_{j} $ indexes the conditional distribution of $ X_{i} $. In this case, $X_j$ is called a parent of $X_i$ and $X_i$ is a child of $X_j$. 

Let $ pa(X_j)$ denote the vector of parents of $X_j$. Thus, conditional on $G$, the joint distribution of $X$ can be factorized as
\begin{equation}
  \begin{aligned}
p(X) &= p(X_1) \, p\big( X_2|pa(X_2) \big) \, 
...p\big( X_d|pa(X_d) \big)
  \end{aligned}
  \label{eq3}
\end{equation}

Whilst the complete network (\ref{eq1}) is the most complex one, in the simplest network, the $d$ variables are mutually independent. The idea of a Bayesian network is to reasonably balance the trade-off between fitting and parsimony to explain the variability of $X$. This paper considers all the variables $X_j$ to be binary. Therefore, a variable with $k$ parents, defines $2^k$ Bernoulli distributions.

Several works dealing with the estimation of Bayesian networks can be found in the literature, under both classical and Bayesian approaches. As it was mentioned before, the latter allows for robust uncertainty quantification through the posterior distribution of $G$. Among all the Bayesian inference methods proposed, the most efficient and general one is the MCMC algorithm proposed by \cite{goudie2016gibbs}. The methodology not only is efficient to sample from the posterior distribution of $G$ for considerably sized datasets as it does not require strong prior assumption in the network structure. The only requirement is that the prior on the $G$ is modular, meaning that it has the form \begin{align}
    p(G) & \propto \prod_jp(pa(X_j)).
\end{align}

Other Bayesian methods for inference in Bayesian networks can be found in 
\citet{friedman_eal2003}, \citet{koivisto_eal2004}, \citet{ellis2008learning}, \citet{niinimaki2016structure}, \cite{pensar2016role} and
\cite{eggeling2019structure}.

Table \ref{Tb1} shows the number of possible DAGs as a function of the number of variables in the network. The rapid increase in the former justifies the prohibitive computational cost of obtaining the posterior distribution of $G$ analytically, even for small values of $d$. In addition to the very large space state, enumerating the states is a complicated and costly task.

\begin{table}[!h]
\begin{center}
    \caption{Number of possible DAGs as a function of the number of nodes. \citep{kjaerulff2008bayesian}.}\label{Tb1}
\begin{tabular}{cr}
\hline
$\#$Nodes & $\#$DAGs \\
\hline
1 & 1 \\
2 & 2 \\
3 & 25 \\
4 & 543 \\
5 & 29,281 \\
6 & 3,781,503 \\
7 & 1,138,779,265 \\
8 & 783,702,329,343 \\
9 & 1,213,442,454,842,881 \\
10 & 4,175,098,976,430,598,143 \\
\hline
\end{tabular}
\end{center}
\end{table}

The MCMC algorithm of \cite{goudie2016gibbs} is a random scan Gibbs sampling on $G$ with blocks defined by the rows of the matrix that represents the DAG. On each iteration of the algorithm, $m$ (usually 3) rows are randomly chosen and sampled from their respective full conditional distribution. In order to circumvent the need to compute the probability of all possible states of that conditional distribution, the algorithm breaks the sampling procedure into two steps.
The set of all admissible (with positive probability) networks, given the configuration of the other rows, is partitioned in a way such that, conditional on a component of the partition, the
parents of each of the $m$ chosen nodes are independent. On the first step, an element of the partition is sampled and, on the second one, a DAG is sampled among the ones in the sampled partition element. This strategy drastically reduces the computational cost for sampling from the full conditional distribution compared to the algorithm that samples directly from it by computing the probability vector of all admissible networks.
The authors recommend that a maximum number of parents $k<d-1$ per variable is fixed to further reduce the computational cost. This is a mild restriction which is in line with the idea of fitting a parsimonious network model.

Another major advantage of the algorithm of \cite{goudie2016gibbs} is that, for the case in which the $X_j$ variables are multinomial, the parameter vector $\theta$ of probabilities that indexes those multinomial distributions can be integrated out so that the MCMC is performed marginally in $G$. Details on how to perform such integration in the Bernoulli case, as well as how to obtain the posterior distribution of $\theta$, are presented in Section \ref{subsecBI}.

The algorithm of \cite{goudie2016gibbs} for multinomial data is available in the R package \textit{structmcmc}. It offers a simplified usage, allowing for the definition of the prior distribution on $G$ and the maximum number of parents per variable.

An important issue regarding the space of DAGs is the redundancy in terms of the model that they imply for $X$. For example, a network of the type $[p(X_1)p(X_2|X_1)\ldots]$ induces the same joint distribution as $[p(X_2)p(X_1|X_2)\ldots]$. Note that redundancies involving conditional distributions with more than one parent are much harder to identify. Formally, this means that the model parameterized in terms of $G$ is not identifiable and may lead to practical problems in the statistical analysis, depending on the adopted methodology. This will not be the case for the full Bayes approach proposed in this paper, since the analyses of synthetic data will be based in the predictive posterior distribution, which automatically averages over all the possible DAGs, weighted by their respective posterior probability.
On the other hand, methods based on the single use of the DAG with the highest posterior probability may lead to poor results. This is not only for the fact that uncertainty is disregarded, but because the highest probability DAG may not be the one that implies the highest probability model. 

%% file: 3.3Priori.tex
\subsection{Priori distribution for the network}

The prior distribution of $G$ plays a crucial role in network estimation. The Bayesian network model has a nested structure in which simpler networks (with more conditional independence relations) are special cases of more complex ones. The complete network is the most general model that encompasses all other networks as special cases, for certain choices of values of the parameters $\theta$. Thus, if a network $G_0$ is a special case of a more general network $G_1$, the value of the likelihood function at $(G_1,\theta)$ will always be greater than or equal to the value of this function at $(G_0,\theta)$, for the respective maximizing values of $\theta$. Therefore, if a uniform prior is adopted for the network, the only source of penalization for more complex networks, under the Bayesian approach, is the higher number of parameters $\theta$. Depending on the number of variables and the size of the dataset, this penalization may not be sufficient to properly balance the trade-off between model fit and parsimony. A clear understanding of how this penalization works is achieved by analyzing the expression of the marginal likelihood function of $G$, presented in expression (\ref{marg_prob}) in section \ref{subsecBI}.

Note that the motivation for using Bayesian networks is precisely to allow for a parsimonious modeling of $X$ and, therefore, achieving a robustly efficient penalization of complexity cannot rely only on the intrinsic penalization in the marginal likelihood function. The natural way to this under the Bayesian approach is through the prior distribution of the network.

As it was mentioned earlier, we adopt a modular prior, as required by the MCMC algorithm of \cite{goudie2016gibbs}. Given the few options of this type of prior in the literature, we propose here a general class of penalizing priors. Specifically,

\begin{equation}
    p(G)  \propto \prod_j  exp \biggl[- h_j(pa(X_j);\gamma_j) \biggl] = exp \biggl[ 
    -\sum_j h_j(pa(X_j);\gamma_j) \biggl],
\end{equation}
for non-negative functions $h_j$, that depend on the network only through the set of parents of each variable. 

This specification allows for the use of various penalty structures. A simple and intuitive particular case is to assume
\begin{equation}\label{Pr_simp}
h_j=\gamma|pa(X_j)|^{\alpha}.
\end{equation}

In Section 4.1.1, we present a calibration study of the hyperparameter \( \gamma \) as a function of the number of variables and of the sample size, when \( \alpha = 1 \).

%% file: 3.1Modelo.tex
\subsection{Bayesian Inference}\label{subsecBI}

The ultimate goal in this paper is to devise an efficient methodology to approximate statistical analyses of an original dataset using only synthetic data, so to preserve the confidentiality of the former. We adopt an approach in which the whole original dataset is confidential and the released data is all synthetic.

Let $\mathbf{X}$ be the original data, consisting of an i.i.d. sample of size $n$ of the random vector $X=(X_1,\ldots,X_d)$ of Bernoulli variables.
$G$ is the Bayesian network that defines the distribution of $X$ along with the respective conditional distributions which are indexed by parameters $\theta$. Define $\mathbf{Y}$ as the synthetic data consisting of $n$ i.i.d. replicates of $(X_1,\ldots,X_d)$. The full Bayesian model on $(\mathbf{X},G,\theta,\mathbf{Y})$ can be factorized as follows.
\begin{align}
p(\mathbf{X},G,\theta,\mathbf{Y})= p(\mathbf{X}|G,\theta)\pi(\mathbf{Y}|G,\theta)\pi(\theta|G) \pi(G)
\end{align}

Mathematically, the aim of the analysis is to obtain the posterior distribution of $(G,\theta,\mathbf{Y})$, i.e., $p(G,\theta,\mathbf{Y}|\mathbf{X})$.
Based on our final goal and for the purposes of optimizing the computational cost, we devise an algorithm that obtains the following distributions (in this particular order):
\begin{enumerate}
    \item the marginal posterior distribution of $G$;
    \item the conditional posterior of $(\theta|G,\mathbf{X})$;
    \item the marginal posterior $p(\mathbf{Y}|\mathbf{X})$, also called the posterior predictive distribution of $\mathbf{Y}$.
\end{enumerate}

The first step above is performed via the MCMC algorithm of \cite{goudie2016gibbs}, meaning that we effective obtain a (approximate) sample of that posterior. In order to implement the marginal MCMC in $G$, we need to obtain the marginal likelihood $p(\mathbf{X}|G)$. By adopting independent Beta$(\alpha_j,\beta_j)$ prior distributions for each $\theta_j$ indexing the networks $G$, we have the following.
\begin{align}\label{marg_prob}
    \pi(\mathbf{X}|G) &= \int \pi(\mathbf{X}|G,\theta_G)\pi(\theta_G)d\theta_G \nonumber \\
    &= \int  \prod_i^n \pi(X_{i.}|G,\theta_G)\prod_{j=1}^{J_G}\pi(\theta_j)d\theta_j \nonumber \\
    &=  \prod_{j=1}^{J_G} \int_{0}^1\theta_j^{z_j} (1-\theta_j)^{n_j-z_j}
        \frac{\Gamma(\alpha_j+\beta_j)}{\Gamma(\alpha_j)\Gamma(\beta_j)}
        \theta_j^{\alpha_j-1}(1-\theta_j)^{\beta_j-1}d\theta_j   \nonumber \\
        \begin{split}
    &= \prod_{j=i}^{J_G} \frac{\Gamma(\alpha_j+\beta_j)}{\Gamma(\alpha_j)\Gamma(\beta_j)}
         \frac{\Gamma(\alpha_j+z_j)\Gamma(\beta_j+n_j-z_j)}{\Gamma(\alpha_j+\beta_j+n_j)} \nonumber \\
        &\quad \times \int_0^1 \frac{\Gamma(\alpha_j+\beta_j+n_j)}{\Gamma(\alpha_j+z_j)\Gamma(\beta_j+n_j-z_j)}
         \theta_J^{\alpha_j+z_j-1}(1-\theta_j)^{\beta_j+n_j-z_j-1}d\theta_j \nonumber 
         \end{split} \\
        &= \prod_{j=1}^{J_G}\frac{\Gamma(\alpha_j+\beta_j)}{\Gamma(\alpha_j)\Gamma(\beta_j)} 
\frac{\Gamma(\alpha_j+z_j)\Gamma(\beta_j+n_j-z_j)}{\Gamma(\alpha_j+\beta_j+n_j)},
\end{align}
where $J_G$ is the total number of parameters $\theta_j$ indexing network $G$, $n_j$ is the number of observations, out of the $n$, where the event upon which the distribution indexed by $\theta_j$ is conditioned on occurs, and $z_j$ is the number of 1's among these $n_j$ observations.

The expression of the marginal likelihood in (\ref{marg_prob}) explicitly shows how the complexity of the Bayesian network is penalized by the dimension of its parametric space. In any sensible problem, the posterior information about the $\theta$ parameters will be dominated by the likelihood information (compared to the prior one), meaning that each term of the product in (\ref{marg_prob}) is dominated by the second ratio. In particular, if $\alpha_j=1$ and $\beta_j=1$, for all $j$, the first ratio equals 1. Now note that, since all of the second ratios are in $(0,1)$ and get smaller for higher values of $n_j$, the marginal likelihood balances model fitting and complexity the following way: the more complex the Bayesian network is, the more terms the likelihood have but these terms have higher values (smaller $n_j$).

Finally, note that the conditional posterior $p(\theta|G,\mathbf{X})$ is independent for all $\theta_j$ indexing $G$ with $(\theta_j|G,\mathbf{X})\sim Beta(\alpha_j+z_j,\beta_j+n_j-z_j)$.

The algorithm to obtain the predictive distribution of $\mathbf{Y}$ is presented in Section \ref{subsecSD} below, along with several relevant practical issues related to this.

%% file: 3.2Simula_d.tex
\subsection{Simulation and analysis of synthetic data}\label{subsecSD}

As our final goal is to approximate statistical analyses of the original data set using only the synthetic data, the core step of the proposed methodology is the one in which the posterior predictive distribution of $\mathbf{Y}$ is obtained.
This is effectively done by sampling from this distribution based on the sample of $(G,\theta)$ previously obtained. The basis of this sampling step is the following equality:
\begin{align}\label{D_pred}
\pi(\mathbf{Y}|\mathbf{X})=\int\pi(\mathbf{Y}|G,\theta)\pi(G,\theta|\mathbf{X})d\theta dG,
\end{align}
meaning that, for each pair $(G,\theta)$ sampled from its respective posterior distribution, a synthetic dataset $\mathbf{Y}$ is sample from the model $p(\mathbf{Y}|G,\theta)$. We  now discuss the advantages of this approach and describe algorithmic strategies to optimize the computational cost.

The methodology for generating synthetic data impacts both the quality of the analyses conducted with them and the security of the original data. Generally, the higher the preserved proportion of the original data is, the more similar the analyses' results is to those obtained with the original data, but the lower is the security, and vice versa. A coherent way to deal with this trade-off between accuracy and security is by replacing the entire original dataset with a synthetic dataset using a methodology that is efficient and robust to approximate the results that would be obtained using the original data. In this sense, a promising approach is to use the posterior predictive distribution of the synthetic data. The Bayesian paradigm not only efficiently and robustly uses the information contained in the data but also allows for efficient quantification of the uncertainty associated to the analysis.

Generally, we can define the result of a statistical analysis of a dataset $\mathbf{X}$ as a set $h_1(\mathbf{X}),h_2(\mathbf{X}),\ldots$ of statistics. These could be, for example:
\begin{enumerate}
    \item Descriptive statistics (mean, median, variance, etc.);
    \item Point and interval estimators of parameters that index the model adopted for the data;
    \item p-values of hypotheses tests about the model.
\end{enumerate}
This way, we can focus our analysis on obtaining the posterior predictive distribution of these statistics for the synthetic data $\mathbf{Y}$, i.e.,
\begin{equation}\label{pdh}
p(h_1(\mathbf{Y}),h_2(\mathbf{Y}),\ldots|\mathbf{X}).    
\end{equation}

In the Monte Carlo context considered here, a sample from the predictive distribution in (\ref{pdh}) is obtained by computing $(h_1(\mathbf{Y}),h_2(\mathbf{Y}),\ldots)$ for each value of $\mathbf{Y}$ sampled from its predictive distribution. From a computational cost perspective, this is particularly convenient as the actual samples of $\mathbf{Y}$ can be discarded after $(h_1(\mathbf{Y}),h_2(\mathbf{Y}),\ldots)$ is computed.
The algorithm below outlines the entire proposed inference process.

\begin{algorithm}[h!]
\caption{generating a sample from $p(h_1(\mathbf{Y}),h_2(\mathbf{Y}),\ldots|\mathbf{X})$}
\textbf{Input:} original dataset $\mathbf{X}$; initial network for the MCMC.\\
Run the MCMC algorithm by \cite{goudie2016gibbs} to obtain a sample from $p(G|\mathbf{X})$;\\
Keep a sample of size $M$ - $(G^{(1)},\ldots,G^{(M)})$, for reasonable choices of burn-in and lag;\\
Set $m$ = 1\\
\While{$m \leq M$}{
    Draw $\theta^{(m)}\sim p(\theta|G^{(m)},\mathbf{X})$; \\
    Draw $\mathbf{Y}^{(m)}\sim p(\mathbf{Y}|G^{(m)},\theta^{(m)})$;\\
    Compute $(h_1(\mathbf{Y}^{(m)}),h_2(\mathbf{Y}^{(m)}),\ldots)$;\\
    Set $m = m + 1$.
}
\label{alg:exemplo}
\end{algorithm}

The values of burn-in, lag, and $M$ should be chosen in order to obtain the largest effective sample size possible for the statistics $h$ with the lowest computational cost possible. Basically, the lag should be chosen to mitigate the autocorrelation of the chain so that the effective sample sizes of the $h_i$ functions are approximately $M$ which, in turn, should be around at least 500.

Note how Algorithm 1 optimizes the computational cost of the whole statistical analysis by breaking this into three main steps and storing the minimum information required. It can be used in different ways regarding the output to be delivered to the final user (who cannot have access the original data). Some examples of these outputs are the following, in increasing order of complexity.

\begin{enumerate}
\item Posterior distribution of $G$ represented by the empirical distribution of the sample $(G^{(1)},\ldots,G^{(M)})$ obtained in Algorithm 1.

\item The $M$ sets of synthetic data drawn in Algorithm 1.

\item Only some (5 to 10) of the $M$ sets of synthetic data from the previous item.

\item The sample of $(h_1(\mathbf{Y}),h_2(\mathbf{Y}),\ldots)$ drawn in Algorithm 1.

\item The mean and credibility interval of $(h_1(\mathbf{Y}),h_2(\mathbf{Y}),\ldots)$ obtained from its sample.
\end{enumerate}

The choice of the output to be provided depends on the objective and expertise of the final user. For example, using the output 1 above, which is the simplest, the final user can reproduce the entire Algorithm 1 by replacing the MCMC step by the simulation of an i.i.d. sample from the provided empirical distribution.

%% file: 4.0main.tex
\section{Numerical examples}

This section presents some analyses with simulated and real data to investigate the efficiency and applicability of the proposed methodology.

The analyses with simulated data have two main objectives: investigating the efficiency of Bayesian network estimation using the algorithm of \cite{goudie2016gibbs} and, primarily, analyzing the efficiency of the proposed methodology to generated and analyze synthetic data. To do that, various scenarios are considered in terms of the number of variables, sample size and prior distribution of the network. The true values of the respective parameters $\theta$ vary between 0.2 and 0.8. For each scenario determined by these factors, 10 replications (datasets) are generated. The simulated scenarios are described in Table \ref{tab:table11}.

\begin{table}[h!]
  \begin{center}
    \caption{Scenarios for simulated data.}
    \label{tab:table11}
\small
    \begin{tabular}{c|c|c}
    \textbf{$d$} & \textbf{Sample size} & \textbf{Network} \\

    \hline
    \multirow{3}{*}{$3$} & 500 & \multirow{3}{*}{$X_1,X_3,(X_2|X_1)$} \\
     & 1000 & \\
     & 5000 & \\
    \hline
    \multirow{2}{*}{$4$} & 1000 & \multirow{2}{*}{$X_1,X_4,(X_2|X_1),(X_3|X_4)$} \\
     & 5000 & \\
    \hline
    \multirow{2}{*}{$7$} & 2000 & \multirow{2}{*}{$X_1,X_3,X_6,X_7,(X_2|X_1),(X_5|X_6,X_7),(X_4|X_3,X_5)$} \\
     & 5000 & \\
    \end{tabular}
  \end{center}
\end{table}

The efficiency of the MCMC algorithm to estimate the Bayesian network is investigated by analyzing some posterior statistics such as the posterior probability of the most likely network and of the true one. Redundancies involving only two variables are identified and treated as the same model (summing the respective posterior probabilities).

Two prior distributions are considered for $G$, a uniform one and an informative with the form in (\ref{Pr_simp}) for $\alpha=1$.

For the synthetic data analysis, we compare three different generation methods, referred to as $S_1$, $S_2$, and $S_3$. $S_1$ refers to the full Bayes approach proposed in this paper, considering the posterior predictive distribution of the synthetic data. $S_2$ refers to a widely used methodology in the literature, consisting of the generation of 5 sets of synthetic data using the Bayesian network with the highest posterior probability and the maximum likelihood estimator (MLE) of $\theta$ under that network. These 5 datasets are then combined according to a standard method for data imputation described below. Method $S_3$ adopts the synthetic data generation technique proposed by \cite{nowok2022package}, available in the R package \textit{synthpop}, to generate a single synthetic dataset. In this package, the authors employ a non-Bayesian model for generating synthetic data, using the Iterative Proportional Fitting (IPF) method. %This approach aims to establish the marginal distributions present in the original dataset. At each iteration, weights are calculated for each record in the original dataset. At each step, IPF evaluates whether the distributions of the adjusted dataset approximate the desired marginal distributions. If convergence is achieved, the process is terminated; otherwise, iterations continu.

Method $S_2$ proceeds as follows. For a statistic $h$ of the synthetic data, and the 5 values $h^{(1)},\ldots,h^{(5)}$ obtained for the 5 datasets, the point and interval estimations for the value of $h$ obtained from the original data are given by:
\begin{eqnarray}
    \bar{h}&=&\frac{1}{5}\sum_{m=1}^5h^{(m)}, \label{est_S2_1}\\
    CI&=&\bar{h}\pm t_{4,\alpha/2}\frac{s}{\sqrt{5}},
\end{eqnarray}
where $s^2=\sum_{m=1}^5 \frac{(h^{(m)} - \hat{h})^2}{5-1}$ and $t_{4,\alpha/2}$ is the $\alpha/2$ percentile of the student-t distribution with 4 degrees of freedom. One major limitation of this approach is the fact that the confidence interval is a crude approximation based on the normal distribution.

The comparison among the three methods aims at providing a comprehensive analysis, exploring different methodologies of synthetic data generation. In particular, it is intended to investigate the efficiency and robustness of the proposed methodology regarding: 1. the approximation of the analyses performed with the original data; 2. the quantification of the uncertainty involved in the inference process.

We consider three statistics of the synthetic data to perform the analyses - $h_1$, $h_2$, and $h_3$. The first statistic is a univariate function comparing the 95\% confidence intervals of some of $\theta_j$ parameters, obtained with the original and synthetic data. Specifically, this function is an overlapping measure between the two intervals. Statistic $h_2$ is the MLE of some $\theta_j$'s. Finally, $h_3$ is the p-value of a chi-square test of independence between two of the $X_j$ variables.

The analysis consists of two main procedures: 1. compare the results obtained with the three synthetic data methods with the respective results obtained with the original data; 2. compare the results obtained with methods $S_2$ and $S_3$ to those obtained with $S_1$ to investigate how much is lost in terms of uncertainty quantification when using $S_2$ and $S_3$.

The measure $h_1$ of overlap of confidence intervals is defined as:
\begin{align}
    O_{IC}=\frac{2\times(\min\{u_1,u_2\}-\max\{l_1,l_2\})}
        {(u_1-l_1)+(u_2-l_2)},
\end{align}
if the two intervals overlap and $O_{IC}=0$, otherwise. Here, $(l_1, u_1)$ is the confidence interval with the original data and $(l_2, u_2)$ is the confidence interval with the synthetic data. The higher this measure is, the better is the method to approximate the results of the corresponding analysis performed with the original data.

Method $S_3$ returns only a point estimate for each statistic $h_i$. In the case of $S_2$, a point estimate and an approximate confidence interval are obtained as previously described.
The proposed method $S_1$, on the other hand, returns the posterior predictive distribution of each $h_i$. These are presented through their posterior mean and the 98\% highest posterior density credibility interval. Note that method $S_1$ quantifies the uncertainty associate to the estimation of $G$ and $\theta$ and to the distribution of the synthetic data given $(G,\theta)$. Method $S_2$, on the other hand, quantifies the uncertainty associated only to the latter.

A reasonable approach to choose the value of the hyperparameter $\gamma$ in the prior distribution for the network is presented in the next section. This is then used to choose the values of $\gamma$ in the analysis with the simulated and real data.

%The analysis of the real dataset will be done similarly to the analysis of the simulated data. The same three methods $S_1$, $S_2$, and $S_3$ will be considered, the same three statistics $h$, and the same two choices of $\alpha$ for the prior distribution of the network.

%% file: 4.1.1Prioris.tex
\subsection{Calibration of the prior distribution of the network}

The aim of this study is to understand the relationship between the hyperparameter $\gamma$ and the level of penalization of the penalizing prior in (\ref{Pr_simp}), as a function of the number of variables and the size of the dataset, when $\alpha=1$. Note that, the higher $\gamma$ is, the stronger is the penalization of network complexity.

The uniform prior on the network corresponds to the prior in (\ref{Pr_simp}) with $\gamma=0$. Since the networks in Table \ref{tab:table11} are considerably simpler than the complete one, it is expected that, starting from $\gamma=0$, the posterior probability of the true network increases as the value of $\gamma$ does. In this sense, the calibration of this hyperparameter will be done by seeking the smallest value of $\gamma$ for which the posterior probability of the true network exceeds 0.85.

The values found for $\gamma$, as a function of the number of variables and the sample size, are expected to be robust choices to address the trade-off between model fit and parsimony whenever networks considerably simpler than the complete one are a good solution.

\begin{figure}[!h]

%\centering
\begin{minipage}{0.32\textwidth}
%\centering
\includegraphics[width=\textwidth]{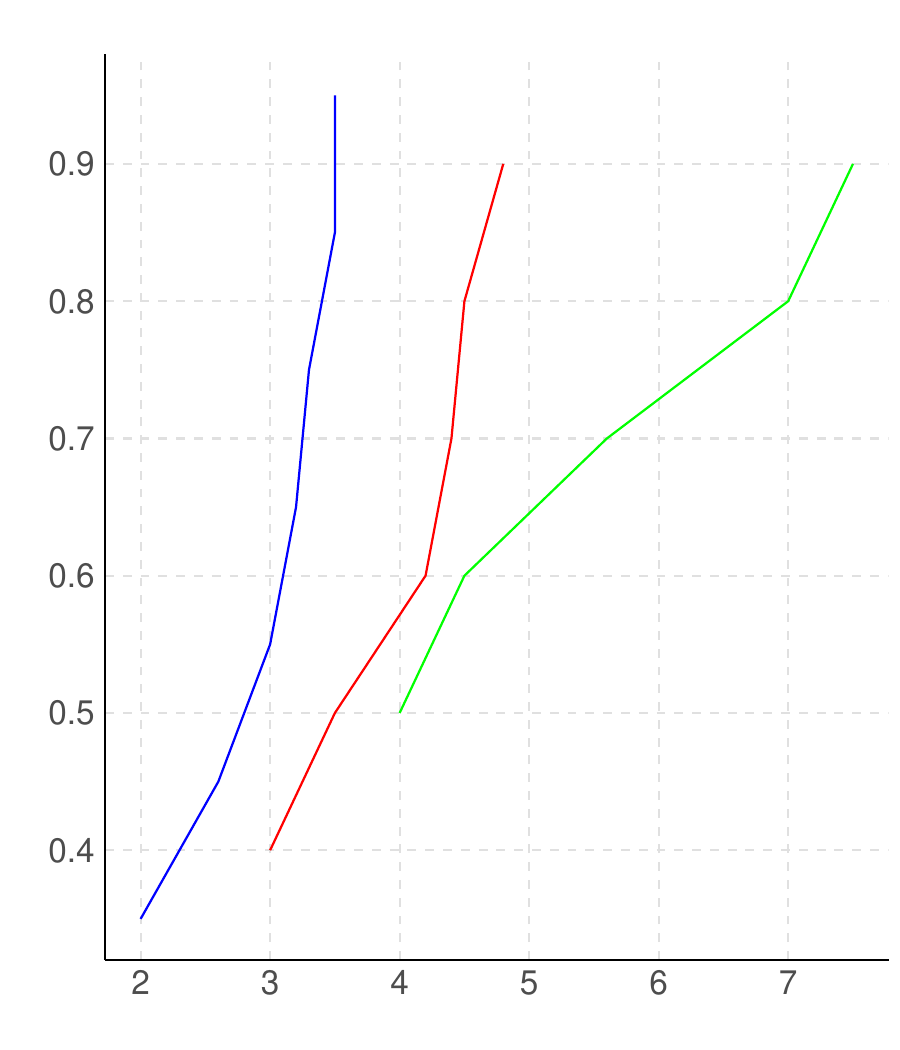}
\end{minipage}
%\hfill
\begin{minipage}{0.32\textwidth}
\centering
\includegraphics[width=\textwidth]{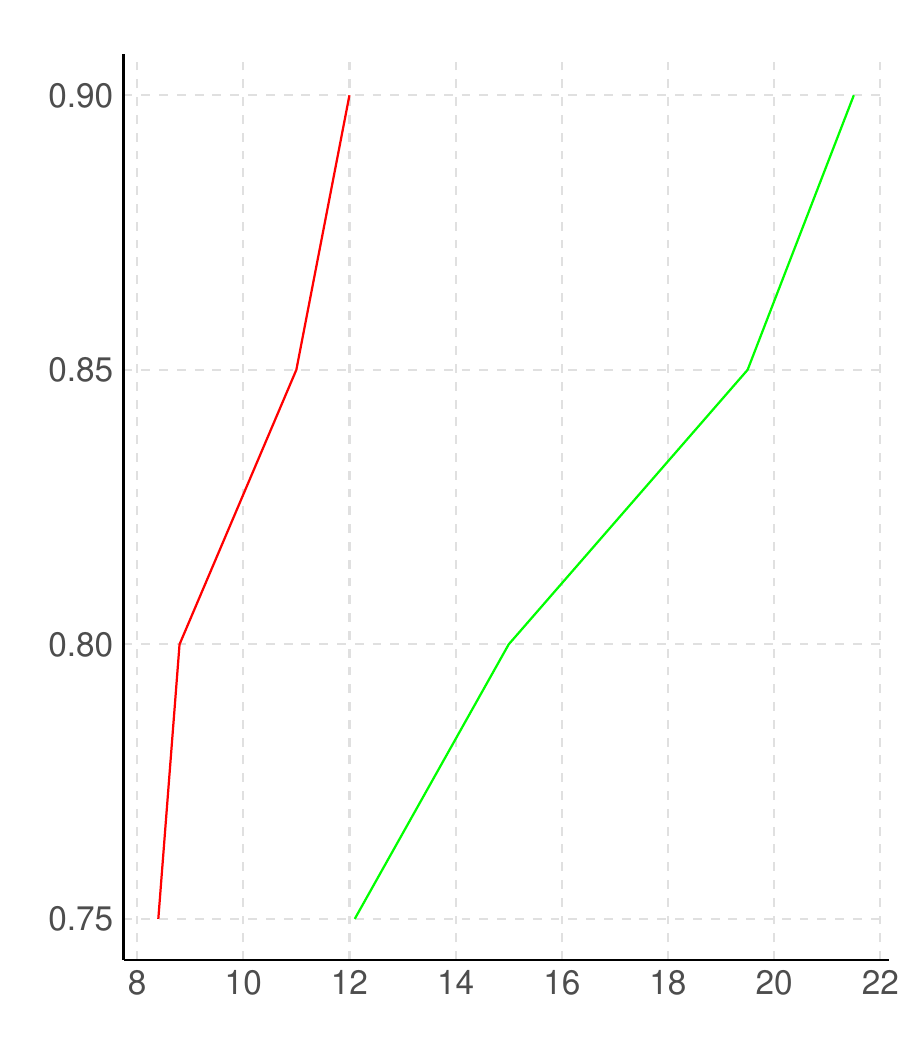}
\end{minipage}
\begin{minipage}{0.32\textwidth}
\centering
\includegraphics[width=\textwidth]{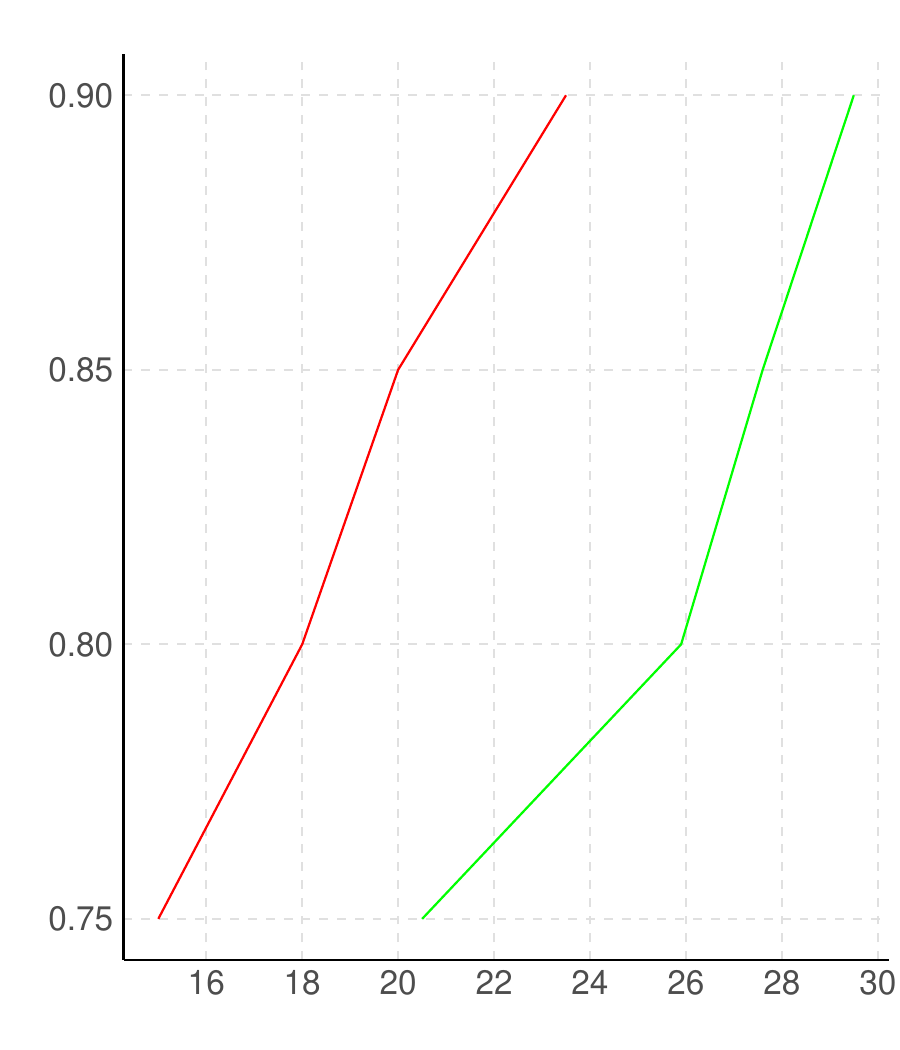}
\end{minipage}

\caption{Hyperparameter $\gamma$ versus the posterior probability of the true network. Left: $d=3$, blue: $n=500$, red: $n=1000$, green: $n=5000$. Center: $d=4$, red: $n=1000$, green: $n=5000$. Right: $d=7$, red: $n=2000$, green: $n=5000$. }
\label{plot_prior}

\end{figure}

Figure \ref{plot_prior} shows the relationship between $\gamma$ and the posterior probability of the true network.

%% file: 4.1.2Estima_rede.tex
\subsection{Network estimation}

For each of the scenarios in Table \ref{tab:table11}, 10 replications are simulated to analyze the accuracy and reliability of the results.
Uniform $U(0,1)$ priors are adopted for the $\theta$ parameters. Two prior distributions for the network are considered, an uniform ($P_2$) and the one proposed in (\ref{Pr_simp}) ($P_1$), with $\alpha=1$ and $\gamma$ calibrated as explained in the previous section. These priors impact the results of methods $S_1$ and $S_2$, which use the Bayesian network model. In the case of method $S_3$, the labels $P_1$ and $P_2$ in the plots refer to two replications (two sets of synthetic data) of the method.

Under the proposed method $S_1$, redundancies involving only two variables are treated as a single network with aggregated posterior probabilities. Although ignoring the other redundancies can potentially compromise the calculation of the most probable posterior model, it does not compromise the analysis of synthetic data, since this is based on the posterior predictive distribution.
In the case of method $S_2$, this can have serious consequences in the analysis.

The size of MCMC chains varies between 5,000 and 80,000 iterations; models with more variables and more data require more iterations. The burn-in varies between 1,000 and 10,000.
For method $S_1$, reasonable values for the lag are considered to select the posterior sample of the network so to mitigate the effects of autocorrelation and work with sample sizes that do not compromise the computational cost. Those values vary between 8 and 140 and aim for an effective sample size of at least 500 for the $h_i$ statistics.

In order analyze the accuracy of the network estimation, we report the number of replications in which the true network had the highest posterior probability and the mean (under these replications) of that probability. We also report the mean of this probability among the other replications. These results are presented in Table \ref{table_21}.

\begin{table}[h!]
\begin{center}
\caption{Statistics for the posterior distribution of the network. Number of replications in which the true network has the highest posterior probability - mean (under these replications) of the posterior probability of the true network / same mean (under the other replications).}
\label{table_21}
\begin{tabular}{c|c|c|c|c}
\hline
\textbf{$p$} & \textbf{n=500} & \textbf{n=1000} & \textbf{n=2000} & \textbf{n=5000} \\
\hline

3 (${P_1}$) & 9 - .93 / .28  & 9 - .93 / .23 & - & 9 - .94 / .15 \\
3 (${P_2}$) & 10 - .87 /   $\quad$  & 10 - .88 /  $\quad$  & - & 9 - .85 / .42 \\
\hline

4 (${P_1}$) & - & 8 - .89 / .19 & - & 8 - .86 / .17\\
4 (${P_2}$) & - & 7 - .82 / .16 & - & 8 - .86 / .20\\
\hline
7 (${P_1}$) & - & - & 5 - .69 / .23 & 7 - .70 / .22 \\
7 (${P_2}$) & - & - & 6 - .71 / .30 & 7 - .72 / .15 \\

\hline

\end{tabular}
\end{center}

\end{table}

It can be noticed that, in almost all the cases, the true network was the one with the highest posterior probability, indicating that the algorithm of \citet{goudie2016gibbs} is efficient to approximate the posterior distribution of the network. Moreover, similar results are observed when comparing the uniform prior $P_2$ to the proposed prior $P_1$. Larger differences are expected for smaller datasets and models with more variables (larger network space).

%% file: 4.1.3Dados_sint.tex
\subsection{Analysis of synthetic data}

\begin{figure}[h!]
\centering
\begin{minipage}{0.49\textwidth}
\centering
\includegraphics[width=\textwidth]{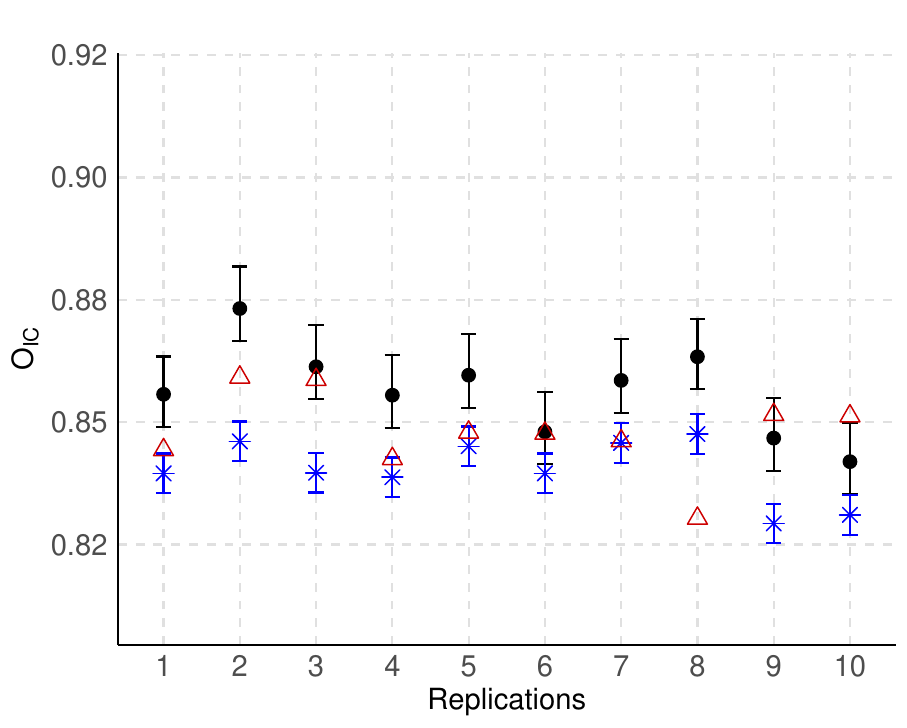}
\end{minipage}
\begin{minipage}{0.49\textwidth}
\centering
\includegraphics[width=\textwidth]{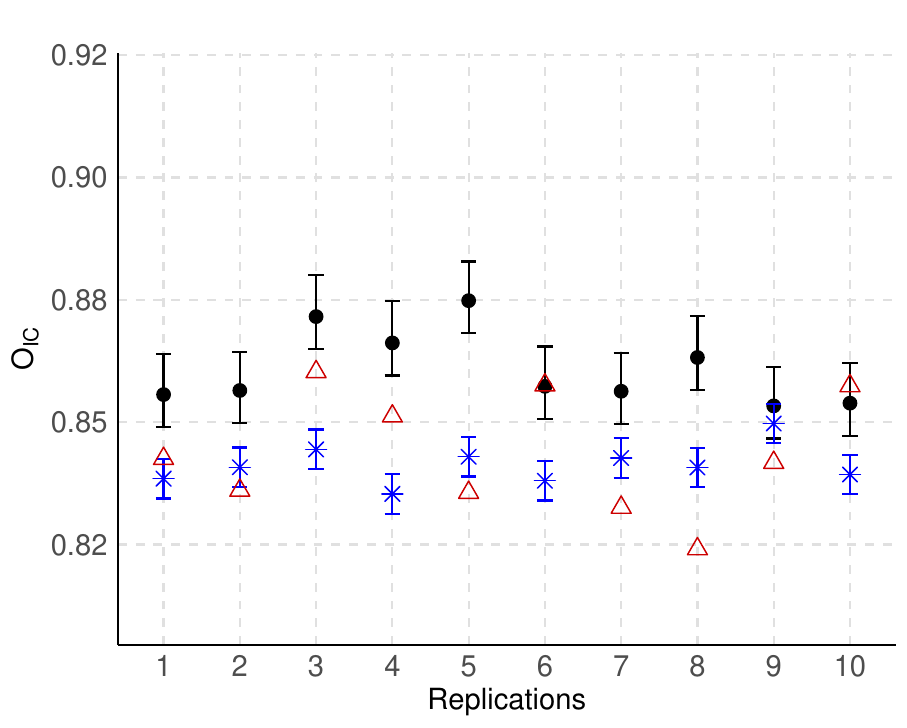}
\end{minipage}

\begin{minipage}{0.49\textwidth}
\centering
\includegraphics[width=\textwidth]{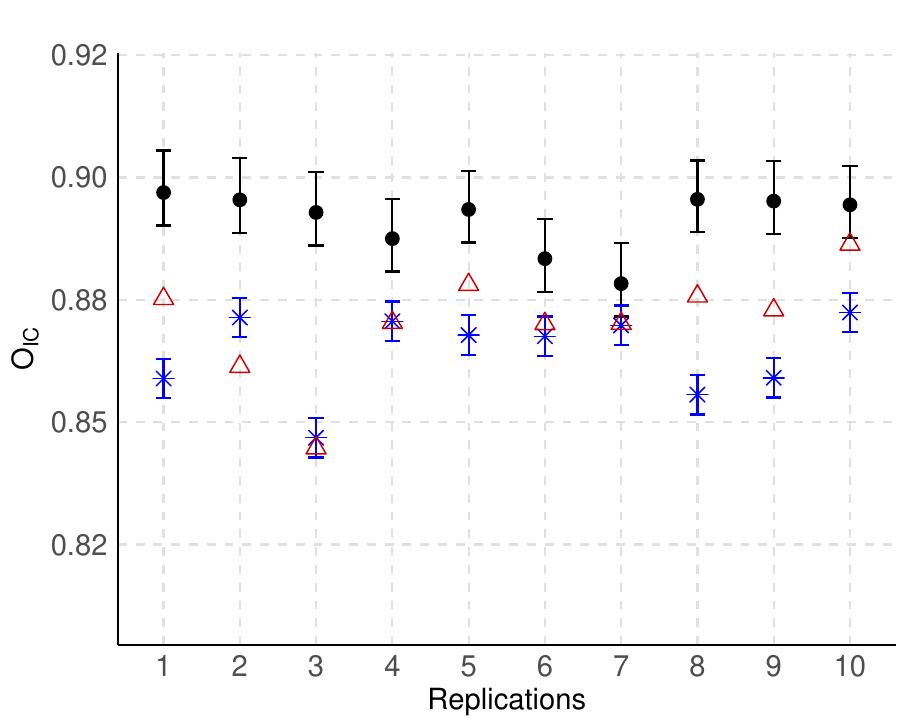}
\end{minipage}
\begin{minipage}{0.49\textwidth}
\centering
\includegraphics[width=\textwidth]{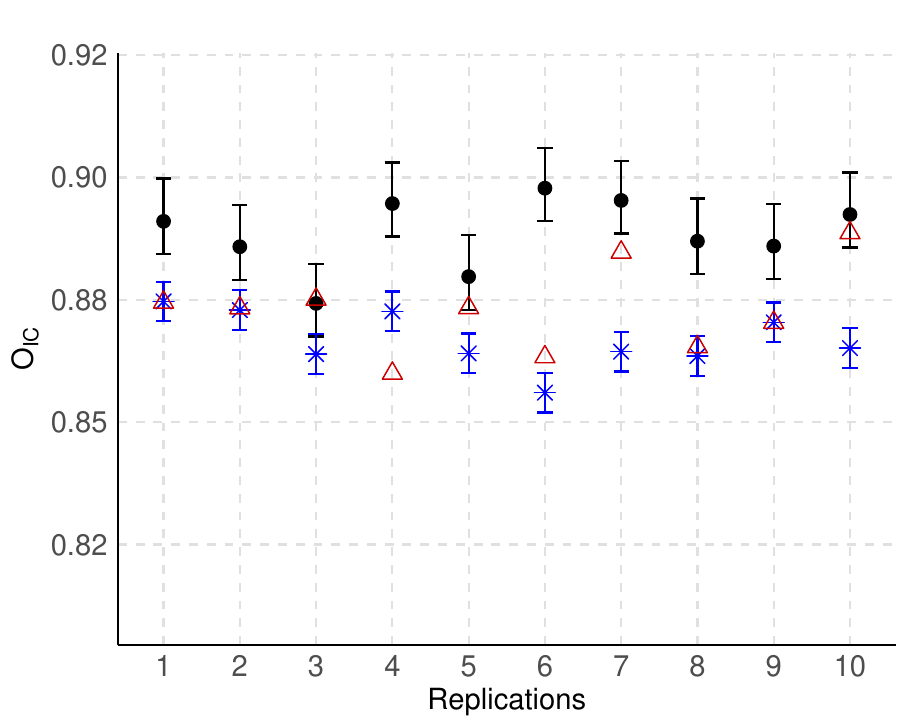}
\end{minipage}

\begin{minipage}{0.49\textwidth}
\centering
\includegraphics[width=\textwidth]{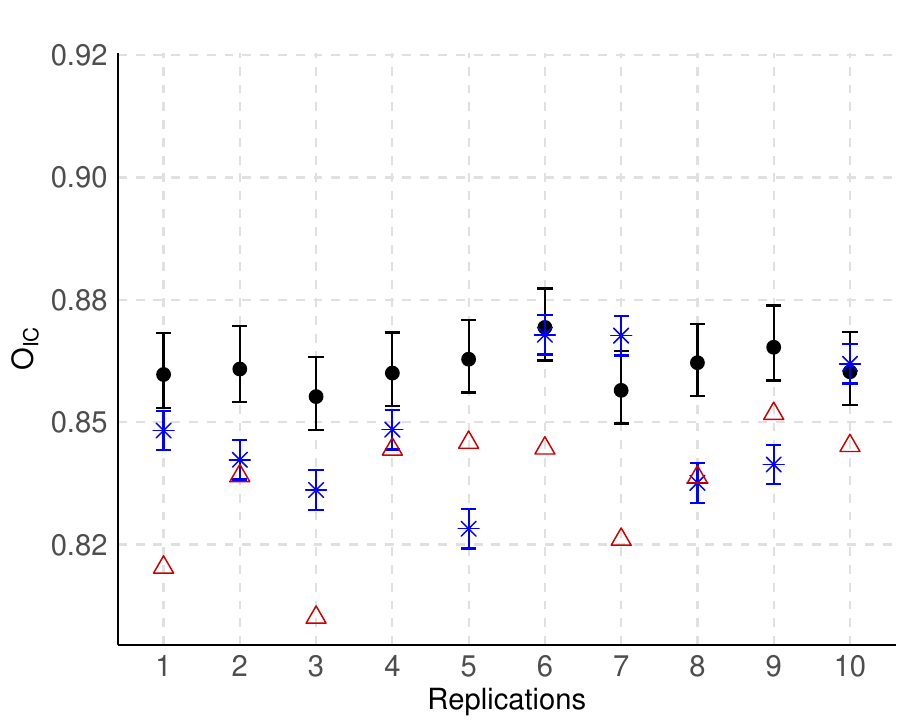}
\end{minipage}
\begin{minipage}{0.49\textwidth}
\centering
\includegraphics[width=\textwidth]{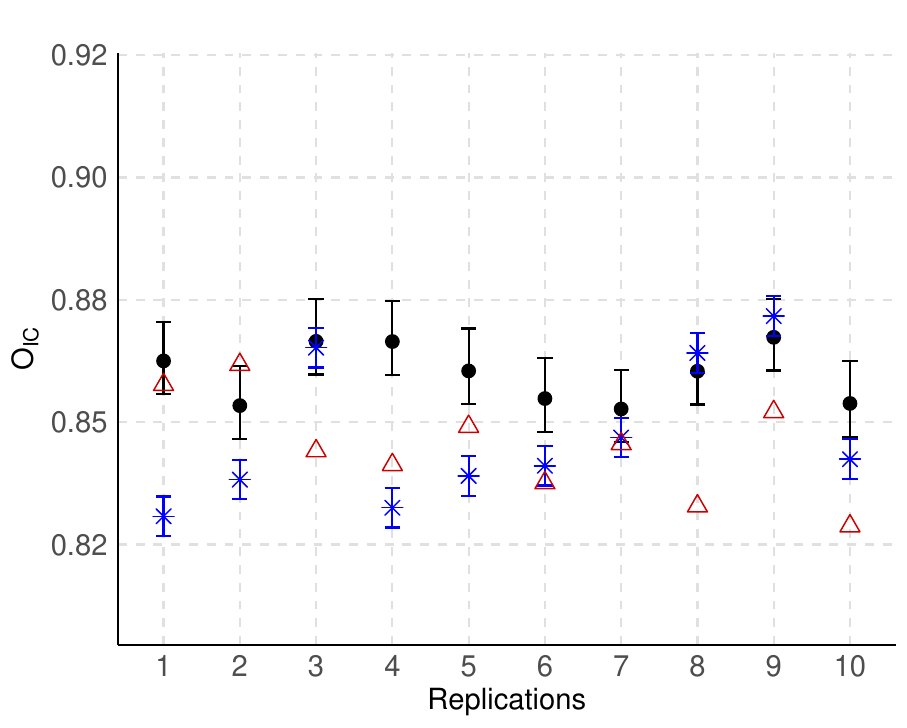}
\end{minipage}
 \caption{$O_{IC}$ measure for $\theta=P(X_2=1|X_1=0)$, with $n=5000$. $S_1$: predictive mean (black point) and $98\%$ predictive credibility interval. $S_2$: point estimate (blue asterisk) and $98\%$ confidence interval. $S_3$: point estimate (red triangle). Top: $d=3$, middle: $d=4$, bottom: $d=7$. $P_1$ on the left and $P_2$ on the right.}
  \label{fig_n=5000_2|1=0_1}
\end{figure}

Figure \ref{fig_n=5000_2|1=0_1} presents the results for the $O_{IC}$ statistic of the parameter $\theta=P(X_2=1|X_1=0)$, with sample size $n=5000$.

Overall, the proposed method shows greater similarity to the results obtained with the original data. In most replications, the $O_{IC}$ values obtained by the other two methods are outside (below) the credibility interval obtained by the proposed method.
In addition to showing a larger overlap with the IC obtained with the original data, method $S_1$ has the additional advantage of efficiently incorporating uncertainty into the analysis. %This characteristic makes it a more robust and informative choice to interpret the results.
The results are similar between the two priors used for the network, for both methods $S_1$ and $S_2$.

%Results for a sample size of 1000 are presented in Appendix A.
Results for $n=1000$ (omitted) are similar to those for $n=5000$. The only relevant difference is that the $O_{IC}$ values are smaller for the former, reflecting the fact that a smaller sample contains less information.

\clearpage
\newpage

\begin{figure}[h!]
\centering
\begin{minipage}{0.49\textwidth}
\centering
\includegraphics[width=\textwidth]{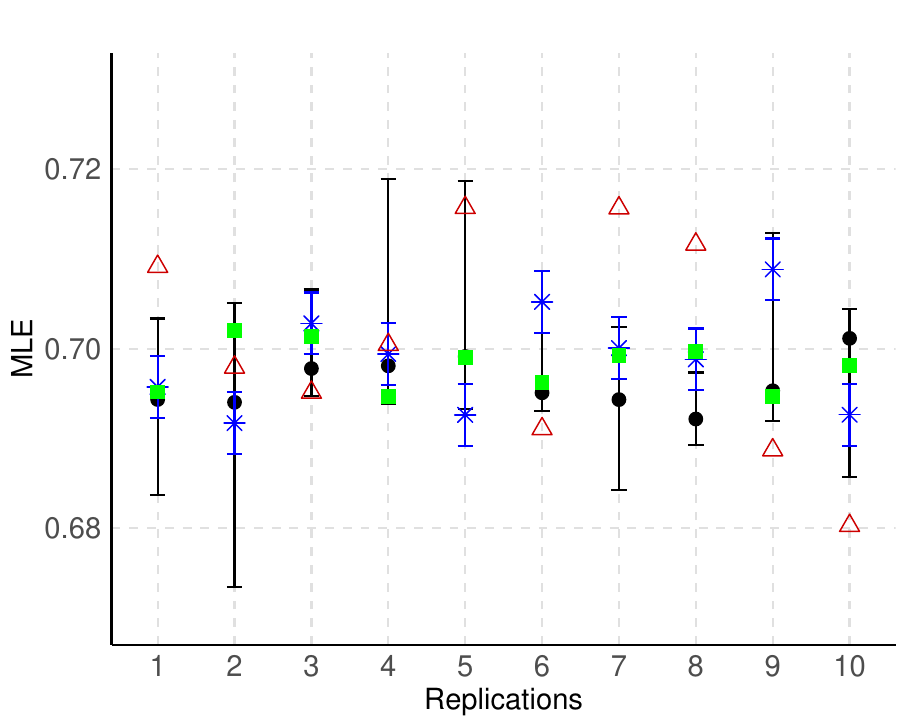}
\end{minipage}
\begin{minipage}{0.49\textwidth}
\centering
\includegraphics[width=\textwidth]{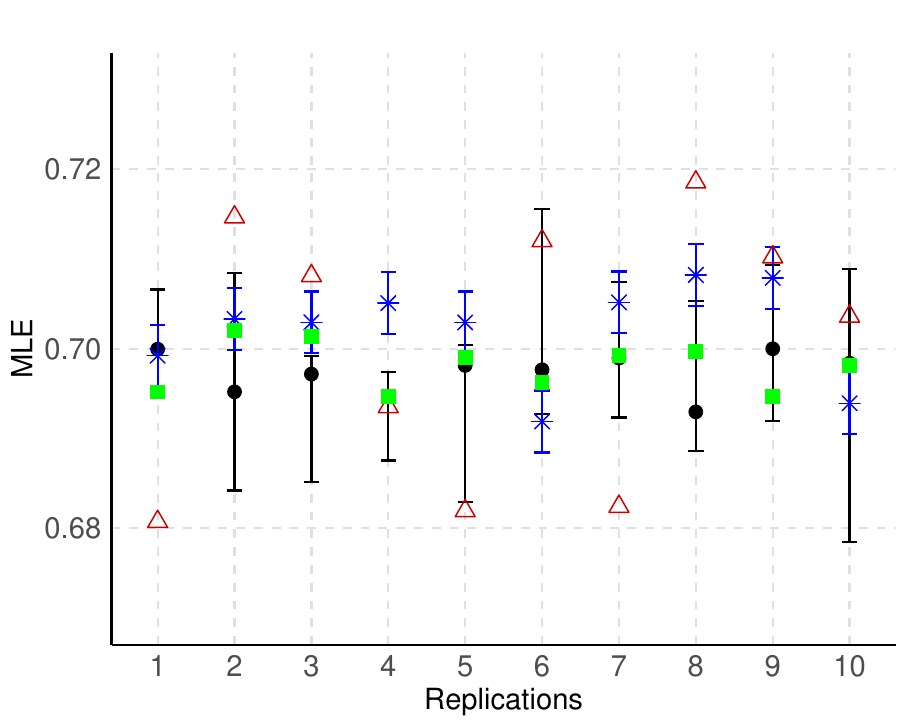}
\end{minipage}

\begin{minipage}{0.49\textwidth}
\centering
\includegraphics[width=\textwidth]{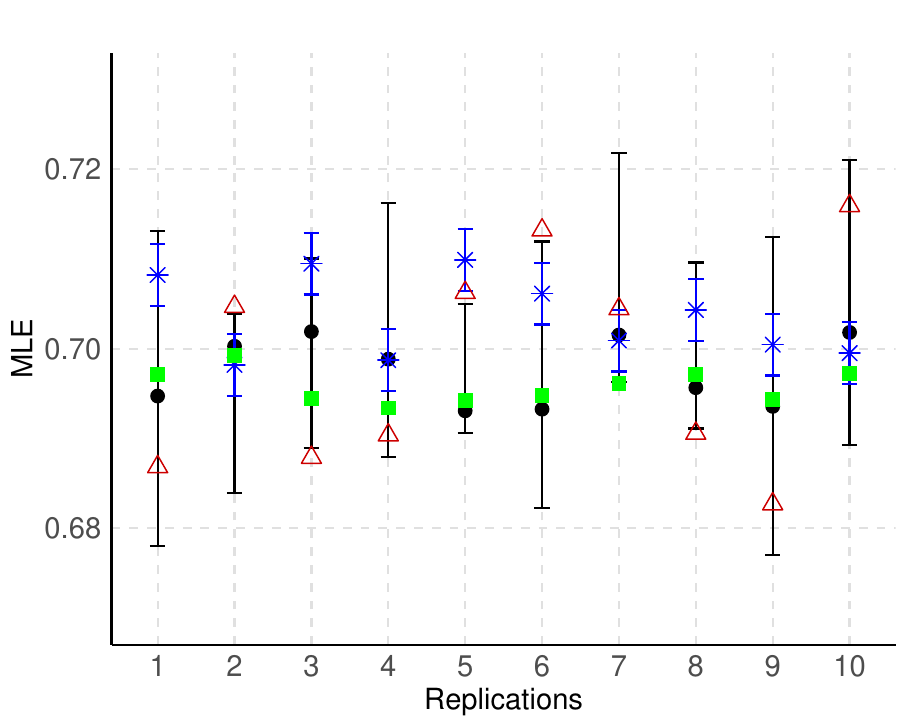}
\end{minipage}
\begin{minipage}{0.49\textwidth}
\centering
\includegraphics[width=\textwidth]{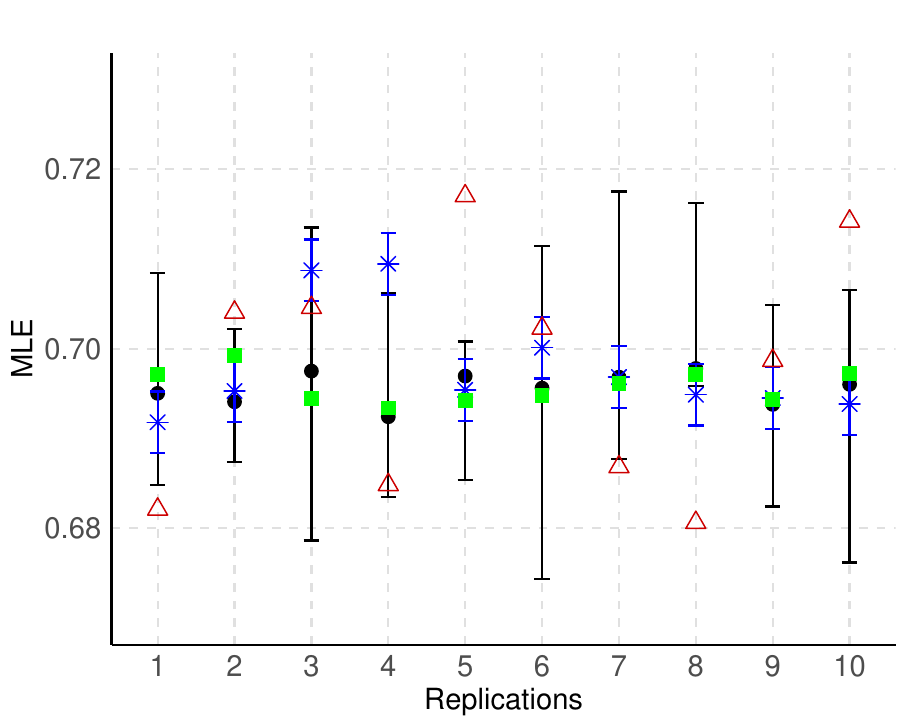}
\end{minipage}

\begin{minipage}{0.49\textwidth}
\centering
\includegraphics[width=\textwidth]{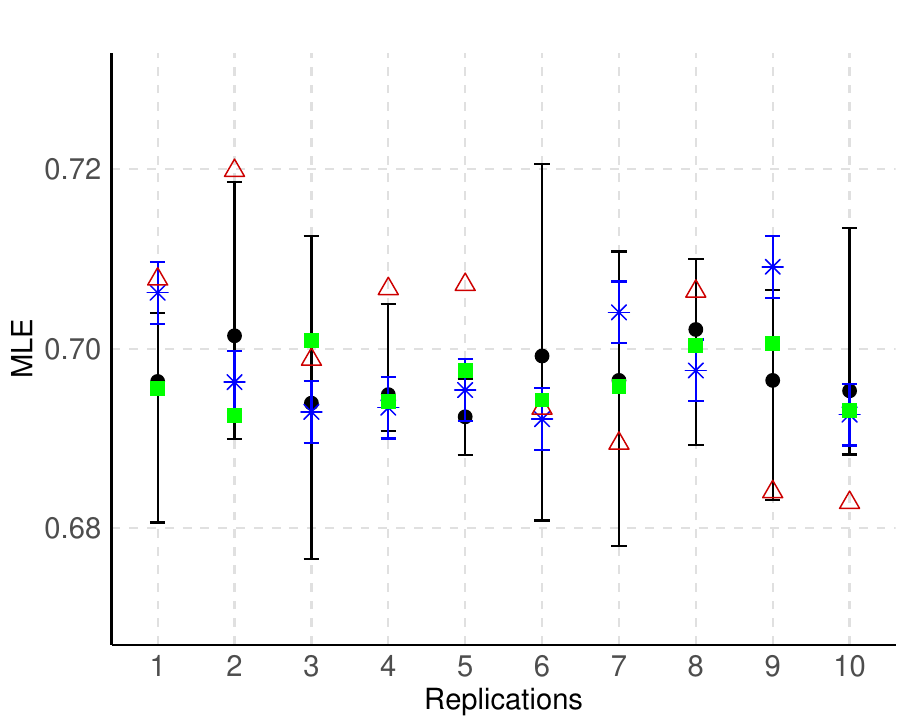}
\end{minipage}
\begin{minipage}{0.49\textwidth}
\centering
\includegraphics[width=\textwidth]{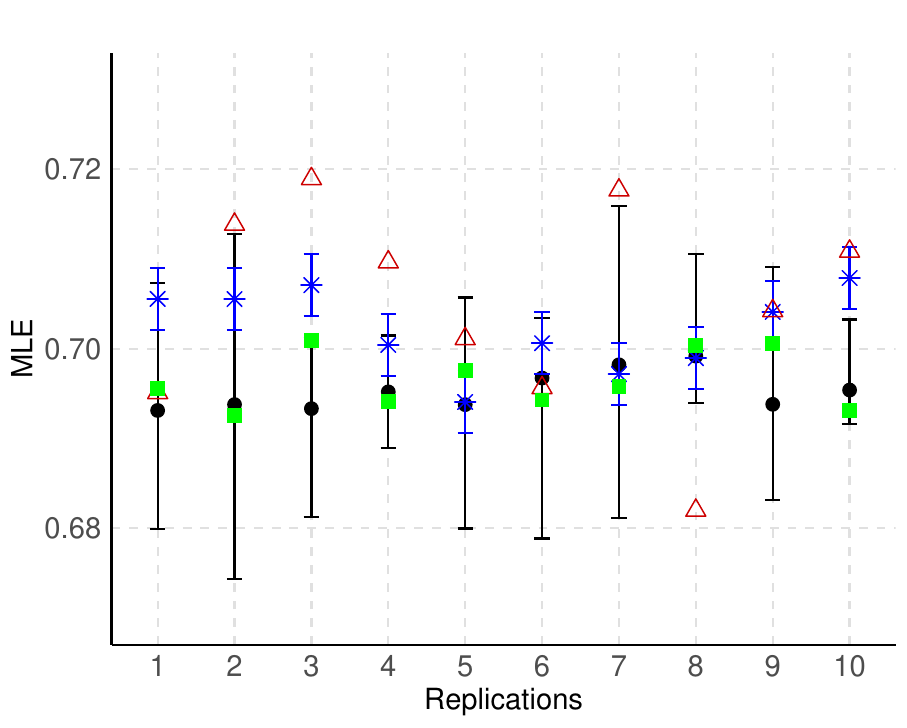}
\end{minipage}
 \caption{MLE of $\theta=P(X_2=1|X_1=0)$, with $n=5000$. True MLE in green. $S_1$: predictive mean (black point) and $98\%$ predictive credibility interval. $S_2$: point estimate (blue asterisk) and $98\%$ confidence interval. $S_3$: point estimate (red triangle). Top: $d=3$, middle: $d=4$, bottom: $d=7$. $P_1$ on the left and $P_2$ on the right.}
  \label{EMV_f1}
\end{figure}

Figure \ref{EMV_f1} presents the results for the MLE of the parameter $\theta=P(X_2=1|X_1=0)$, for $n=5000$. The differences between the results for the two priors are discrete.
Overall, method $S_3$ has the worst performance. In the case of the proposed method $S_1$, the true MLE falls within the credibility interval in almost all cases.

\begin{figure}[h!]
\centering
\begin{minipage}{0.49\textwidth}
\centering
\includegraphics[width=\textwidth]{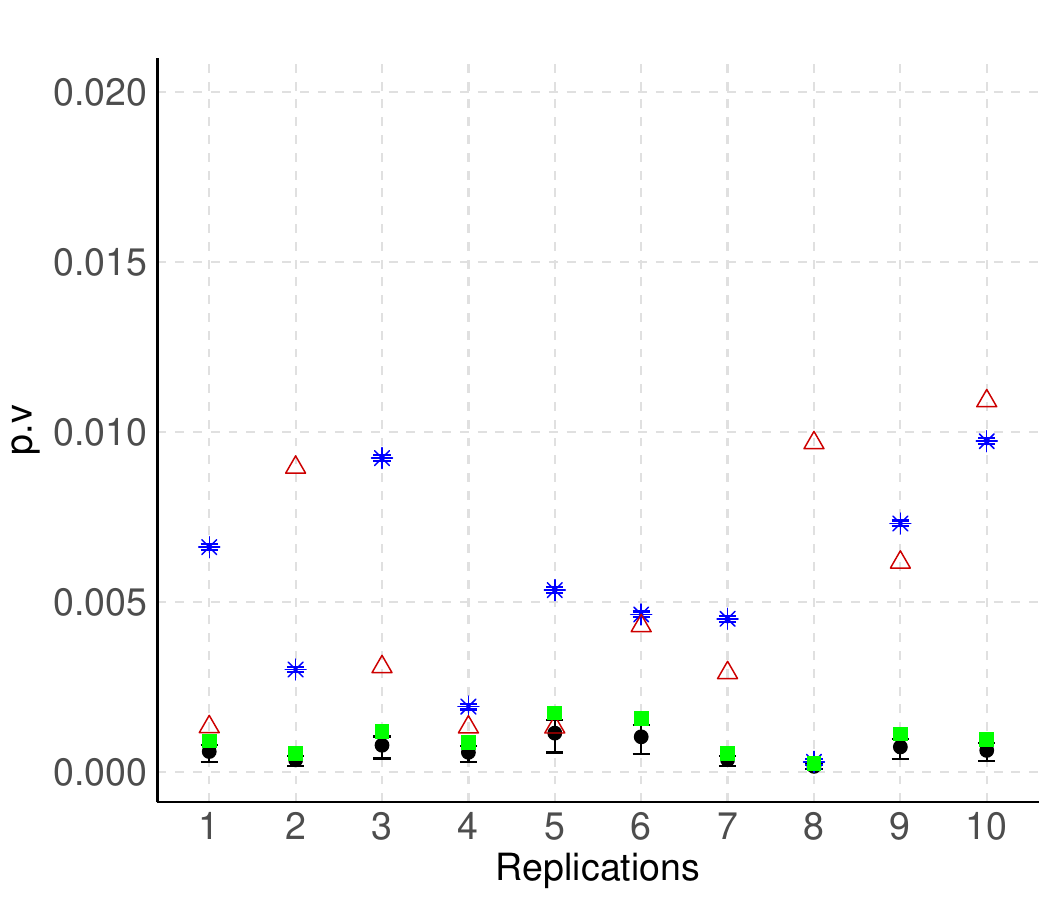}
\end{minipage}
\begin{minipage}{0.49\textwidth}
\centering
\includegraphics[width=\textwidth]{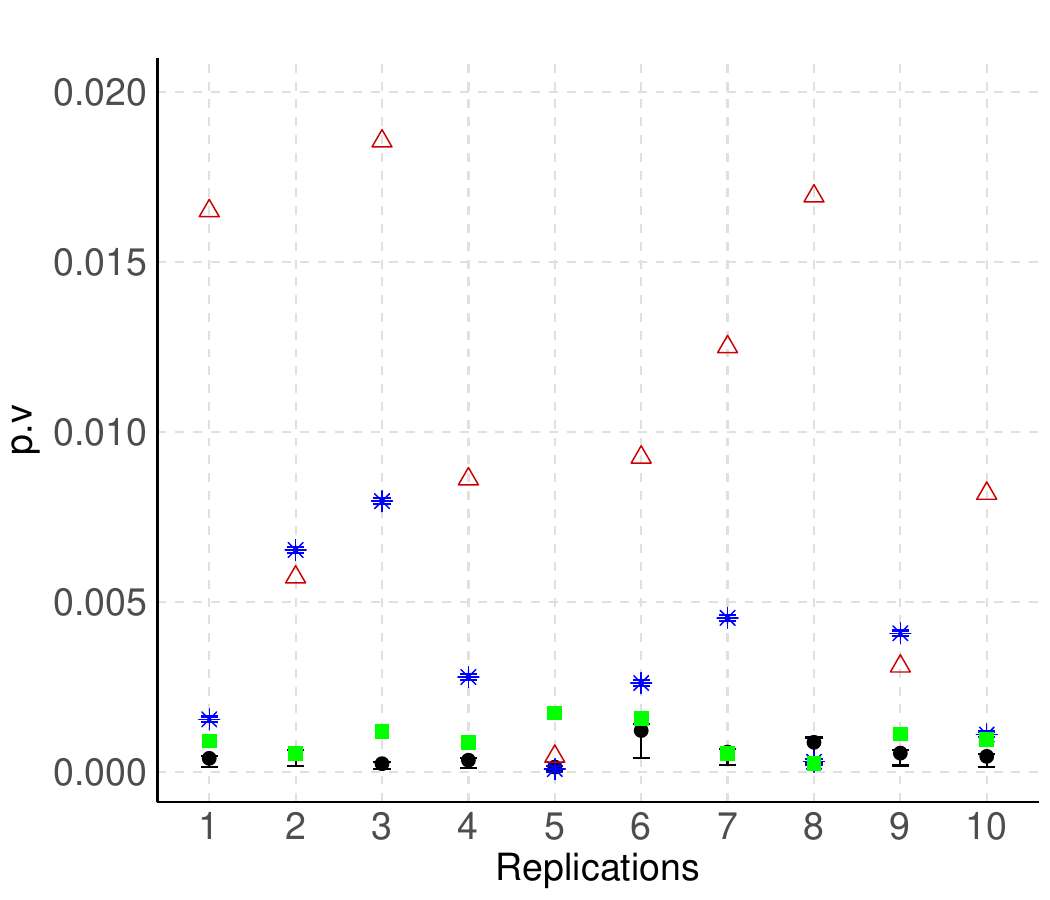}
\end{minipage}

\begin{minipage}{0.49\textwidth}
\centering
\includegraphics[width=\textwidth]{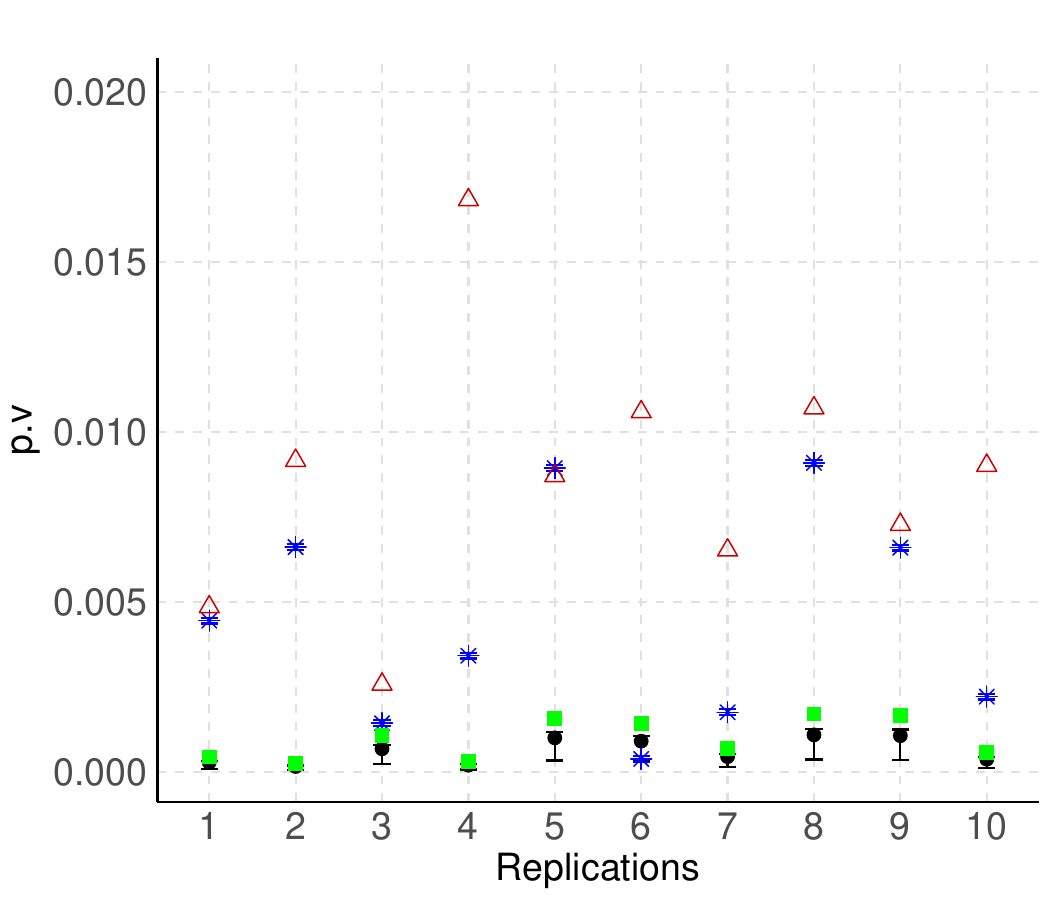}
\end{minipage}
\begin{minipage}{0.49\textwidth}
\centering
\includegraphics[width=\textwidth]{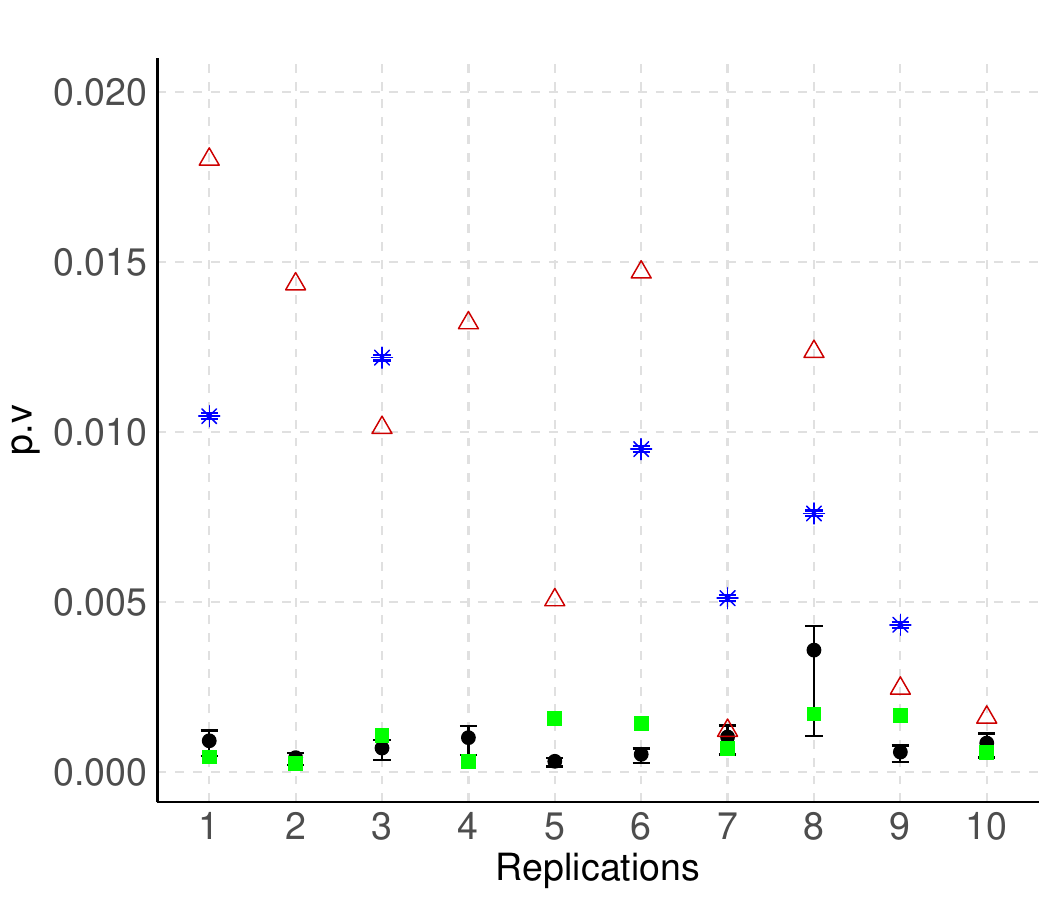}
\end{minipage}

\begin{minipage}{0.49\textwidth}
\centering
\includegraphics[width=\textwidth]{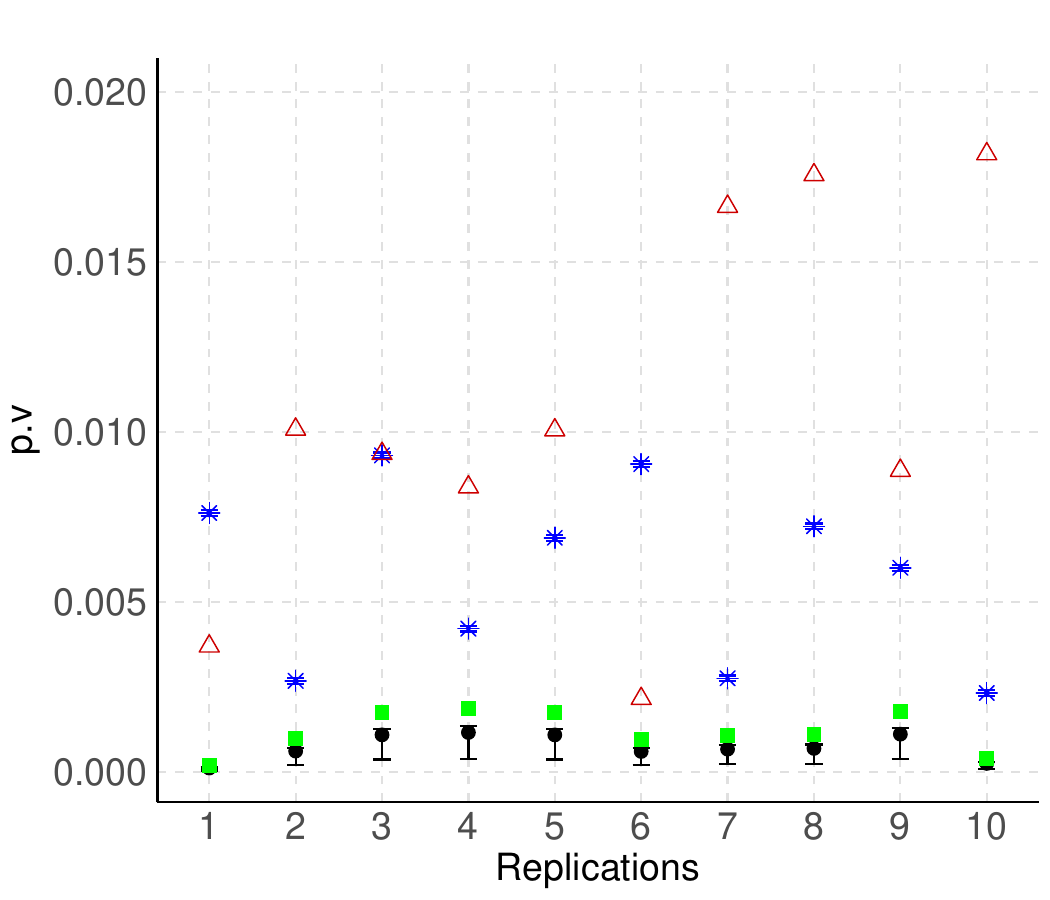}
\end{minipage}
\begin{minipage}{0.49\textwidth}
\centering
\includegraphics[width=\textwidth]{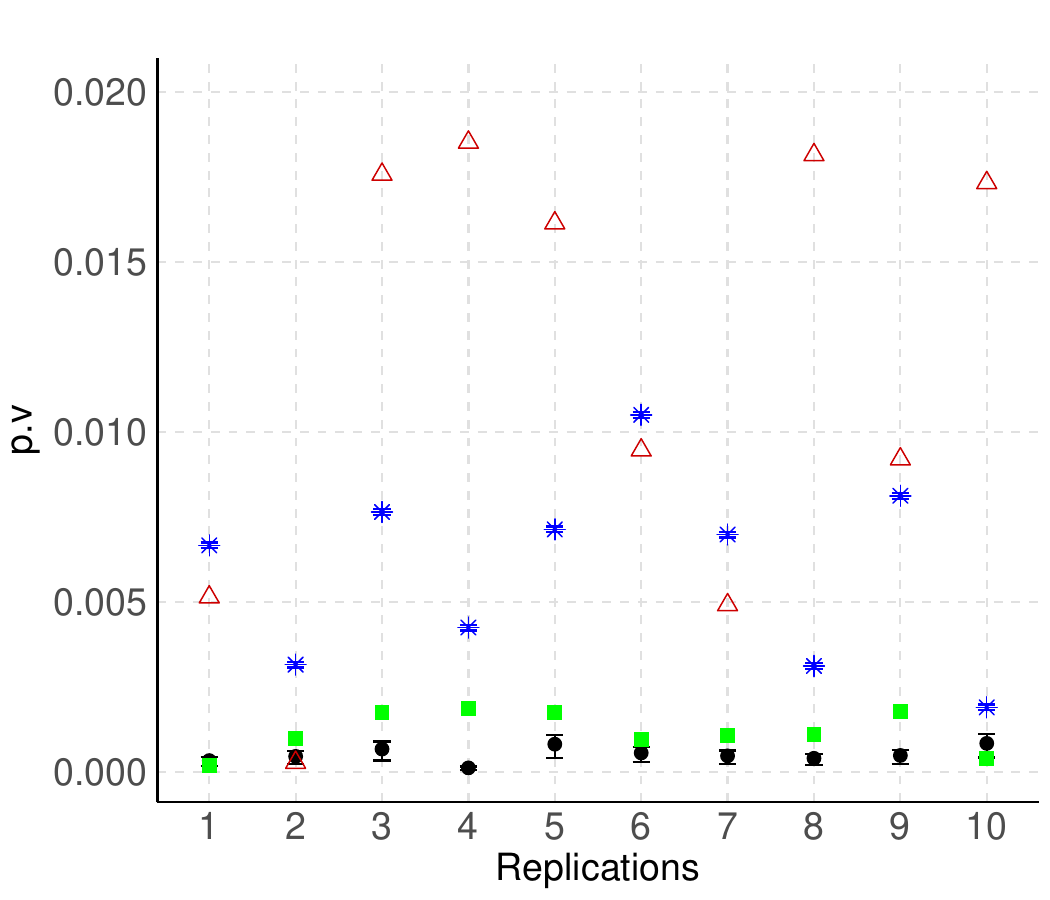}
\end{minipage}
 \caption{P-value (p.v)  of the independence test between $X_2$ and $X_1$, with $n=5000$. True p-value in green. $S_1$: predictive mean (black point) and $98\%$ predictive credibility interval. $S_2$: point estimate (blue asterisk) and $98\%$ confidence interval. $S_3$: point estimate (red triangle). Top: $d=3$, middle: $d=4$, bottom: $d=7$. $P_1$ on the left and $P_2$ on the right.}
  \label{vp_f1}
\end{figure}

\begin{figure}[h!]
\centering
\begin{minipage}{0.49\textwidth}
\centering
\includegraphics[width=\textwidth]{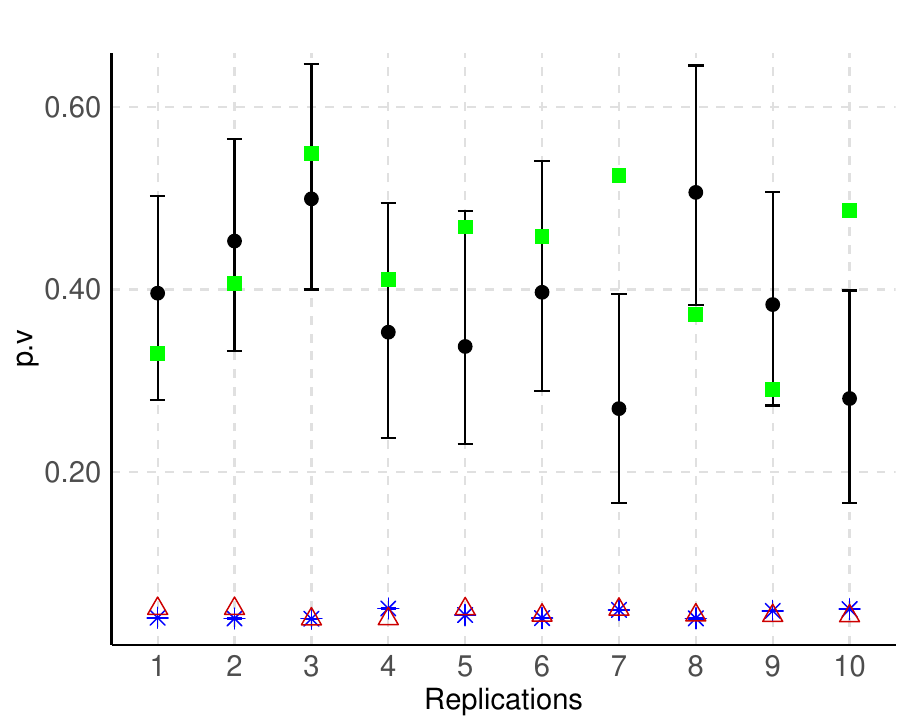}
\end{minipage}
\begin{minipage}{0.49\textwidth}
\centering
\includegraphics[width=\textwidth]{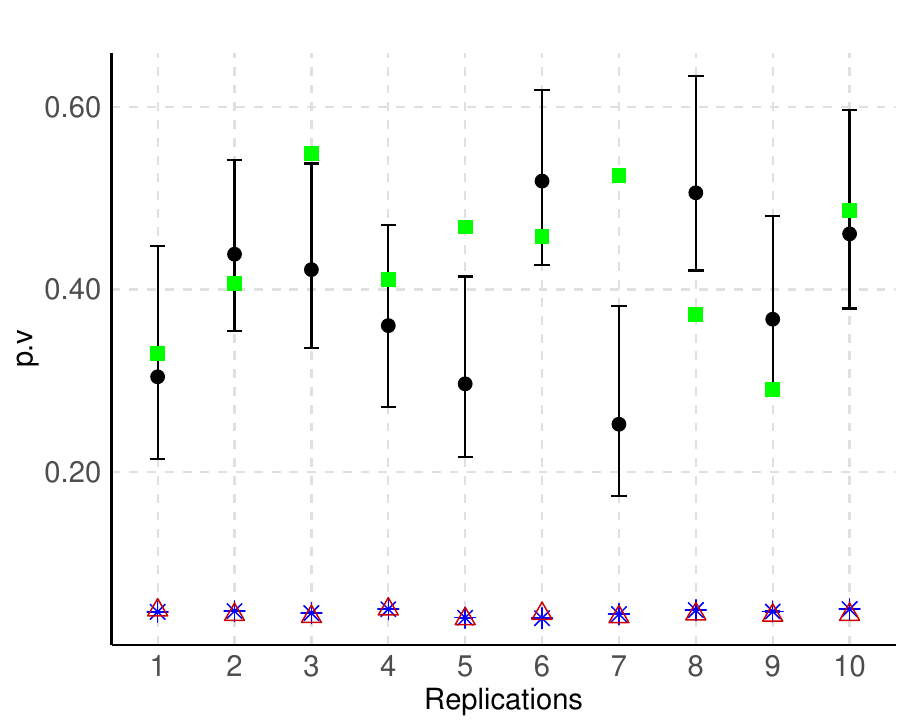}
\end{minipage}

\begin{minipage}{0.49\textwidth}
\centering
\includegraphics[width=\textwidth]{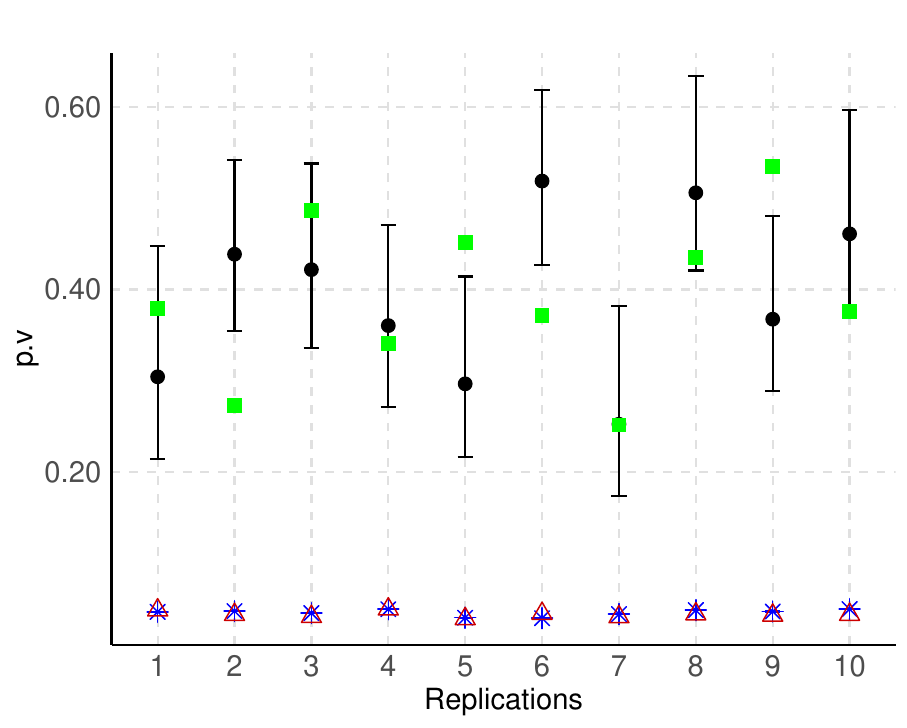}
\end{minipage}
\begin{minipage}{0.49\textwidth}
\centering
\includegraphics[width=\textwidth]{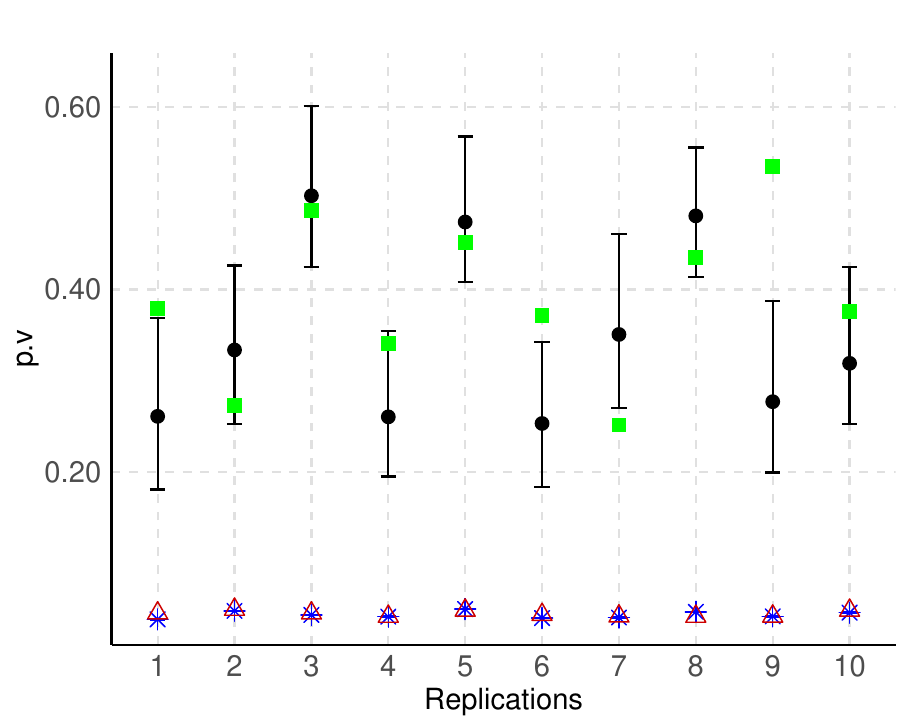}
\end{minipage}
 \caption{P-value (p.v) of the independence test between $X_1$ and $X_5$, with $p=7$. True p-value in green. $S_1$: predictive mean (black point) and $98\%$ predictive credibility interval. $S_2$: point estimate (blue asterisk) and $98\%$ confidence interval. $S_3$: point estimate (red triangle). Top: $n=2000$, bottom: $n=5000$. $P_1$ on the left and $P_2$ on the right.}
  \label{ap8}
\end{figure}

Figures \ref{vp_f1} and \ref{ap8} show the results for the p-value of the chi-square independence test between $X_1$ and $X_2$ and demonstrate a much superior performance of the proposed method $S_1$. The mean of the predictive distribution of this statistic is much closer to the true value and always agrees with the conclusion of the original test, unlike the other two methods.
The confidence intervals for method $S_2$ are practically degenerated.

Omitted results for other parameters $\theta$ and sample sizes follow the same pattern as those presented in this section.

%A collection of results for other parameters $\theta$ and sample sizes are presented in Appendix A. The results follow the same pattern as those presented in this section.

\clearpage
\newpage

\clearpage
\newpage

\clearpage
\newpage

%% file: 4.2Dados_reais.tex
\subsection{Application}

This section presents a real data analysis with data from the The National Household Sample Survey (PNAD) from Brazil. This is a nationwide household survey conducted by the Brazilian Institute of Geography and Statistics (IBGE). The survey collects information about the socioeconomic characteristics of the Brazilian population, including information on employment, education, income, health, and housing. PNAD data is available in the R package \emph{PNADcIBGE} \citep{braga2018pnadcibge}.

We consider a database from 2023 with information on 5651 households and select 5 (out of the 237) variables, which are chosen to exemplify information that may be confidential. Table \ref{tab:table4.3} presents the selected variables along with their categories and Table \ref{tab:table_real} shows the observed proportions for each variable.

\begin{table}[!h]
  \begin{center}
    \caption{Selected variables from the PNAD 2023 database.}
    \label{tab:table4.3}
\small
    \begin{tabular}{l|c}
    Variable & Categories \\
    \hline
    $X_1:$ House location & Rural or Urban \\
    $X_2:$ Gender & Female or Male \\
    $X_3:$ Age & $>$50 or $<=$50 \\
    $X_4:$ Race & White or Brown/Black \\
    $X_5:$ Income & $>$ Minimum Wage 2023 or $<=$ Minimum Wage 2023 \\
    \end{tabular}
  \end{center}
  \footnotesize{Note: 
minimum wage in Brazil in 2023 = $R\$ 1,320.00$.}
\end{table}

\begin{table}[!h]
  \begin{center}
    \caption{Observed proportions in the PNAD 2023 dataset.}
    \label{tab:table_real}
\small
    \begin{tabular}{l|c}
    Variable & $\%$ \\
    \hline
    $X_1=$ Urban & 0.79 \\
    $X_2=$ Male & 0.59  \\
    $X_3=$ $<=$50 &  0.75\\
    $X_4=$ Brown/Black & 0.74 \\
    $X_5=$ $<=$ Minimum Wage & 0.34  \\
    \end{tabular}
  \end{center}
\end{table}

Table \ref{tab:table45} presents the probabilities of the two networks with the highest posterior probabilities.

\begin{table}[!h]
  \begin{center}
    \caption{Posterior probabilities of the two most likely networks for the PNAD data.}
    \label{tab:table45}
\small
    \begin{tabular}{c|c|c}
    \textbf{Prior} & \textbf{Estimated Network} & \textbf{Probability} \\
    \hline
    $P_1$ & \multirow{2}{*}{$X_1, X_3|X_1, X_4|X_3,   (X_5|X_1,X_4),(X_2|X_1,X_5)$} & 0.256   \\
    $P_2$  & & 0.131  \\
    \hline
    $P_1$ & \multirow{2}{*}{$X_5, X_4|X_5, X_3|X_4, (X_1|X_3,X_5), (X_2|X_1,X_5)$} &  0.057  \\
    $P_2$  & &  0.007 \\
    \end{tabular}
  \end{center}
\end{table}

As expected, the posterior probabilities of the two networks with the highest posterior probabilities are higher for the informative prior, indicating a higher posterior concentration. The fact that the those probabilities are not high raises important issues related to the problem studied in this paper. Firstly, this indicates a reasonable level of uncertainty about the network and that a large number of different networks are needed to encompass most of the posterior probability mass. Therefore, a method that uses all these networks, properly weighted by their respective probabilities, is meant to be considerably more efficient and robust for quantifying the uncertainty of the system. This is exactly the case of the $S_1$ method proposed in this paper. The second important point is that the specific conditional independence relationships of the network with the highest posterior probabilities should not be overemphasized.

\begin{figure}[h!]

\begin{minipage}{0.49\textwidth}
\centering
\includegraphics[width=\textwidth]{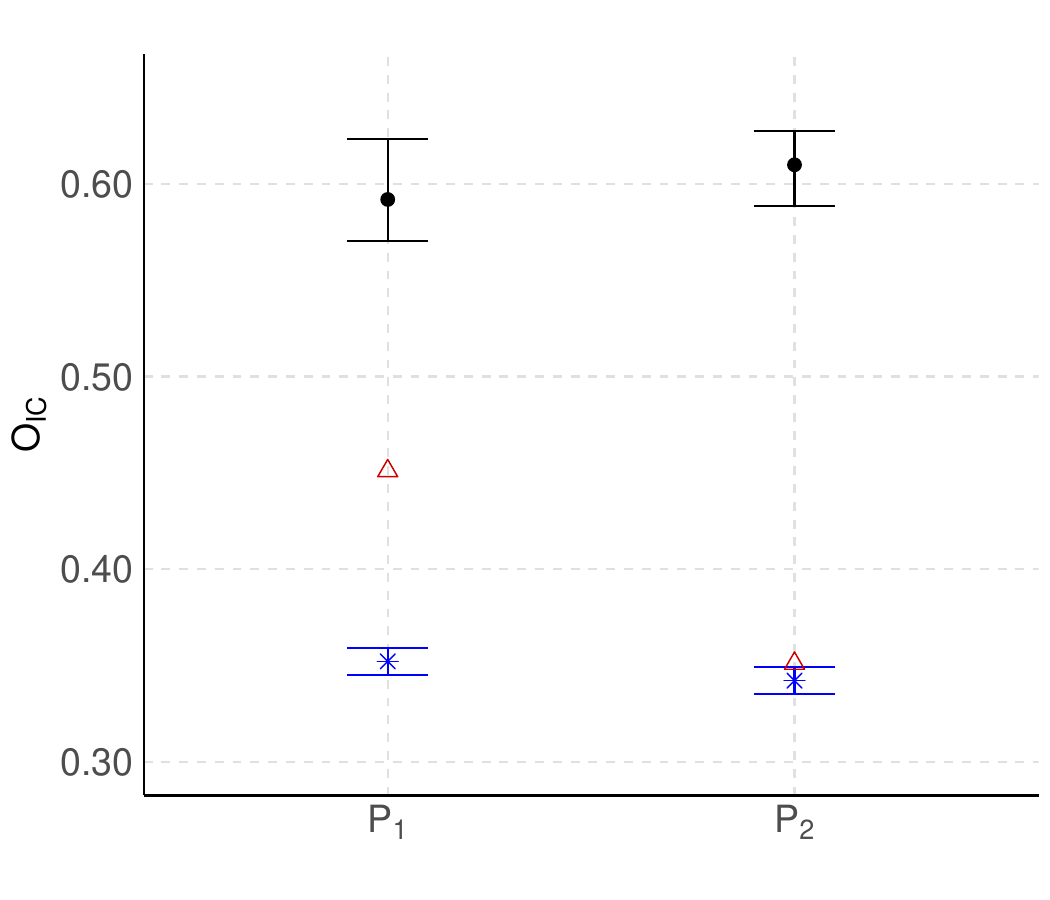}
\end{minipage}
\begin{minipage}{0.49\textwidth}
\centering
\includegraphics[width=\textwidth]{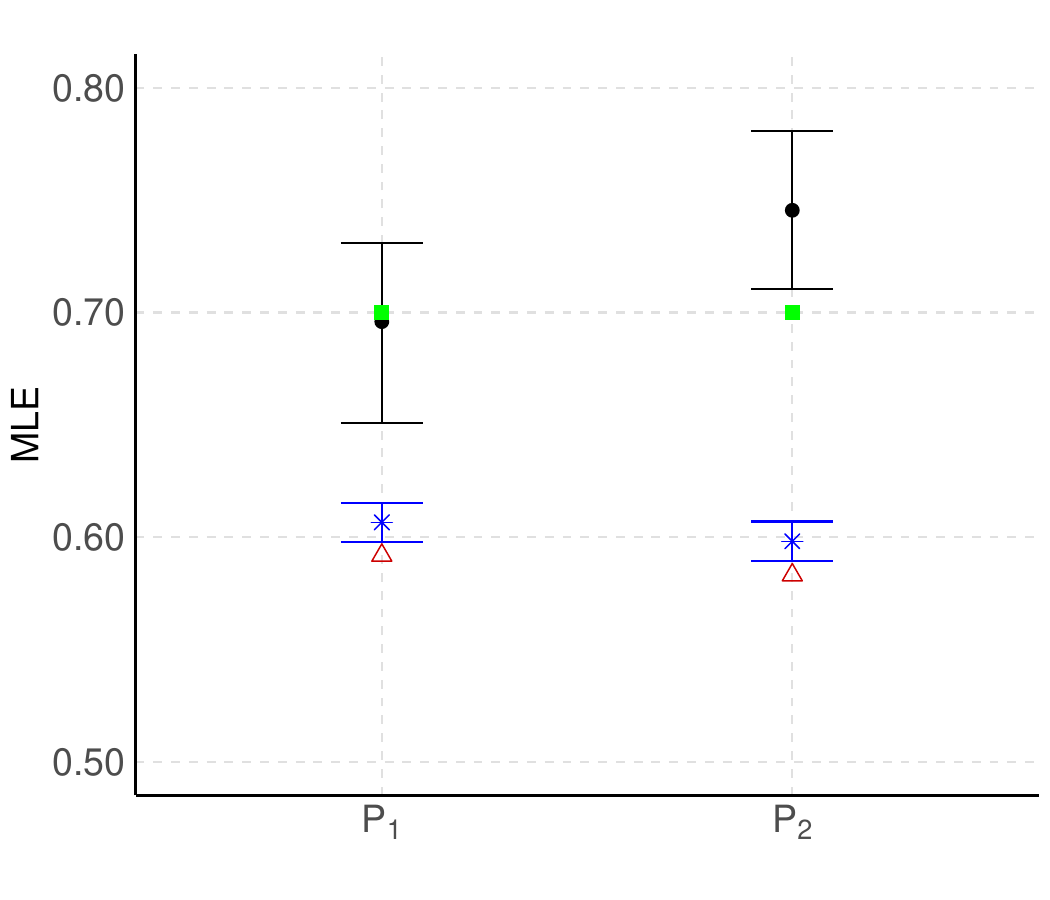}
\end{minipage}

\begin{minipage}{0.49\textwidth}
\centering
\includegraphics[width=\textwidth]{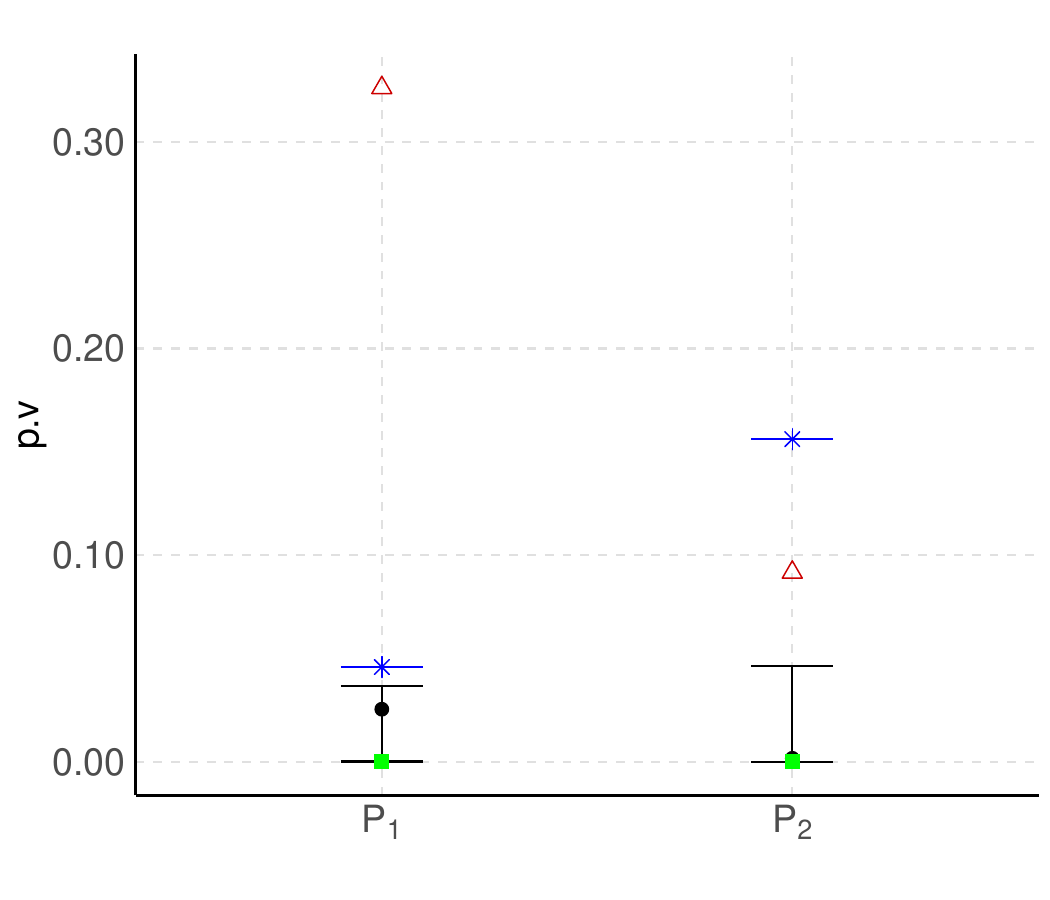}

\end{minipage}

 \caption{$O_{IC}$,  MLE of $\theta=P(X_3=1|X_1=1)$, and p-value (p.v) of the independence test between $X_1$ and $X_3$. True p-value is $0.00001$ and true MLE is $0.70$ (green). $S_1$: predictive mean (black point) and $98\%$ predictive credibility interval. $S_2$: point estimate (blue asterisk) and $98\%$ confidence interval. $S_3$: point estimate (red triangle).}
  \label{fig1_real}
\end{figure}

\begin{figure}[h!]

\begin{minipage}{0.49\textwidth}
\centering
\includegraphics[width=\textwidth]{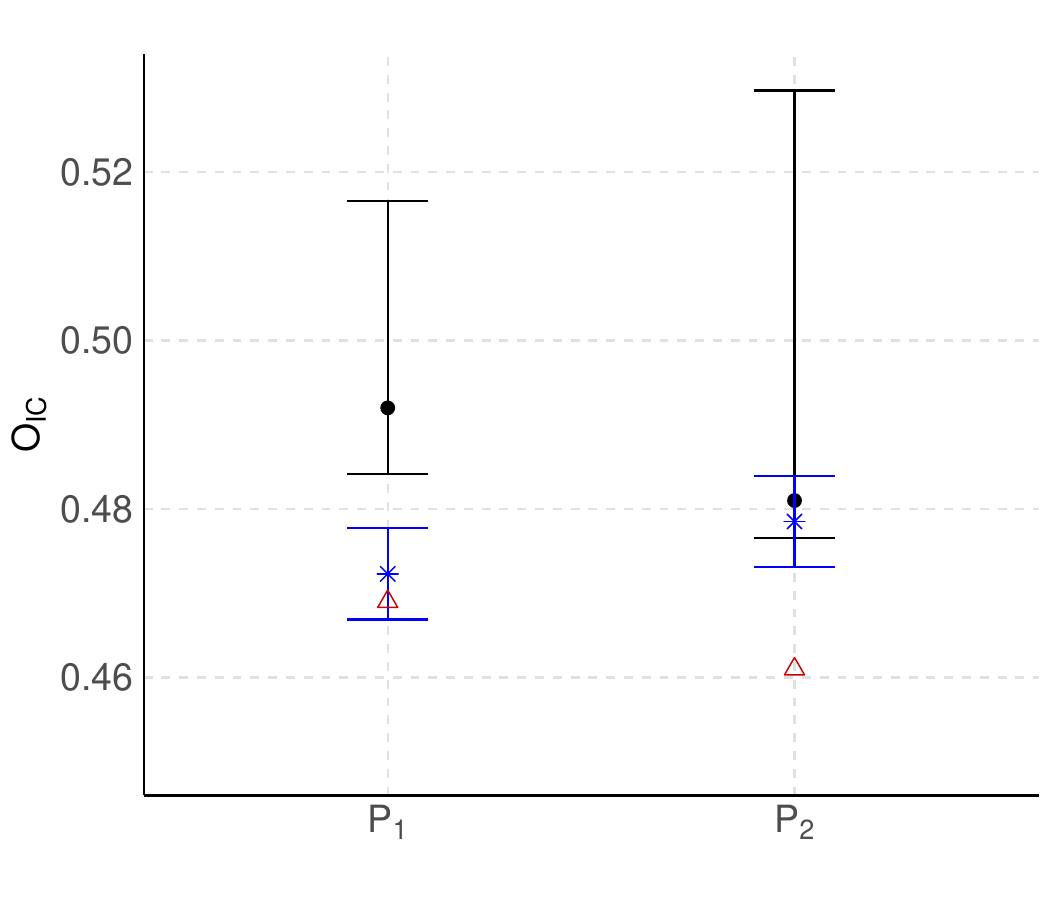}
\end{minipage}
\begin{minipage}{0.49\textwidth}
\centering
\includegraphics[width=\textwidth]{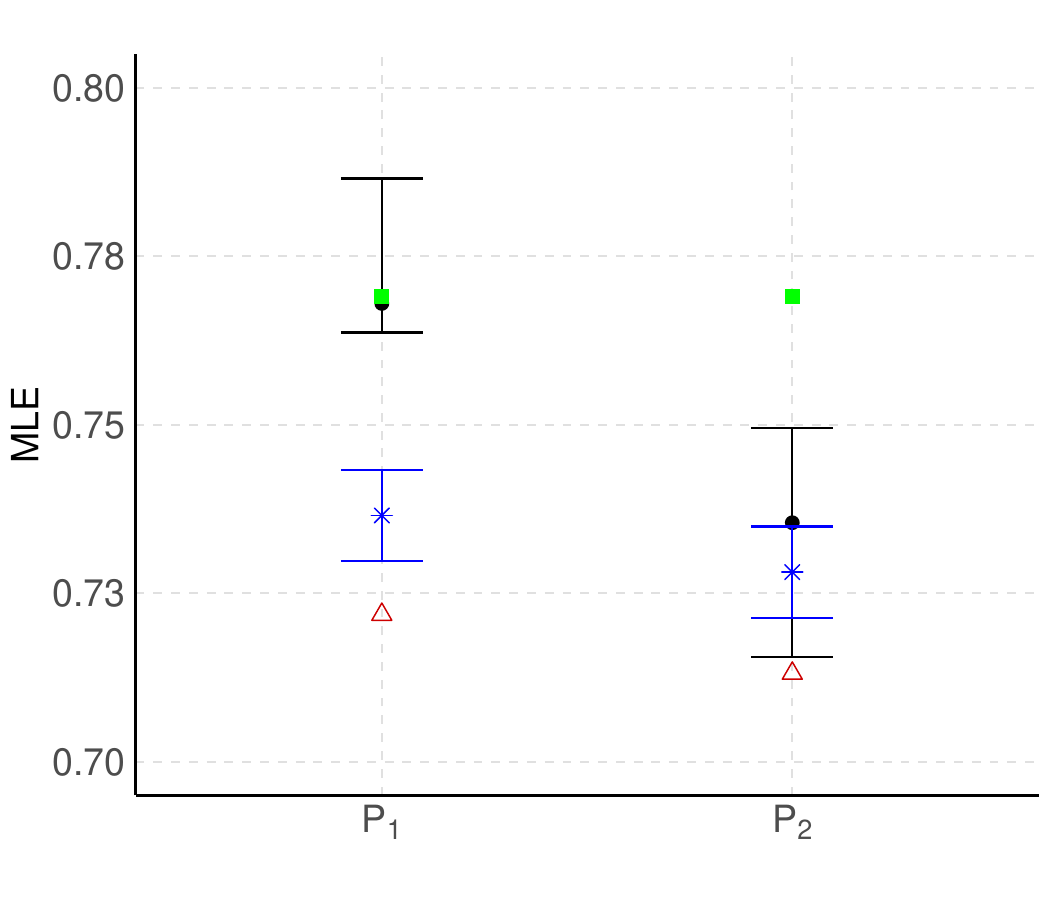}
\end{minipage}

\begin{minipage}{0.49\textwidth}
\centering
\includegraphics[width=\textwidth]{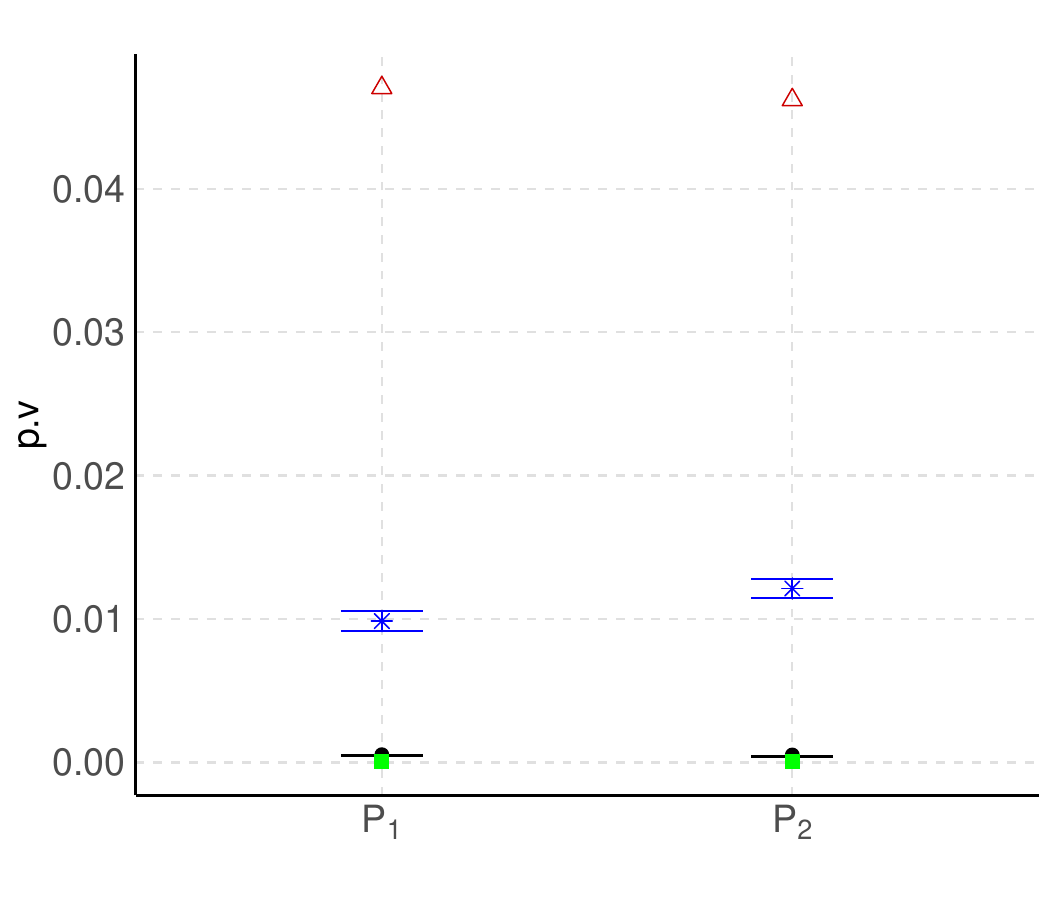}

\end{minipage}

 \caption{$O_{IC}$,  MLE of $\theta=P(X_4=1|X_3=1)$, and p-value (p.v) of the independence test between $X_3$ and $X_4$. True p-value is $0.0015$ and true MLE is $0.769$ (green). $S_1$: predictive mean (black point) and $98\%$ predictive credibility interval. $S_2$: point estimate (blue asterisk) and $98\%$ confidence interval. $S_3$: point estimate (red triangle).}
  \label{fig1_real2}
\end{figure}

Figures \ref{fig1_real} and \ref{fig1_real2} present the results for the same three statistics considered in the simulated data studies for the same three methods. Results are considerably better for the proposed method $S_1$ for all the three statistics. Additionally, for method $S_1$, the informative prior ($P_1$) performed much better than the uniform prior ($P_2$) in the case of the EMV.

\clearpage
\newpage

\clearpage
\newpage

%% file: Consideracoes.tex
\section{Final remarks}

This paper proposed an efficient and robust methodology to generate and analyze synthetic data via Bayesian networks under a full Bayes approach. The methodology involves the estimation of the Bayesian network to fit the observed data and the generation and analysis of synthetic data. A new class of modular penalizing priors was proposed to provide more robustness to the analyses.

As discussed, the Bayesian network is an important tool to estimate joint distributions. In the context of synthetic data, this methodology brings significant advantages over existing approaches. The higher computational cost of the proposed approach does not compromise its use in examples where other approaches are computationally feasible.

The MCMC proposed by Goudie et al. (2016) has shown to be efficient in approximating the posterior distribution of the network. This algorithm is used to propose a method for generating synthetic data and obtaining the predictive distribution of the statistics of interest. This produces different outputs to be delivered to the final user, depending on their objectives and statistical expertise.

The comparison between the proposed methodology and two other commonly used ones in the literature was conducted through the analysis of three statistics of the synthetic data, with the main objective being to approximate the same statistic obtained with the original data. In addition to demonstrating better performance, the proposed method allows for efficient uncertainty quantification.

A synthetic data analysis was conducted for a real dataset corresponding to the 2023 PNAD to illustrate the applicability of the propose methodology. This example highlighted the advantages of the proposed methodology when there is greater uncertainty regarding the posterior distribution of the Bayesian network.

Ideas for future work include other types of variables, from multinomial to continuous ones, to improve the scope and applicability of univariate models. Also, considering databases that contain variables of different types, providing a more comprehensive and realistic view of the relationships between different types of data. Finally, scenarios with missing in the original dataset are also an interesting extension.

%% file: Z_apendice.tex
\begin{figure}[h!]
\centering
\begin{minipage}{0.49\textwidth}
\centering
\includegraphics[width=\textwidth]{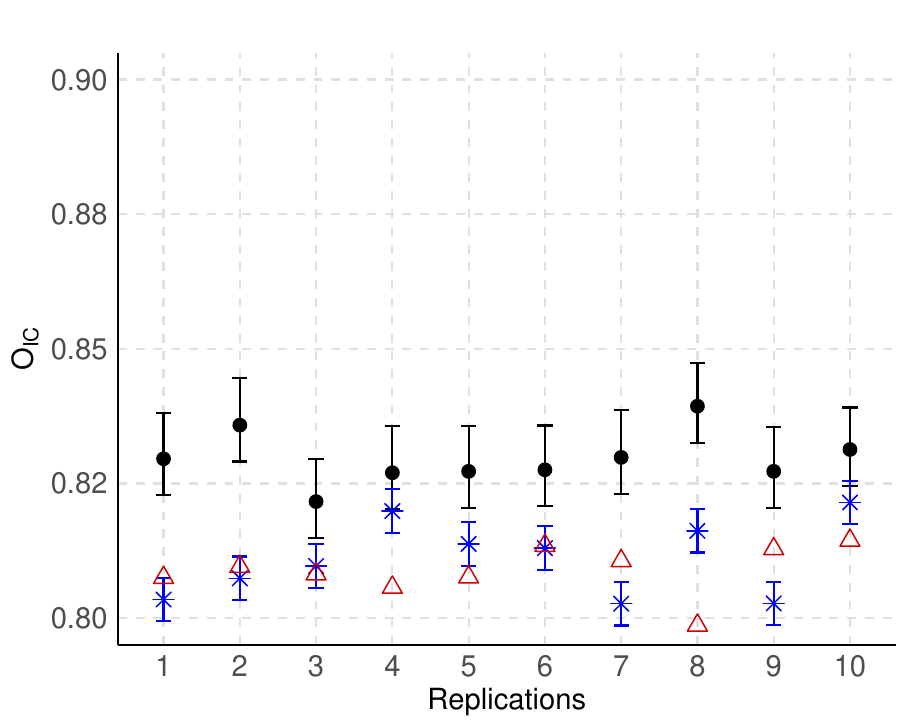}
\end{minipage}
\begin{minipage}{0.49\textwidth}
\centering
\includegraphics[width=\textwidth]{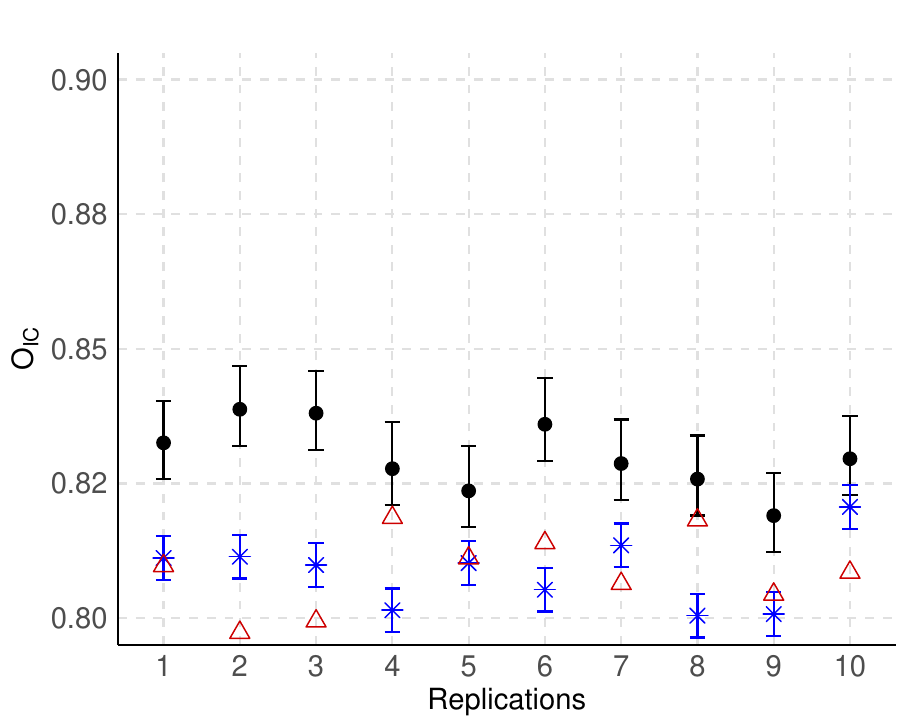}
\end{minipage}

\begin{minipage}{0.49\textwidth}
\centering
\includegraphics[width=\textwidth]{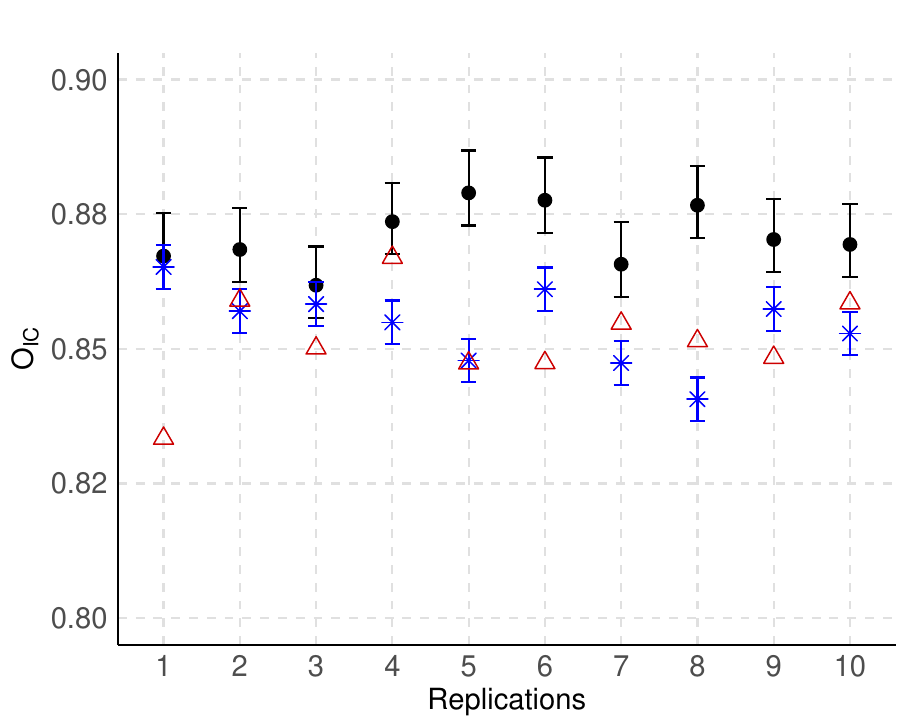}
\end{minipage}
\begin{minipage}{0.49\textwidth}
\centering
\includegraphics[width=\textwidth]{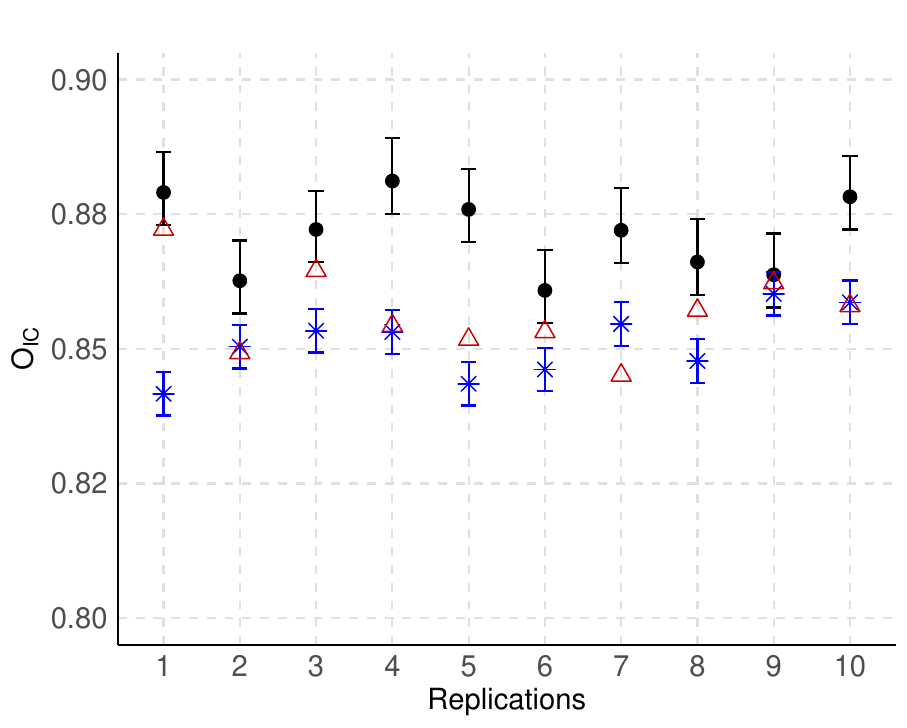}
\end{minipage}

 \caption{$O_{IC}$ measure for $\theta=P(X_2=1|X_1=0)$, with $n=1000$. $S_1$: predictive mean (black point) and $98\%$ predictive credibility interval. $S_2$: point estimate (blue asterisk) and $98\%$ confidence interval. $S_3$: point estimate (red triangle). Top: $d=3$, bottom: $d=4$. $P_1$ on the left and $P_2$ on the right.}
  \label{ap_1}
\end{figure}
\clearpage
\newpage

\begin{figure}[h!]
\centering
\begin{minipage}{0.49\textwidth}
\centering
\includegraphics[width=\textwidth]{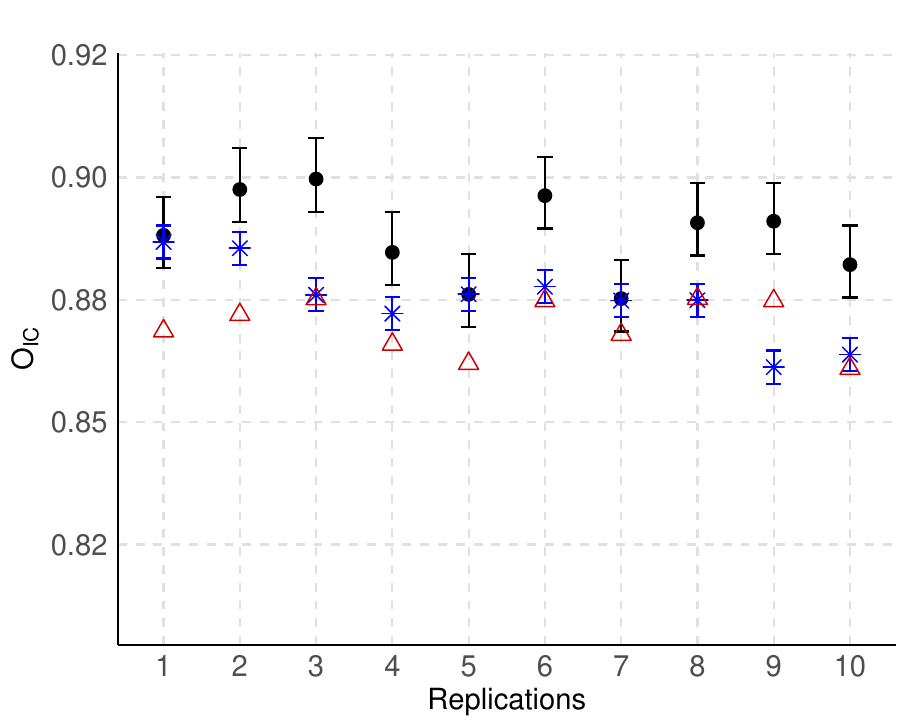}
\end{minipage}
\begin{minipage}{0.49\textwidth}
\centering
\includegraphics[width=\textwidth]{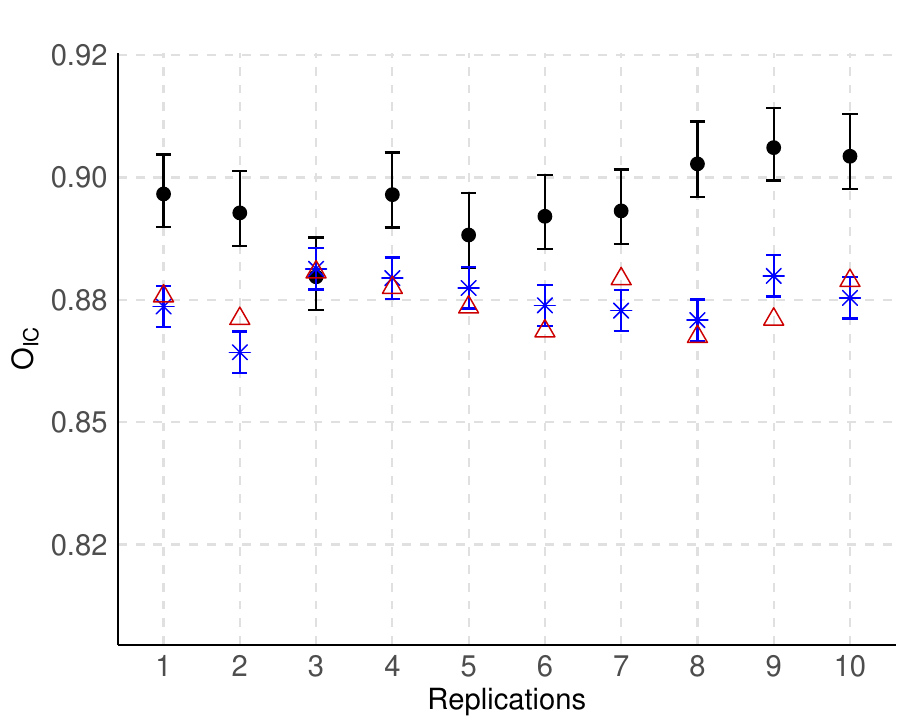}
\end{minipage}

\begin{minipage}{0.49\textwidth}
\centering
\includegraphics[width=\textwidth]{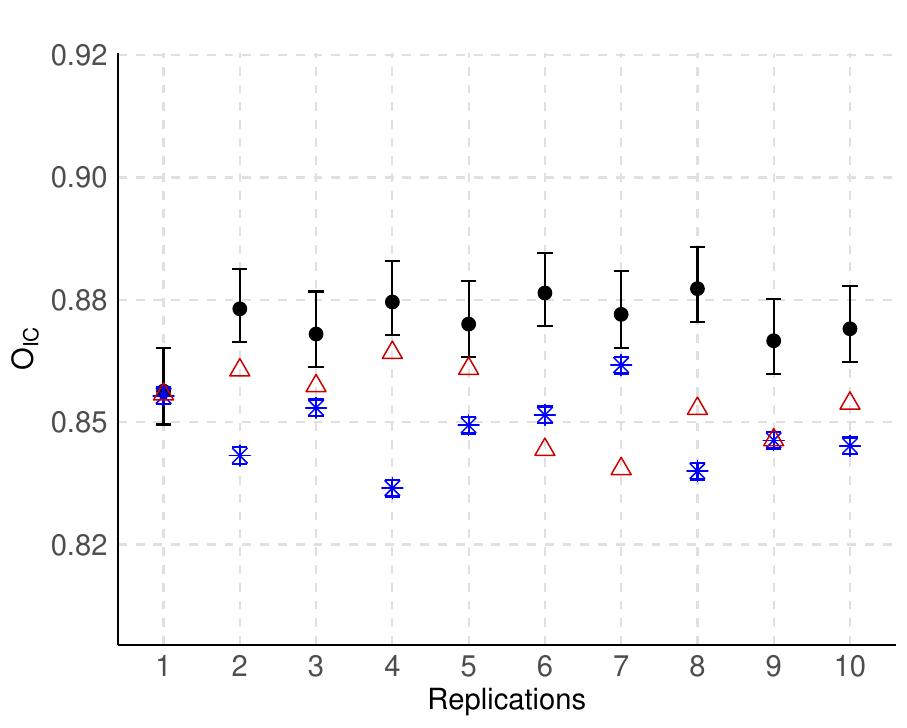}
\end{minipage}
\begin{minipage}{0.49\textwidth}
\centering
\includegraphics[width=\textwidth]{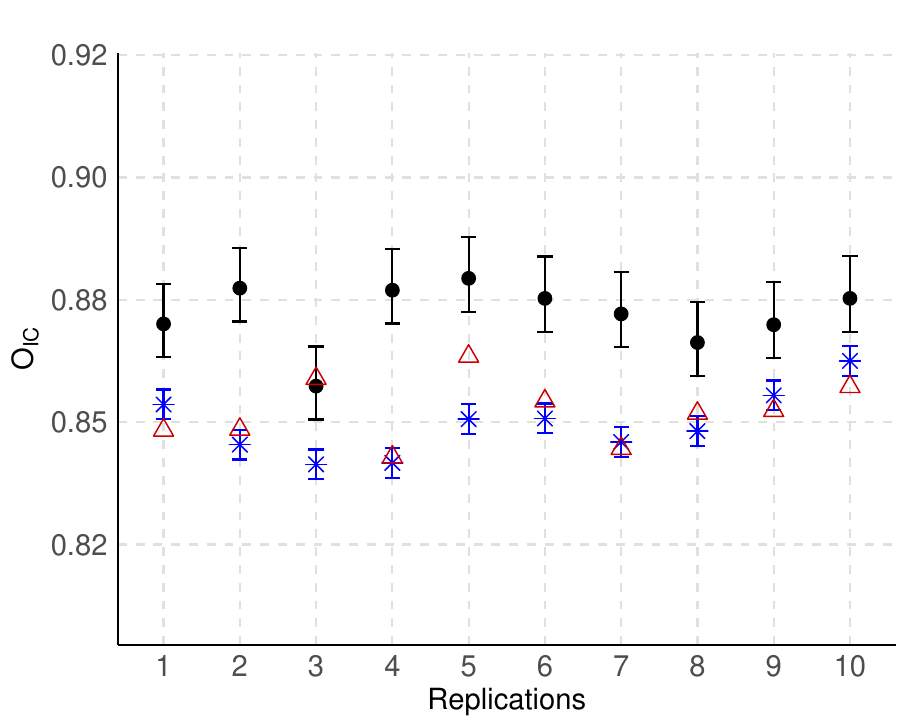}
\end{minipage}

\begin{minipage}{0.49\textwidth}
\centering
\includegraphics[width=\textwidth]{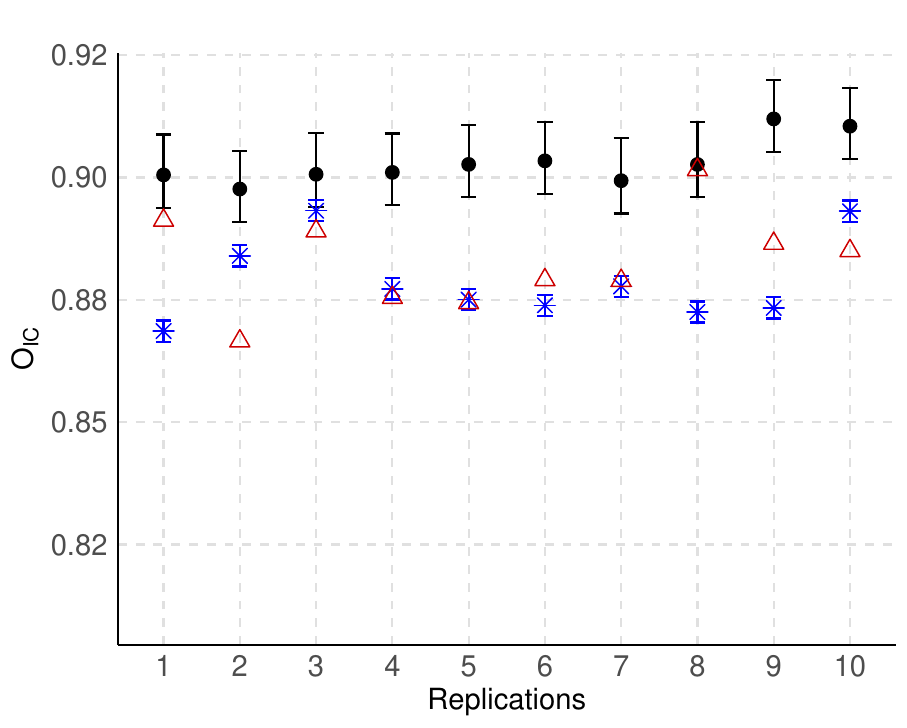}
\end{minipage}
\begin{minipage}{0.49\textwidth}
\centering
\includegraphics[width=\textwidth]{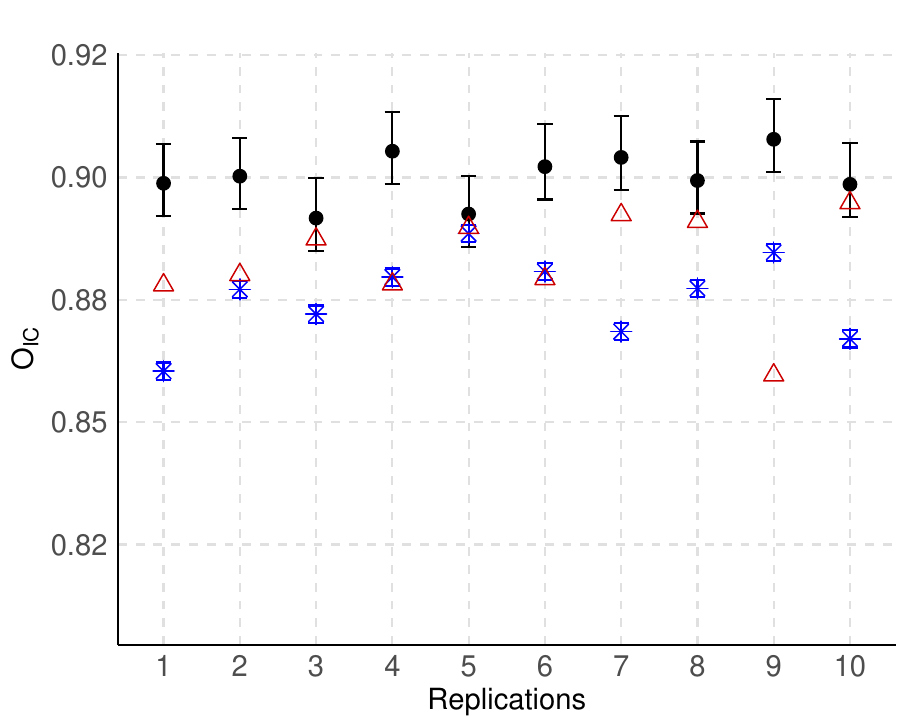}
\end{minipage}

 \caption{$O_{IC}$ measure for $\theta=P(X_3=1)$, with $d=3$. $S_1$: predictive mean (black point) and $98\%$ predictive credibility interval. $S_2$: point estimate (blue asterisk) and $98\%$ confidence interval. $S_3$: point estimate (red triangle). Top: $n=500$, middle: $n=1000$, bottom: $n=5000$. $P_1$ on the left and $P_2$ on the right.}
  \label{ap_2}
\end{figure}
\clearpage
\newpage

\begin{figure}[h!]
\centering
\begin{minipage}{0.49\textwidth}
\centering
\includegraphics[width=\textwidth]{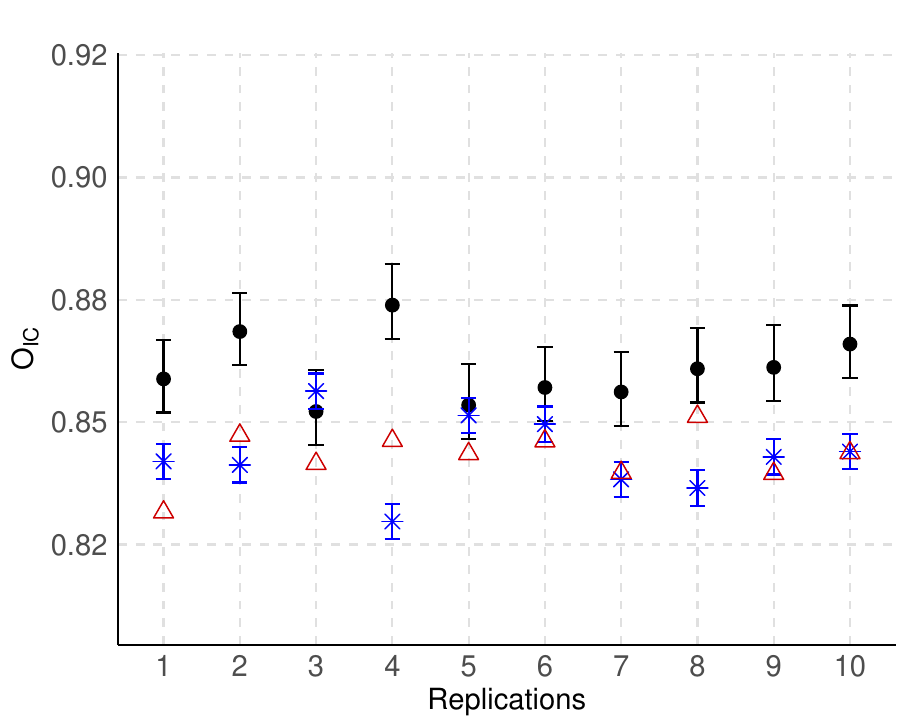}
\end{minipage}
\begin{minipage}{0.49\textwidth}
\centering
\includegraphics[width=\textwidth]{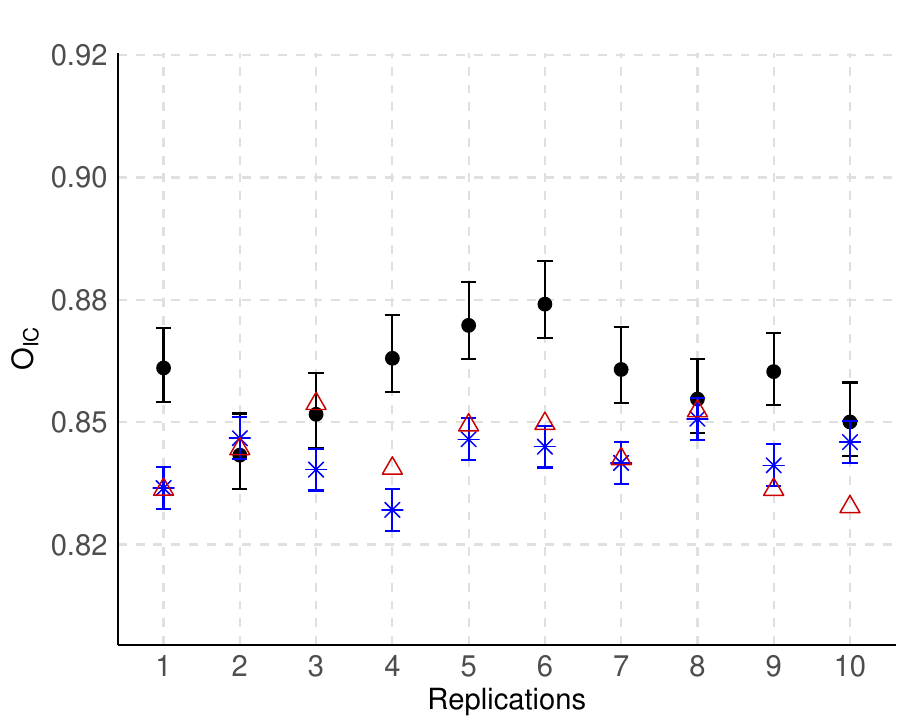}
\end{minipage}

\begin{minipage}{0.49\textwidth}
\centering
\includegraphics[width=\textwidth]{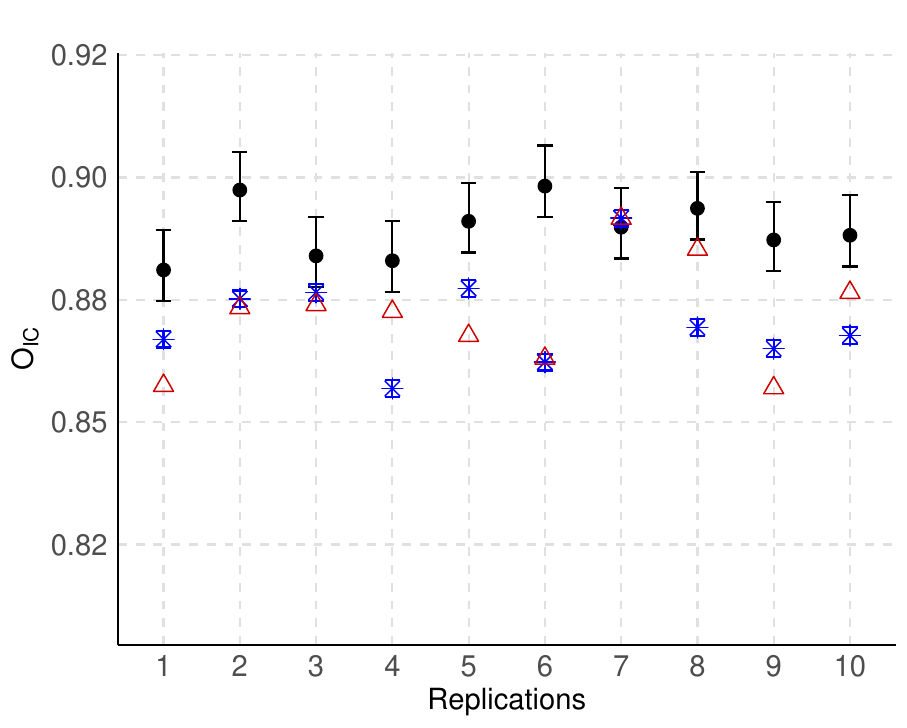}
\end{minipage}
\begin{minipage}{0.49\textwidth}
\centering
\includegraphics[width=\textwidth]{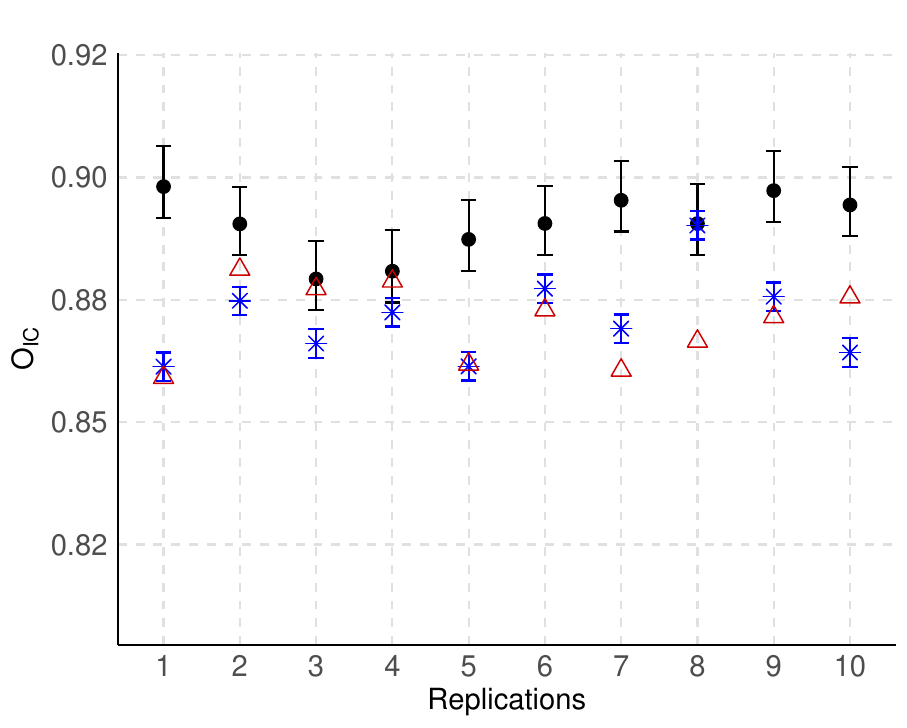}
\end{minipage}

 \caption{$O_{IC}$ measure for $\theta=P(X_3=1|X_4=1)$, with $d=4$. $S_1$: predictive mean (black point) and $98\%$ predictive credibility interval. $S_2$: point estimate (blue asterisk) and $98\%$ confidence interval. $S_3$: point estimate (red triangle). Top: $n=1000$, bottom: $n=5000$. $P_1$ on the left and $P_2$ on the right.}
  \label{ap_3}
\end{figure}
\clearpage
\newpage

\begin{figure}[h!]
\centering
\begin{minipage}{0.49\textwidth}
\centering
\includegraphics[width=\textwidth]{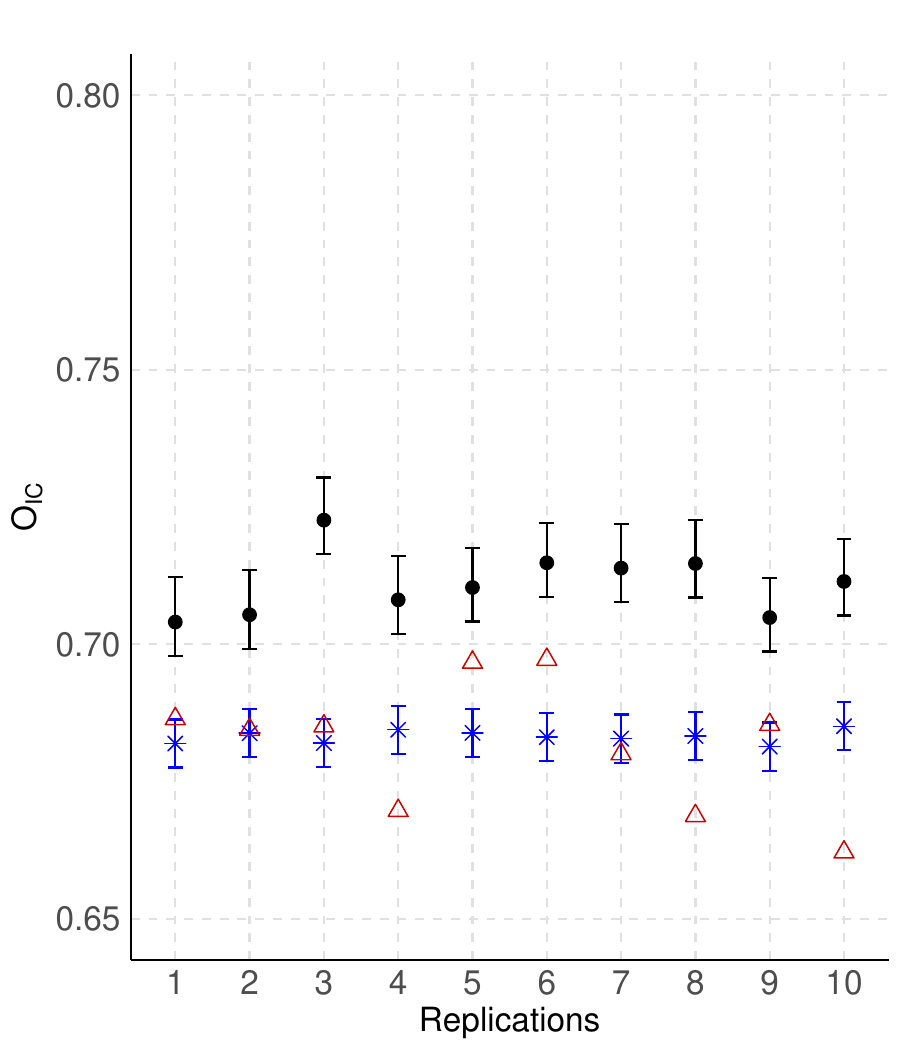}
\end{minipage}
\begin{minipage}{0.49\textwidth}
\centering
\includegraphics[width=\textwidth]{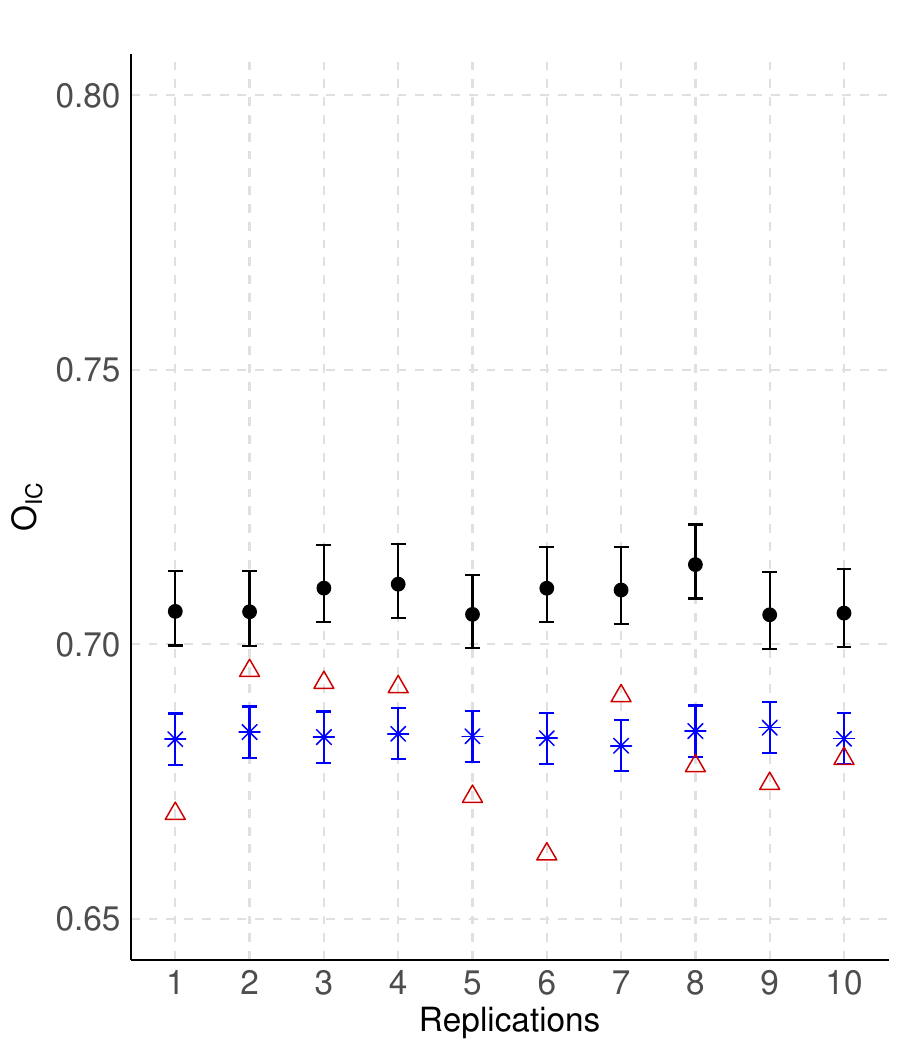}
\end{minipage}

\begin{minipage}{0.49\textwidth}
\centering
\includegraphics[width=\textwidth]{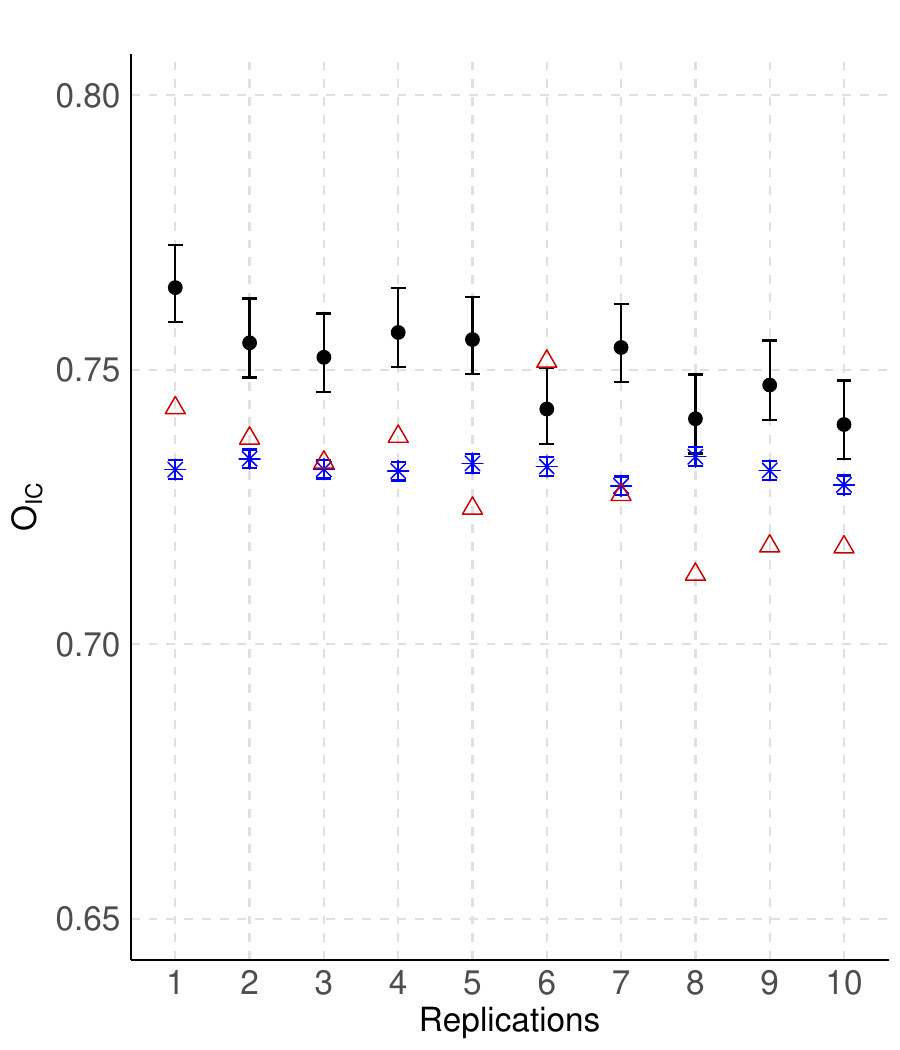}
\end{minipage}
\begin{minipage}{0.49\textwidth}
\centering
\includegraphics[width=\textwidth]{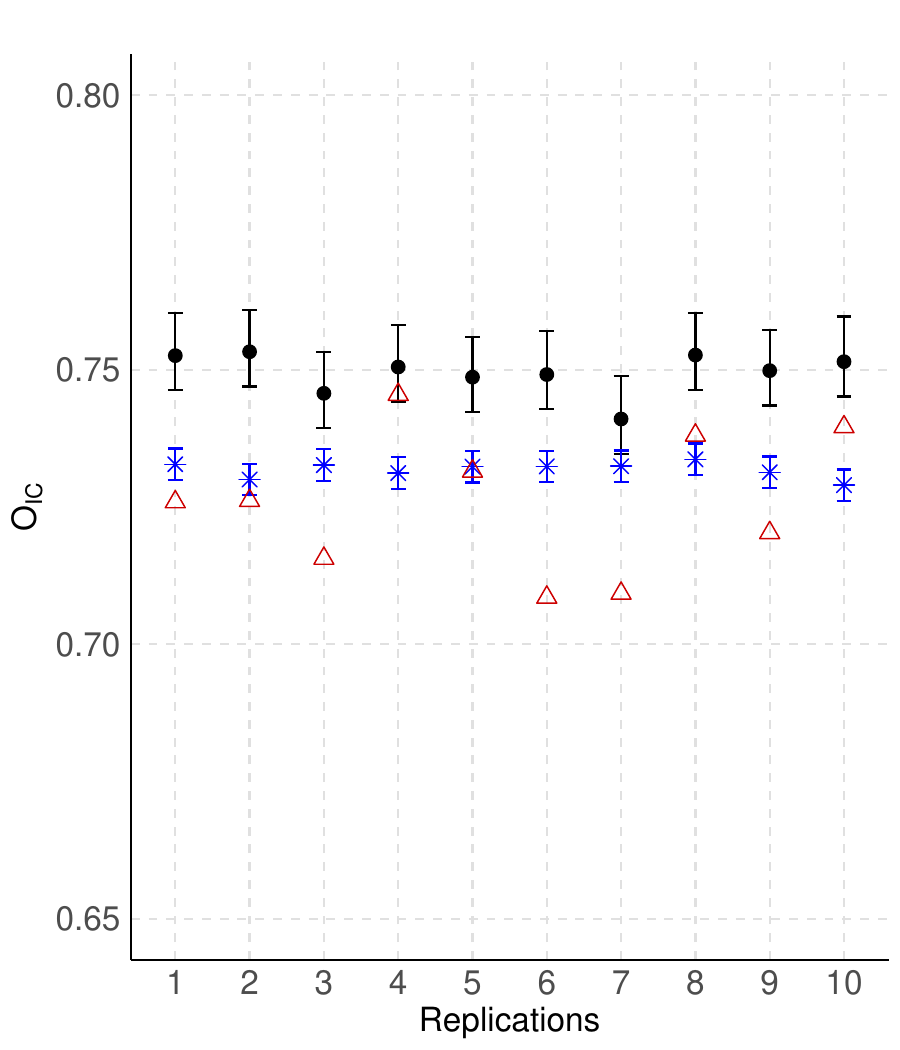}
\end{minipage}

 \caption{$O_{IC}$ measure for $\theta=P(X_4=1|X_3=1,X_5=1)$, with $d=7$. $S_1$: predictive mean (black point) and $98\%$ predictive credibility interval. $S_2$: point estimate (blue asterisk) and $98\%$ confidence interval. $S_3$: point estimate (red triangle). Top: $n=2000$, bottom: $n=5000$. $P_1$ on the left and $P_2$ on the right.}
  \label{ap_4}
\end{figure}
\clearpage
\newpage

\begin{figure}[h!]
\centering
\begin{minipage}{0.49\textwidth}
\centering
\includegraphics[width=\textwidth]{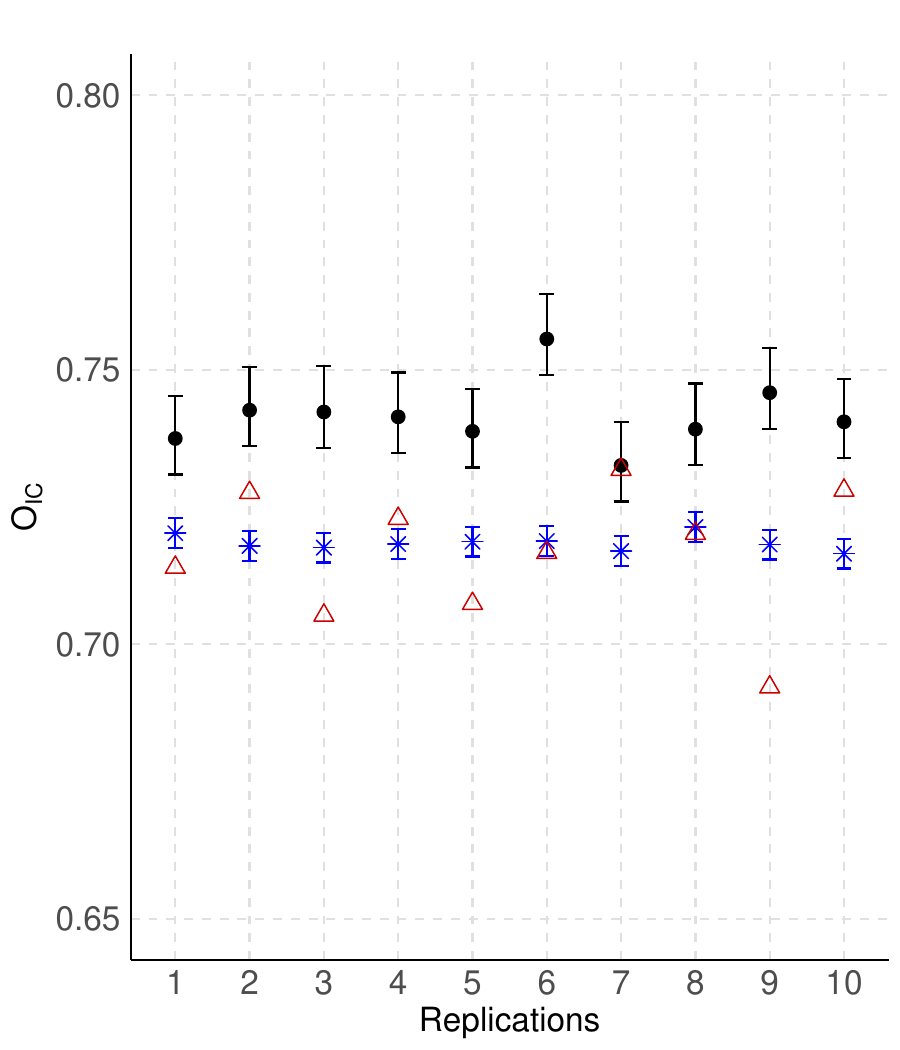}
\end{minipage}
\begin{minipage}{0.49\textwidth}
\centering
\includegraphics[width=\textwidth]{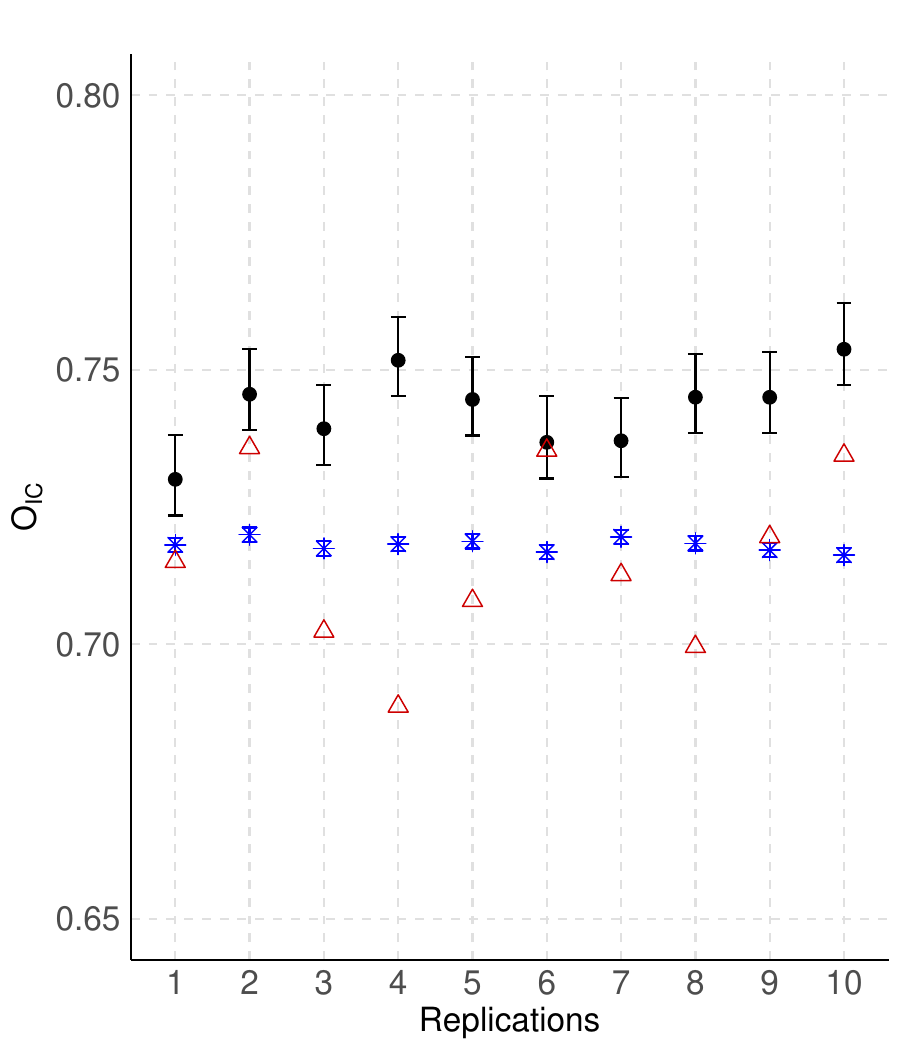}
\end{minipage}

\begin{minipage}{0.49\textwidth}
\centering
\includegraphics[width=\textwidth]{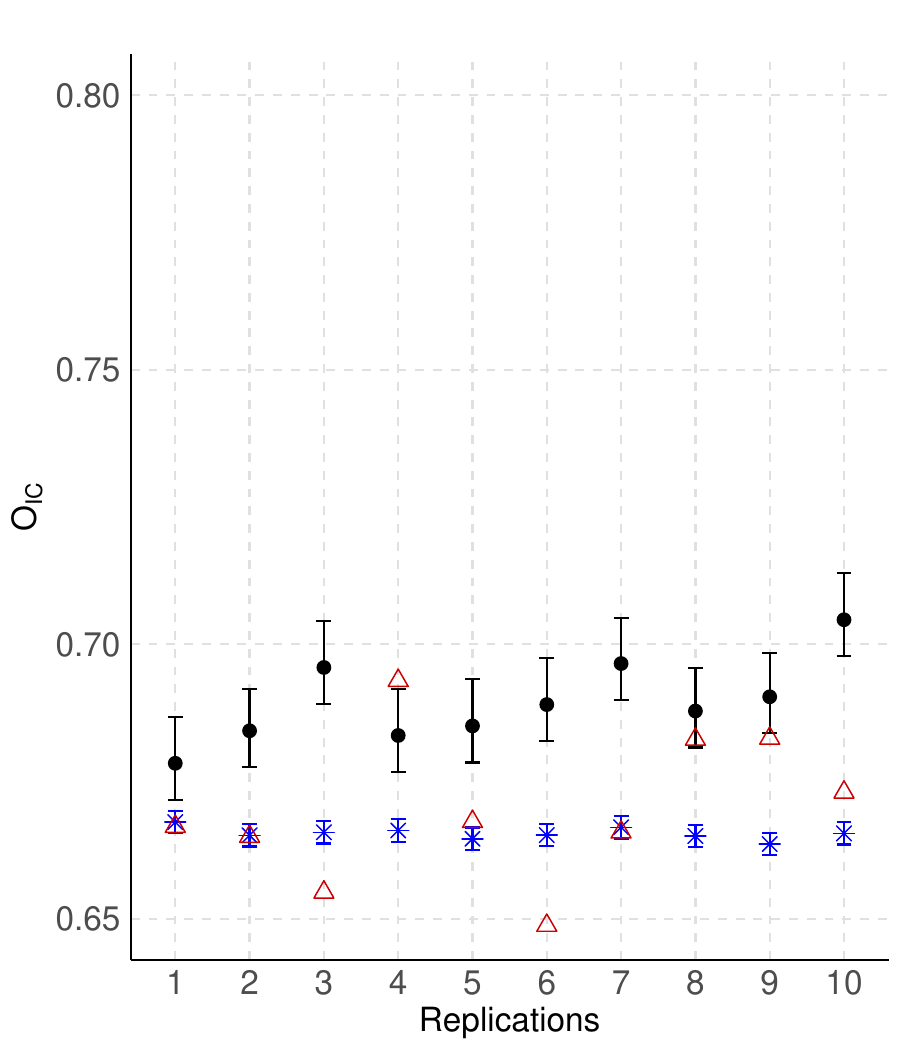}
\end{minipage}
\begin{minipage}{0.49\textwidth}
\centering
\includegraphics[width=\textwidth]{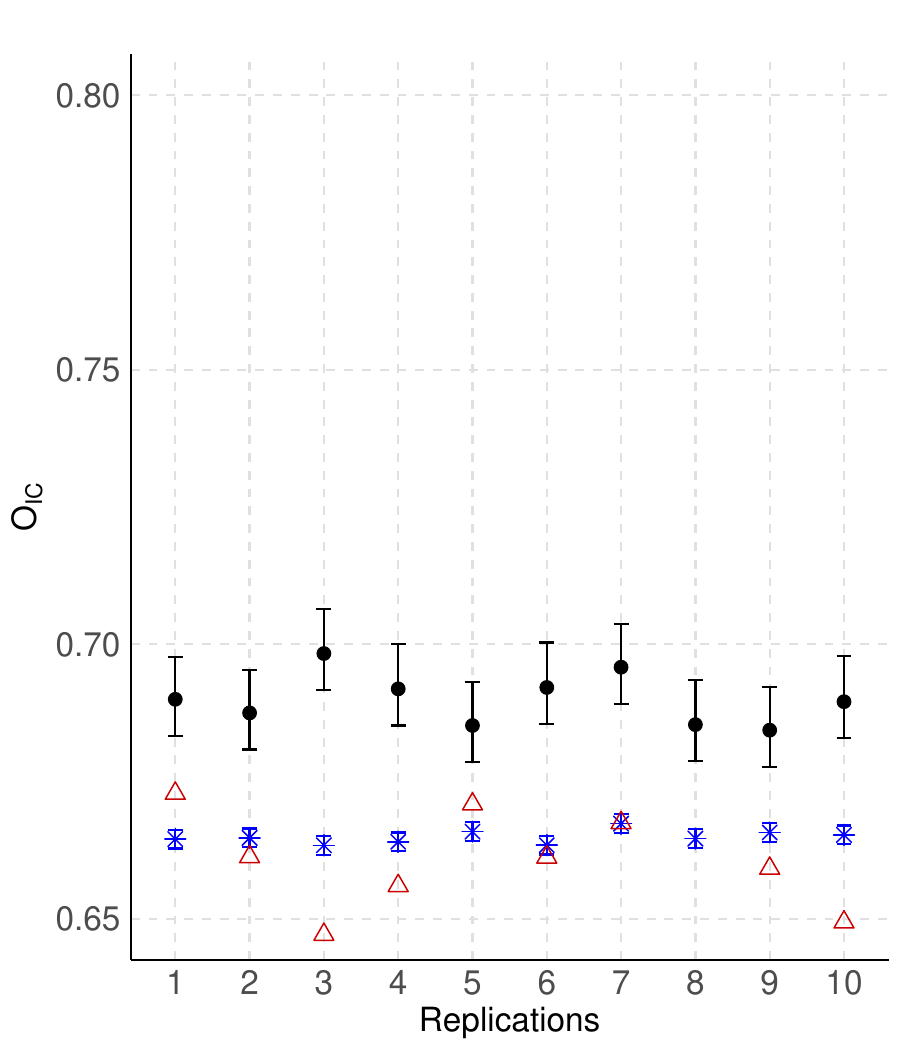}
\end{minipage}

 \caption{$O_{IC}$ measure for $\theta=P(X_5=1|X_6=0,X_7=1)$, with $d=7$. $S_1$: predictive mean (black point) and $98\%$ predictive credibility interval. $S_2$: point estimate (blue asterisk) and $98\%$ confidence interval. $S_3$: point estimate (red triangle). Top: $n=2000$, bottom: $n=5000$. $P_1$ on the left and $P_2$ on the right.}
  \label{ap_5}
\end{figure}
\clearpage
\newpage

\begin{figure}[h!]
\centering
\begin{minipage}{0.40\textwidth}
\centering
\includegraphics[width=\textwidth]{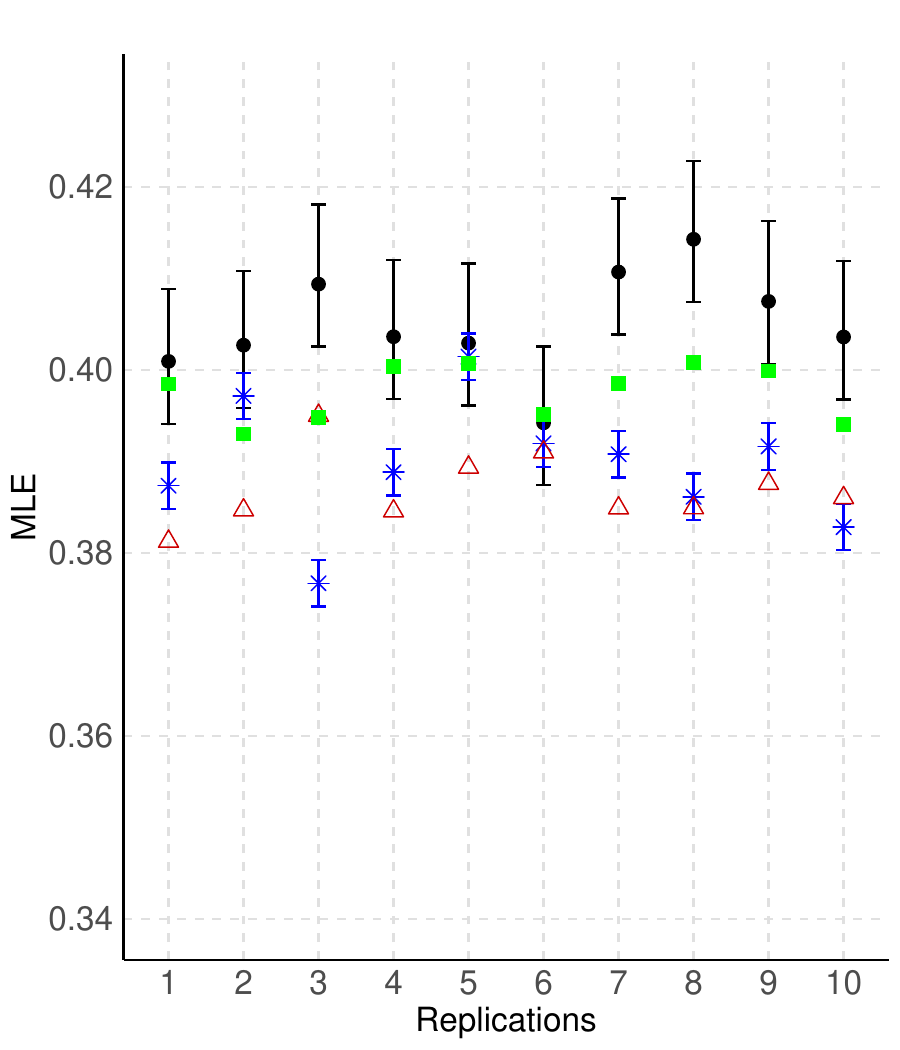}
\end{minipage}
\begin{minipage}{0.40\textwidth}
\centering
\includegraphics[width=\textwidth]{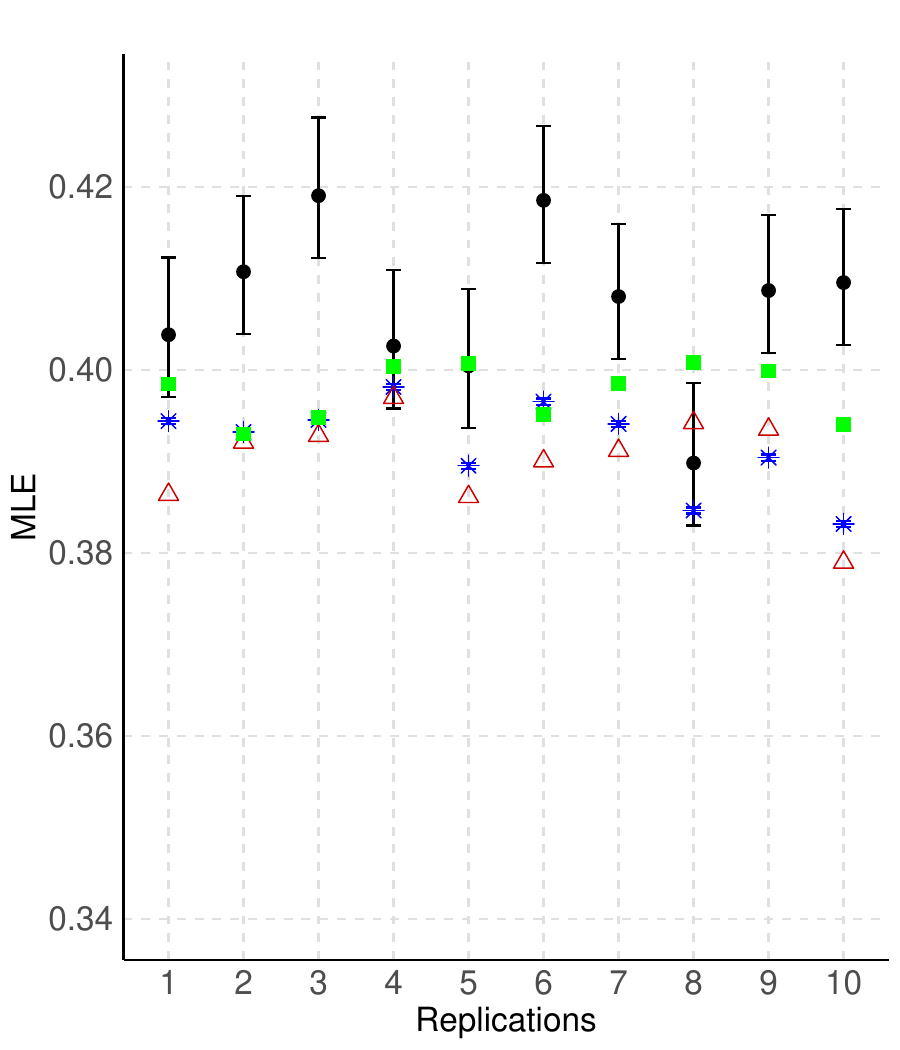}
\end{minipage}

\begin{minipage}{0.40\textwidth}
\centering
\includegraphics[width=\textwidth]{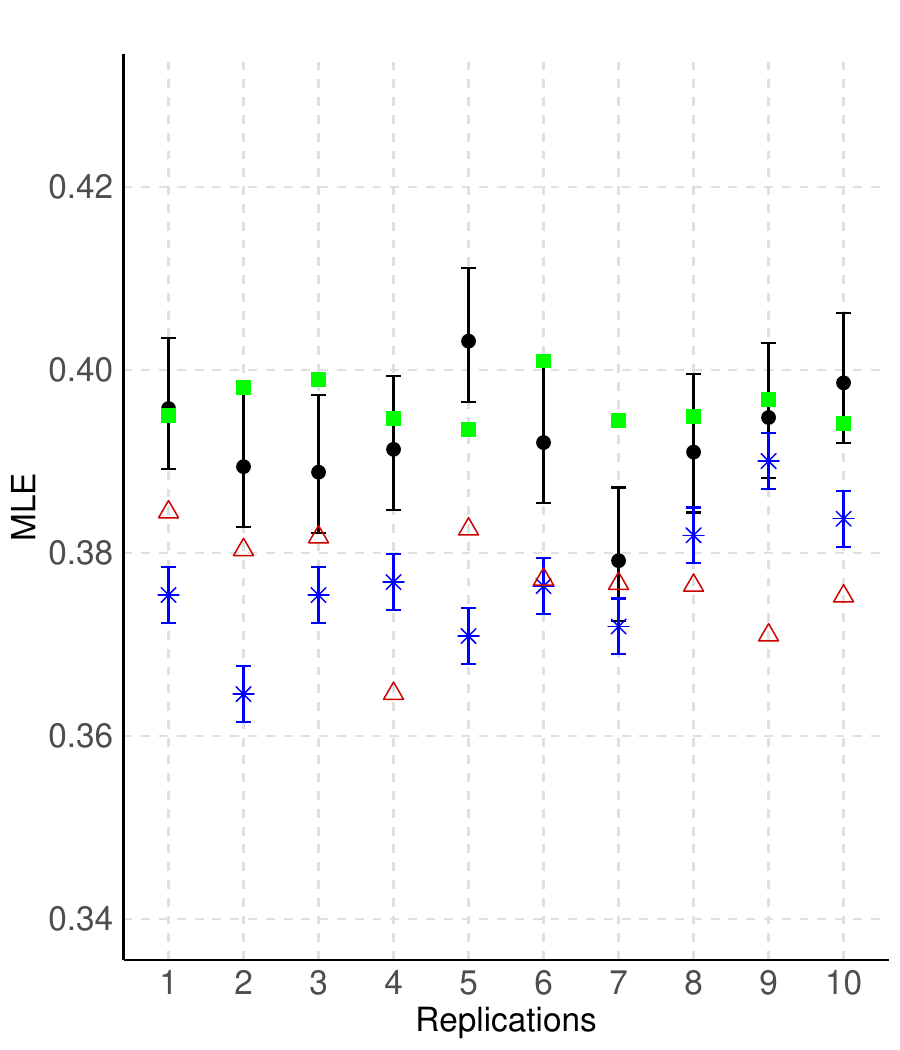}
\end{minipage}
\begin{minipage}{0.40\textwidth}
\centering
\includegraphics[width=\textwidth]{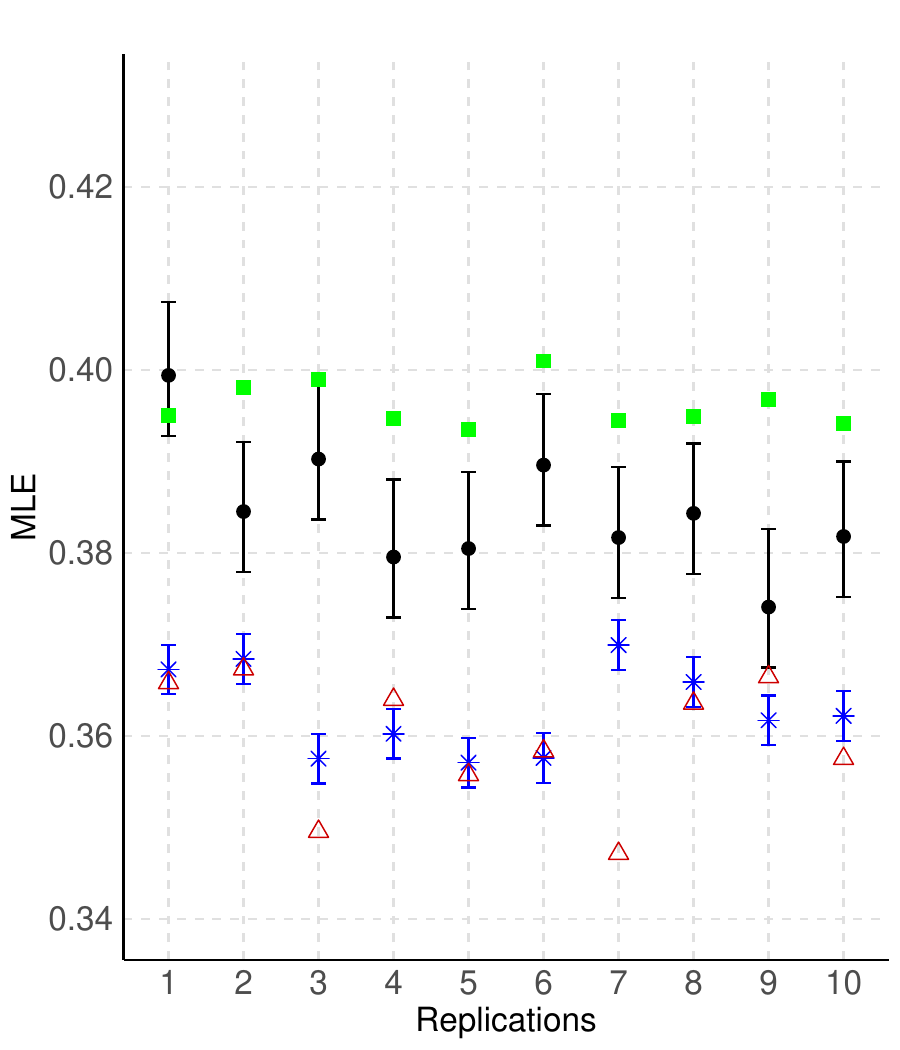}
\end{minipage}

\begin{minipage}{0.40\textwidth}
\centering
\includegraphics[width=\textwidth]{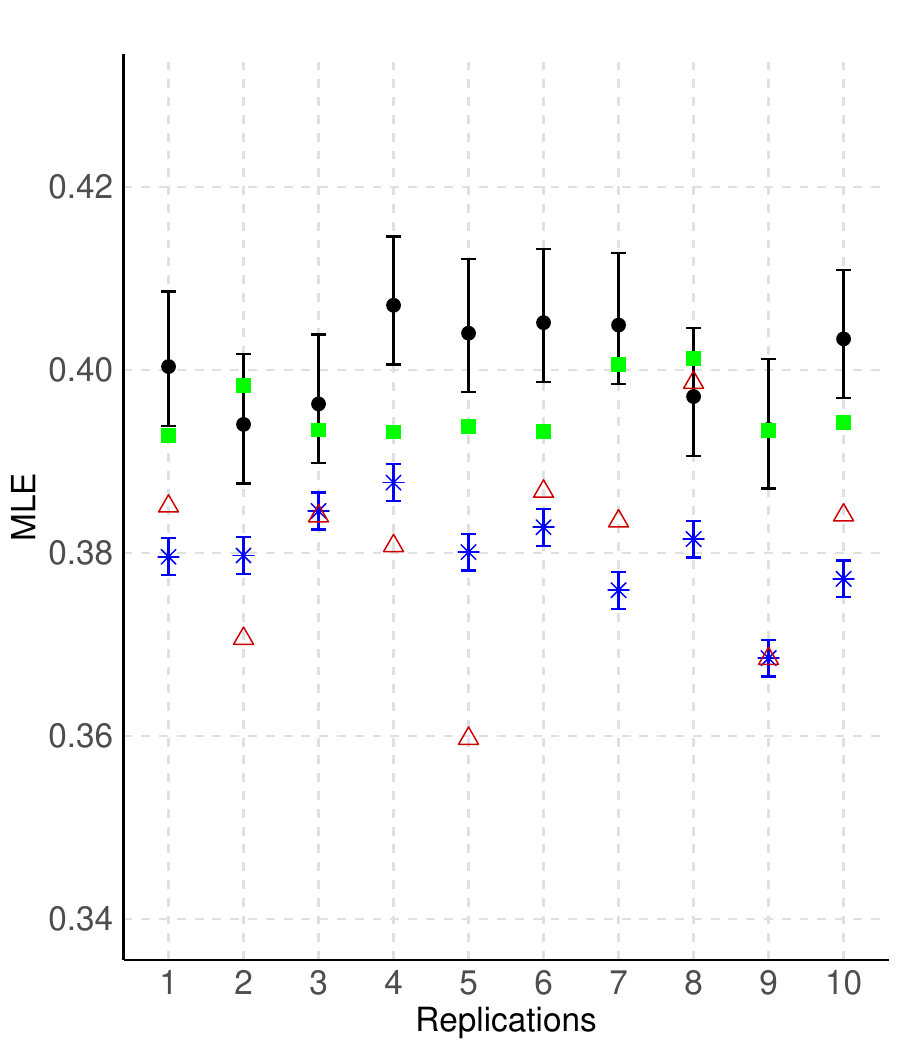}
\end{minipage}
\begin{minipage}{0.40\textwidth}
\centering
\includegraphics[width=\textwidth]{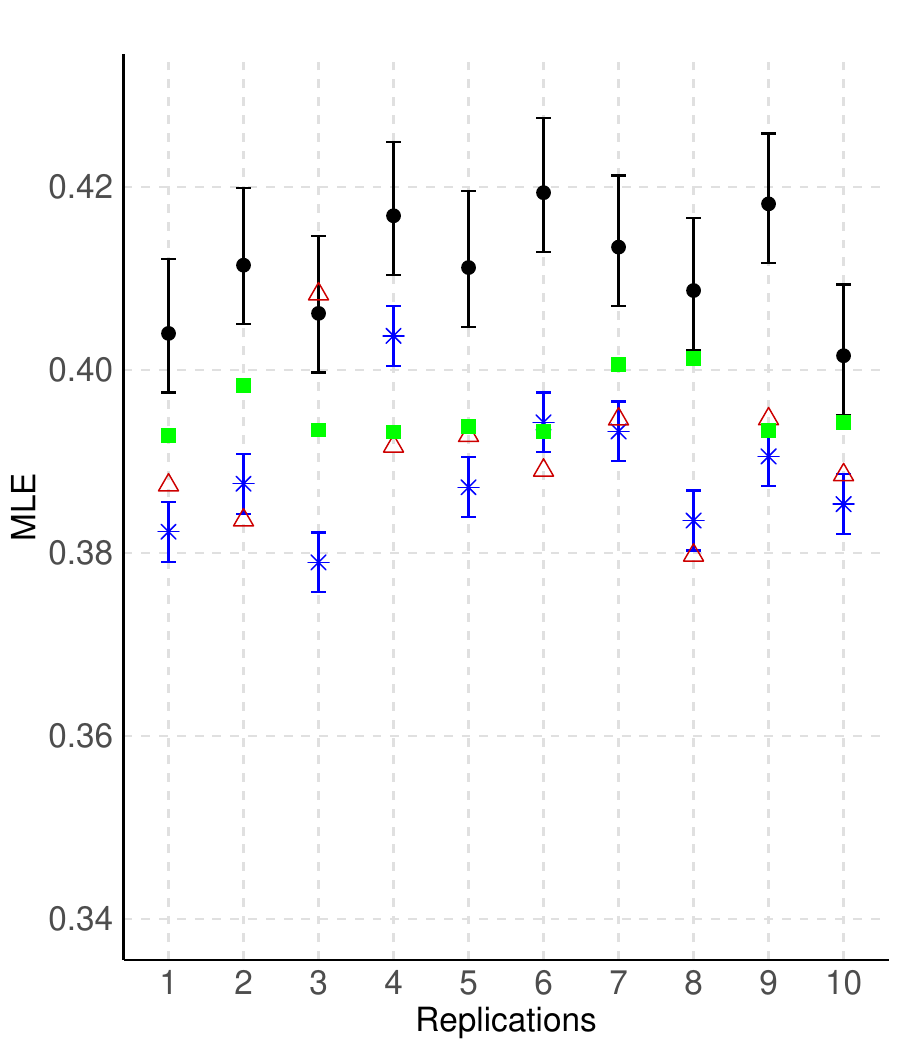}
\end{minipage}

 \caption{$MLE$ measure for $\theta=P(X_1=1)$, with $d=3$. $S_1$: predictive mean (black point) and $98\%$ predictive credibility interval. $S_2$: point estimate (blue asterisk) and $98\%$ confidence interval. $S_3$: point estimate (red triangle). Top: $n=500$, middle: $n=1000$, bottom: $n=5000$. $P_1$ on the left and $P_2$ on the right.}
  \label{ap_6}
\end{figure}
\clearpage
\newpage

\begin{figure}[h!]
\centering
\begin{minipage}{0.40\textwidth}
\centering
\includegraphics[width=\textwidth]{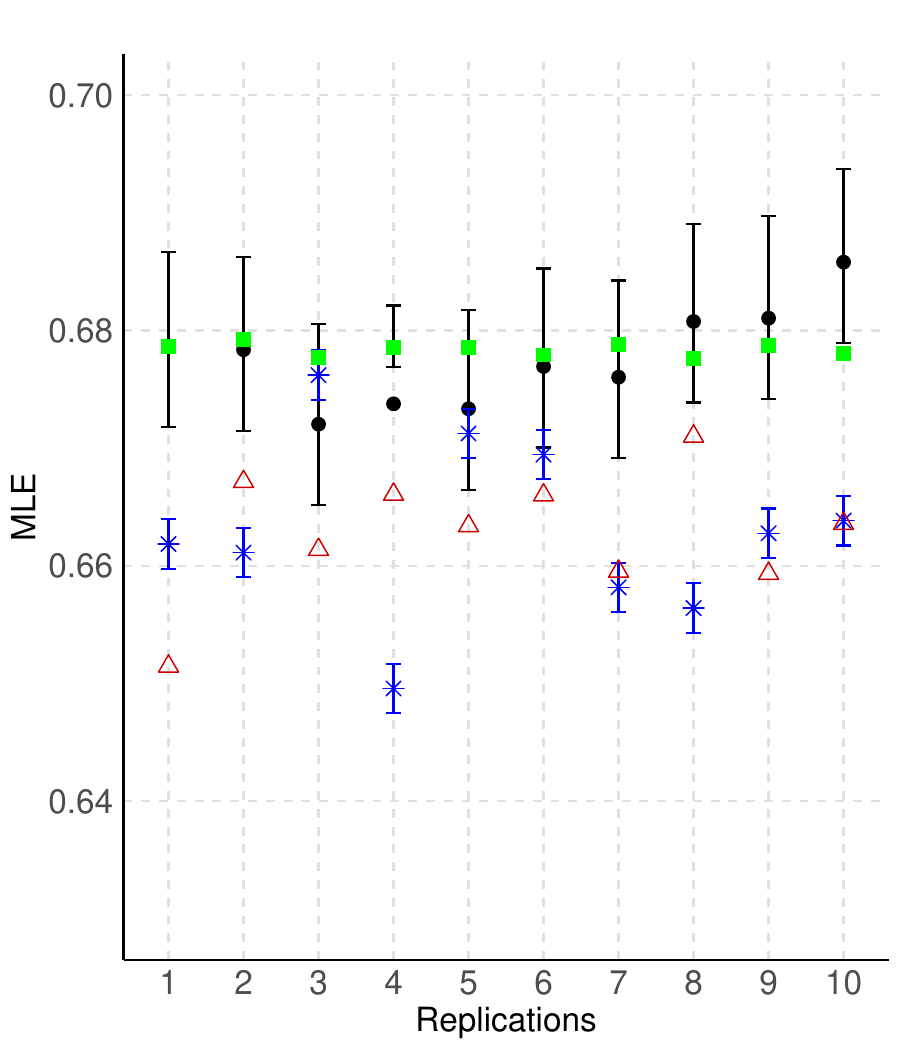}
\end{minipage}
\begin{minipage}{0.40\textwidth}
\centering
\includegraphics[width=\textwidth]{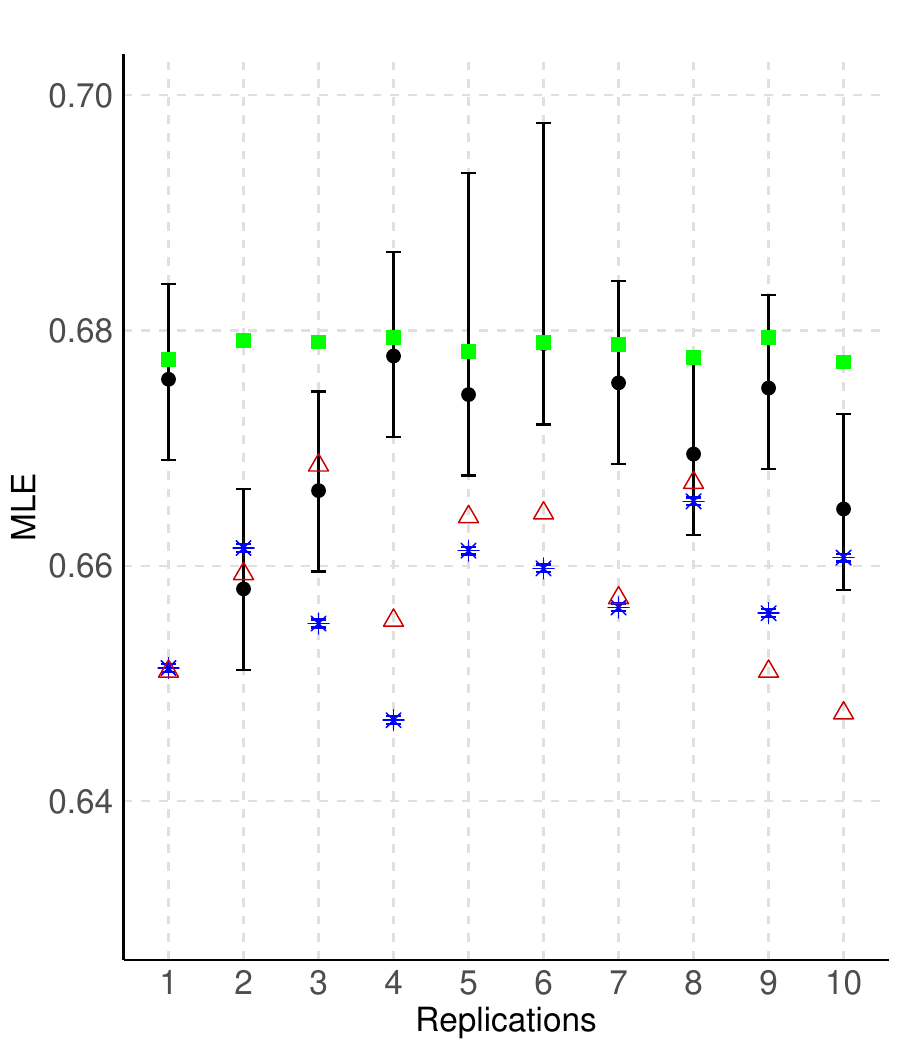}
\end{minipage}

\begin{minipage}{0.40\textwidth}
\centering
\includegraphics[width=\textwidth]{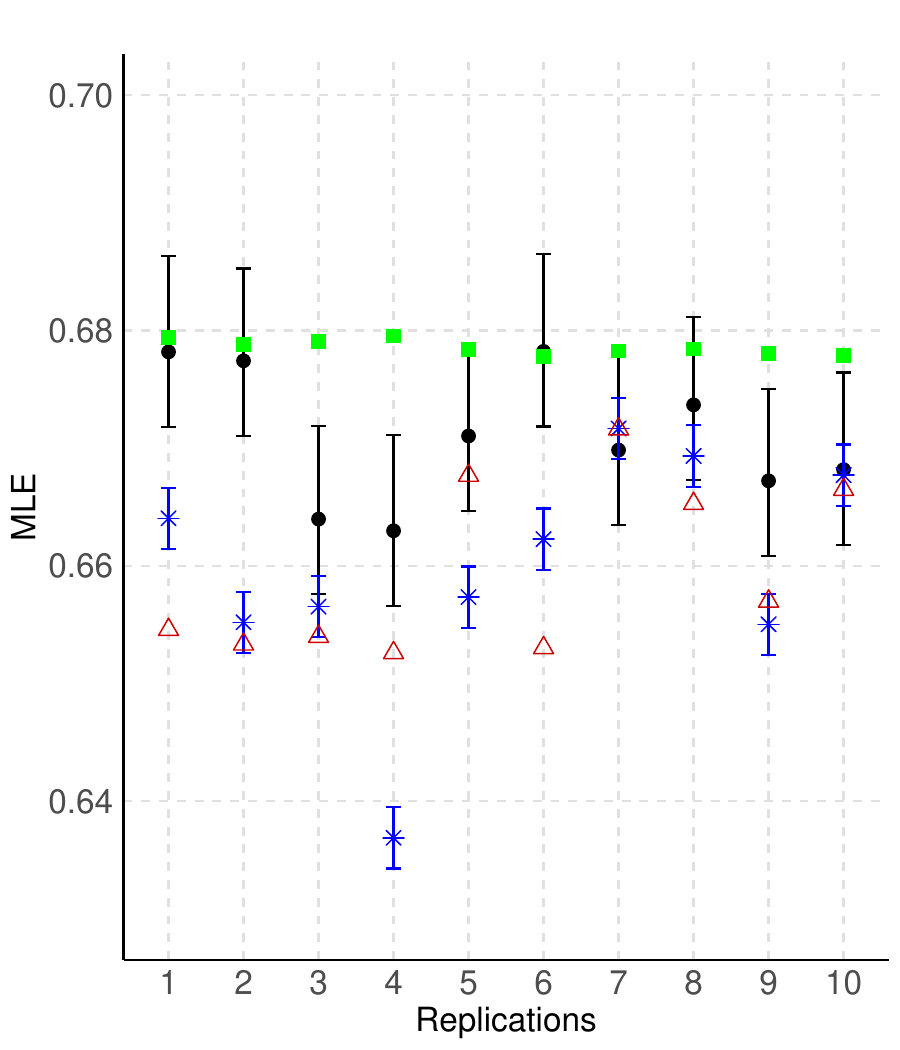}
\end{minipage}
\begin{minipage}{0.40\textwidth}
\centering
\includegraphics[width=\textwidth]{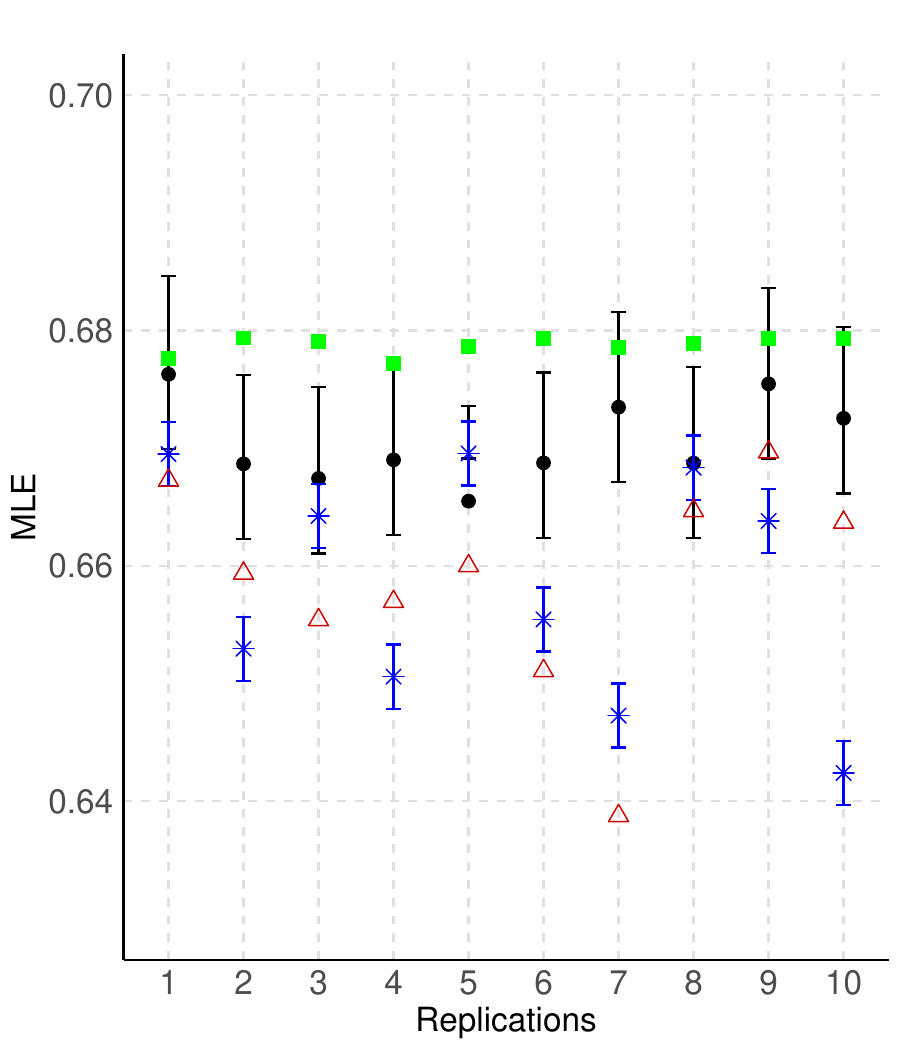}
\end{minipage}

 \caption{$MLE$ measure for $\theta=P(X_3|X_4=1)$, with $d=4$. $S_1$: predictive mean (black point) and $98\%$ predictive credibility interval. $S_2$: point estimate (blue asterisk) and $98\%$ confidence interval. $S_3$: point estimate (red triangle). Top: $n=100$, bottom: $n=5000$. $P_1$ on the left and $P_2$ on the right.}
  \label{ap_7}
\end{figure}
\clearpage
\newpage